\documentclass[galaxies,review,accept,moreauthors,pdftex]{Definitions/mdpi} 
\usepackage{aas_macros}
\usepackage{suffix}
\usepackage{acronym}
\usepackage[shortcuts]{extdash}

\firstpage{1} 
\pubvolume{1}
\issuenum{1}
\articlenumber{0}
\pubyear{2022}
\copyrightyear{2022}
\externaleditor{Academic Editor: Gabriele Vajente}
\datereceived{17 March 2022} 
\dateaccepted{09 June 2022} 
\datepublished{} 
\hreflink{https://doi.org/} 

\definecolor{tiffany}{RGB}{79, 166, 158}
\definecolor{seagreen}{rgb}{0.190, 0.525, 0.361}
\definecolor{amber(sae/ece)}{rgb}{1.0, 0.49, 0.0}

\newcommand{\rev}[1]{{\color{black} #1}}

\newcommand{\msun}{\ensuremath{\mathrm{M}_{\odot}}}
\newcommand{\zsun}{\ensuremath{\mathrm{Z}_{\odot}}}
\newcommand{\rsun}{\ensuremath{\mathrm{R}_{\odot}}}
\newcommand{\mzams}{\ensuremath{\mathrm{M}_{\rm ZAMS}}}
\newcommand{\gpcyr}{\ensuremath{\mathrm{~Gpc}^{-3}\mathrm{yr}^{-1}}}
\newcommand{\gyr}{\ensuremath{\mathrm{~Gyr}}}
\newcommand{\au}{\ensuremath{\mathrm{~au}}}
\newcommand{\valpm}[3]{\ensuremath{#1^{+ #2}_{- #3}}}
\newcommand{\cago}{$^{12}$C($\alpha$, $\gamma$)$^{16}$O}
\newcommand{\mathplus}{+}
\newcommand{\dlog}[2]{\ensuremath{\frac{\mathrm{d}\log #1}{\mathrm{d}\log #2}}}
\extrafloats{100}

\Title{Compact Binary Coalescences: Astrophysical Processes and Lessons Learned}

\TitleCitation{{Compact Binary Coalescences: Astrophysical Processes and Lessons Learned}}

\Author{{Mario} Spera $^{1,2,3,}${*}\orcidA{}, 
Alessandro Alberto Trani $^{4,5,6,}${*}\orcidB{} and 
Mattia Mencagli $^{1}$\orcidC{}}

\AuthorCitation{Spera, M.; Trani, A.A.; Mencagli, M.}

\address{$^{1}$ \quad International School for Advanced Studies - SISSA, Via Bonomea 265, I-34136 Trieste, Italy; \\

$^{2}$ \quad National Institute for Nuclear Physics - INFN, Sezione di Trieste, I-34127 Trieste, Italy\\ 
$^{3}$ \quad National Institute for Nuclear Physics - INFN, Sezione di Padova, Via Marzolo 8, I-35131 Padova, Italy\\
$^{4}$ \quad Research Center for the Early Universe, Graduate School of Science, The University of Tokyo, 7-3-1 Hongo, Bunkyo-ku, Tokyo 113-0033, Japan\\
$^{5}$ \quad Department of Earth Science and Astronomy, College of Arts and Sciences, The University of Tokyo, 3-8-1 Komaba, Meguro-ku, Tokyo 153-8902, Japan\\
$^{6}$ \quad Okinawa Institute of Science and Technology, 1919-1 Tancha, Onna-son, {Okinawa} 904-0495, Japan}

\corres{Correspondence: mario.spera@sissa.it (M.S.);  aatrani@gmail.com (A.A.T.)}

\abstract{On 11 February 2016, the LIGO and Virgo scientific collaborations announced the first direct detection of gravitational waves, a signal caught by the LIGO interferometers on 14 September 2015, and produced by the coalescence of two stellar-mass black holes. The discovery represented the beginning of an entirely new way to investigate the Universe. The latest gravitational-wave catalog by LIGO, Virgo and KAGRA brings the total number of gravitational-wave events to 90, and the count is expected to significantly increase in the next years, when additional ground-based and space-born interferometers will be operational. From the theoretical point of view, we have only fuzzy ideas about where the detected events came from, and the answers to most of the five Ws and How for the astrophysics of compact binary coalescences are still unknown. In this work, we review our current knowledge and uncertainties on the astrophysical processes behind merging compact-object binaries. Furthermore, we discuss the astrophysical lessons learned through the latest gravitational-wave detections, paying specific attention to the theoretical challenges coming from exceptional events (e.g., GW190521 and GW190814).}


\begin{document}

\section{Introduction}\label{sec:introduction}

\textbf{{Merging compact-object binaries}} are binary systems composed of two compact objects that are so close to each other to merge via {gravitational wave (GW)} emission within the age of the Universe. The members of such binaries can be {white dwarfs (WDs)}, neutron
stars (NSs), black holes (BHs), and their combinations, e.g., neutron star-black hole binary
(NSBH) systems. These systems have been investigated for decades by many authors, who predicted their existence through theoretical studies that go from the formation and evolution of the stellar progenitors to accurate numerical relativity simulations of the final merger phase \citep{zeldovich1966,rees1974,lattimer1974,clark1977,clark1979,hils1990,narayan1991,phinney1991,tutukov1993}.


From the observational point of view, proving the existence of merging compact-object binaries has always been challenging. While such systems are potentially loud GW sources, catching their GW signal is not straightforward. The passage of a GW produces a relative change in the distance between two points which is \rev{
\begin{equation}
\frac{\Delta L}{L} \propto \frac{m_1 m_2}{rl_0}\frac{G^2}{c^4},
\end{equation}
where $r$ is the distance from the GW source, $l_0$ is the orbital separation of the binary, $L$ is the reference distance, $m_1$ and $m_2$ are the masses of the GW source, $G$ is the gravitational constant, and $c$ is the speed of light \citep{einstein1918}. The factor $G^2 c^{-4}$ is minuscule ($\sim 5\times 10^{-55} \rm\,m^2\,kg^{-2}$), thus, when a GW reaches the Earth, it causes an extremely small perturbation, which is very hard to detect. }

Even without direct evidences of GWs, the loss of orbital energy of a compact binary via GWs was verified through radio observations of the binary pulsar PSR B1913~+~16 \citep{hulse1975}. 
The observed orbital decay of the \textbf{{Hulse--Taylor binary}} is remarkably consistent with a GW-induced shrinking. This system, which is expected to merge in ${\sim} 300$ Myr, provided not only an additional confirmation of the Einstein's theory of general relativity, but it also suggested to us that there might be not just one, but a population of binary neutron
stars (BNSs) that can merge in relatively short times via GW emission.


For the first direct evidence of merging compact-object binaries and their GW's fingerprint, we had to wait until 14 September 2015, when the two ground-based interferometers of the Laser Interferometer Gravitational-wave Observatory (LIGO) were able to \textit{measure} the effect of a passing GW. The signal, named \textbf{{GW150914}}, was attributed to the coalescence of two stellar-mass BHs with masses $m_1 = \valpm{36}{5}{4}\,\msun{}$ and $m_2 = \valpm{29}{4}{4}\,\msun{}$ \citep{abbott_150914_disc, abbott_150914_astro}\endnote{Throughout this work, we will use the symbol $\msun$ to refer to the Sun's mass.}. \rev{The event carried many scientific implications with itself and it laid the foundations of a new way to investigate the Universe by allowing us to access data never collected before.}

\rev{\textls[-25]{The initial identification of GW150914 was made through an unmodelled, low-latency search for GW bursts, which is a search procedure that does not assume any particular morphology of the GW signal, i.e., it is agnostic with respect to the source's properties~\citep{klimenko2008,abbott_150914_disc,lynch2015}. Later, the event was recovered also by other matched-filter pipelines \citep{abbott2016_150914_firstres}}.}
GW150914 established the existence of binary black holes (BBHs) and that stellar-mass BHs can merge in a Hubble time, becoming detectable sources of GWs. However, the biggest surprise came from \textbf{{the masses of the BHs}}: we did not expect to detect stellar BHs with masses $\gtrsim$$20\,\msun{}$.
 
Prior to GW150914, our knowledge of stellar-mass BHs was limited to electromagnetic observations of Galactic BH X-ray binaries. At the time of GW150914 discovery, there were only a handful of known BHs with confirmed dynamical mass measurements, most of them with mass $\lesssim$$15\,\msun{}$ \citep{ozel2010,farr2011,fishbach2021}. Theoretical models did not predict the existence of BHs with masses $\gtrsim$$20\,\msun{}$, with a few remarkable exceptions \citep{heger2003,mapelli2009,spera2015,mapelli2013,belc2010_maximum,woosley2002,fryer2012,ziosi2014}; thus, we could have only approximate ideas about where the BHs of GW150914 came from. One of the very few clues we were able to obtain was that the heavy compact objects likely formed in a low-metallicity environment, where stellar winds are quenched and stars can retain enough mass to turn into heavy BHs \citep{abbott_150914_astro}. 

\rev{The only solid conclusion was that GW150914 marked a new starting point for the astrophysical community. It gave an unprecedented boost to the development of new theoretical models to study the formation and evolution of compact-object binaries and their progenitor stars, with a new goal: providing an astrophysical interpretation to GW~sources.}

From the theoretical point of view, two main formation channels have been proposed so far for the formation of merging compact objects. In the \textbf{isolated binary channel}, two progenitor stars are bound since their formation; they evolve, and then turn into (merging) compact objects at the end of their life, without experiencing any kind of external perturbation \cite{webbink1975,paczynski1976,vanDenHeuvel1976,hut1981,webbink1984,bethe1998,belc2002,hurley2002,demink2008a,dominik2012,postnov2014,demink2016,mandel2016a,marchant2016,stevenson2017,giacobbo2018a,zevin2021}. \rev{This scenario is driven by single and binary stellar evolution processes, and it is sometimes referred to as the ``field'' scenario, because it assumes that binaries are born in low-density environments, i.e., that they evolve in isolation}. In contrast, in the \textbf{dynamical channel}, two compact objects get very close to each other after one (or more) gravitational interactions with other stars or compact objects. This evolutionary scenario is quite common in dense stellar environments (e.g., star clusters), and it is driven mainly by stellar dynamics \cite{heggie1975,lightman1978,hills1980,mcmillan1991,hut1992,sigurdsson1993,portegieszwart2000,miller2002,oleary2006,downing2010,ziosi2014,rodriguez2015,antonini2016a,mapelli2016,kimpson2016,banerjee2017,samsing2017,zevin2021,trani2022}. In reality, the two formation pathways might have a strong interplay. In star clusters, the orbital parameters of binaries might be perturbed by many passing-by objects. Dynamical interactions might be strong enough to eject the stellar binary from the cluster and to trigger the merger event in the field. Such an apparently isolated merger would not have occurred if the progenitor stars had evolved in isolation. \rev{Such \textbf{hybrid scenarios} blur the line between the dynamical and the isolated binary channel, and they have already been investigated by various authors \citep{kumamoto2019,dicarlo2020,trani2021,arcasedda-bpop}}.

Our theoretical knowledge of the formation scenarios is hampered by the \textbf{uncertainties and degeneracies of the astrophysical models}. Single-star evolutionary tracks, the strength of stellar winds (especially for massive stars at low metallicity), core-collapse and pair-instability supernova (PISN), the orbital parameters of binary stars at birth, binary mass transfer, compact-object birth kicks, stellar mergers, tidal interactions, common envelope
(CE), and GW recoil, are only part of the uncertain ingredients of the unknown recipe of merging binaries. In contrast, stellar dynamics is {{simple and elegant}}, but developing accurate and fast algorithms for the long-term evolution of tight binaries is challenging. Furthermore, studying the evolution of small-scale systems (2 bodies within $\sim$$10^{-6}$ pc) in large star clusters ($\gtrsim$$10^5$ objects within a few pc) is computationally intensive \cite{mikkola1998,aarseth2003,konstantinidis2010,hypki2012,dolcetta2013, dolcetta2013_b, rodriguez2016b,maureira2017,wang2020,mencagli2022}.

Therefore, disentangling different shades of flavors by tasting the final result and going back to the responsible ingredients is very challenging. The consequence is that \textbf{the GW catalog is growing faster than our theoretical understanding} of merging compact-object binaries. At the time of writing, we have already achieved an historic breakthrough: we have just started talking about a \textit{population} of BHs. Indeed, the latest Gravitational Wave
Transient Catalog (GWTC) reports $\sim$$90$ events\endnote{LIGO-Virgo-KAGRA (LVK)-independent analyses have even found a few additional GW candidates (e.g., \citep{nitz2021_catalog, olsen2022, zackay2019}).}, mostly BBH mergers, and the count is expected to significantly increase in the next years, at even faster rates than ever because new ground-based interferometers will be operational and the existing ones will increase their sensitivity \citep{gwtc3_disc_2021,lvcgwtc3_2021}.

The catalog already contains many flavors of BBHs that challenge even up-to-date theoretical models.
For instance, \textbf{GW190814} (see Section~\ref{subsection:gw190814}) is an event with very asymmetric masses, a merger that most theoretical models find very difficult to explain \citep{LVK190814_2020}. Furthermore, the lightest member is a mystery compact object with an uncertain nature: it can be the heaviest NS or the lightest BH ever observed and its mass falls right into the lower mass gap (see Section~\ref{subsec:compactremnants}).  \textbf{GW190521} (see Section~\ref{subsection:gw190521}) is the event with the heaviest BHs, with at least one of the two falling in the upper mass gap (see Section~\ref{sec:uppermassgap}) \citep{lvk_190521_2020, lvk_190521_astro2020}. Its merger product, a BH with mass $\valpm{148}{28}{16}\,\msun{}$, is the first confirmation of the existence of intermediate-mass BHs. \textbf{GW200105\_162426} and \textbf{GW200115\_042309} (see Section~\ref{subsection:gw200105_gw200115}) are the first NSBHs ever observed \citep{lvk_nsbh_2021}.  \textbf{GW170817} is associated with a merger of two NSs and it is the only event observed not only through GWs but also throughout the whole electromagnetic spectrum, a crucial milestone for \textbf{multi-messenger astronomy} \citep{lvk_170817_2017}. There are also 5 events with preference for \textbf{negatively aligned spins} with respect to the orbital angular momentum of the binary, including the mentioned GW200115\_042309. Spins and their in-plane components might provide important insights on the formation channels (see Section~\ref{subsection:spins}). The BH mass distribution \rev{and the inferred BBH merger rate make the current scenario even more complex. The former seems to have statistically significant \textbf{substructures}, that is, it shows up as clumpy, with BHs that tend to accumulate at chirp masses\endnote{$\mathcal{M}=\frac{\left(m_1m_2\right)^{3/5}}{\left(m_1+m_2\right)^{1/5}}$} $\mathcal{M}=8\,,14\,,27\, \msun{}$, whereas the latter increases with redshift \citep{lvcgwtc3_2021, tiwari2021, karathanasis2022, mukherjee2021}.}

Rather than presenting new results, in this work we review our knowledge of the main astrophysical processes that lead to merging compact-object binaries, focusing mainly on BHs. Furthermore, we discuss the clues we can currently collect on the astrophysical origin of some exceptional GW events, and we discuss the main astrophysical lessons learned so far. 
This is surely a rapidly evolving field (see also the reviews by \citep{mapelli2021_book,mandel2021_review}), and most of the topics reported in this work would deserve a review on their own right. Here, we just give an overview of the main aspects that are relevant for the formation of compact objects. 

In Section \ref{sec:singlestars} we discuss the evolution of single stars and their relation to compact remnants, Section \ref{sec: binarystars} deals with binary stellar evolution processes, in Section \ref{sec:dynchannel} we examine the effects of stellar dynamics, Section \ref{sec:lessonslearned} presents the astrophysical lessons learned through GW events, and Section \ref{sec:conclusions} contains a summary and a brief outline of future prospects.

\section{Single Stars}\label{sec:singlestars}

Throughout this work, we will often refer to \textbf{population-synthesis simulations}. To conduct statistical studies on stellar populations and their compact objects, we should follow self-consistently the evolution of any possible type of single and/or binary star from its formation to its death, and possibly beyond. This is prohibitive if we consider that simulating the evolution of just {one} star from the main sequence until core collapse might take days (if the complex underlying algorithms converge). Thus, for {{fast population-synthesis studies}}, the evolution of single stars is approximated through either (i) fitting formulas to detailed stellar evolution calculations (e.g., \citep{hurley2002}) or (ii) the interpolation of look-up tables containing pre-evolved stellar evolution tracks for different stars at various metallicity (e.g.,~\citep{spera2015}). Binary stellar evolution (see Section~\ref{sec: binarystars}) and other additional processes (e.g., supernova explosions) are generally added through analytical prescriptions on top of the single-star approximations. Fast population-synthesis codes are currently the main resource available to study compact objects from single and binary stars. 
\subsection{Overview}\label{sec:overview}

The life of a star can be thought as a series of gravitational contractions of the whole structure, and expansions under the influence of thermonuclear fusions of increasingly heavy elements in the core, until the formation of the nuclides of the iron group. Each gravitational contraction increases the central temperature until the heaviest element is ignited. After the exhaustion of input elements in the core, the burning process continues in an outer shell while the core contains the heavier products of the previous thermonuclear~reactions. 

The fusion of elements lighter than iron is an exothermic reaction, which means that it \textit{releases} energy, balancing the thermal energy that stars lose via radiation. However, the average binding energy per nucleon starts to decrease for elements heavier than iron-56, thus forming these elements is an endothermic process, i.e., it \textit{requires} energy. In reality, (i) the chain of nuclear reactions could continue until the formation of nickel-62, which is the nuclide with the highest binding energy per nucleon, but photodisintegration suppresses its formation, and (ii) iron-56 forms as nickel-56 decays ($^{56}_{28}\mathrm{Ni}\rightarrow\, ^{56}_{27}\mathrm{Co}+e^++\nu_{\rm e}+\gamma$ and $^{56}_{27}\mathrm{Co}\rightarrow\,^{56}_{26}\mathrm{Fe}+e^++\nu_{\rm e}+\gamma$), therefore, \textbf{nickel-56} is the heaviest element that stars can produce efficiently through nuclear fusion ($^{52}_{26}\mathrm{Fe}+^{4}_{2}\mathrm{He}\rightarrow\,^{56}_{28}\mathrm{Ni}+\gamma$).

\begin{figure}[H]
    
    \includegraphics[width=0.7\columnwidth]{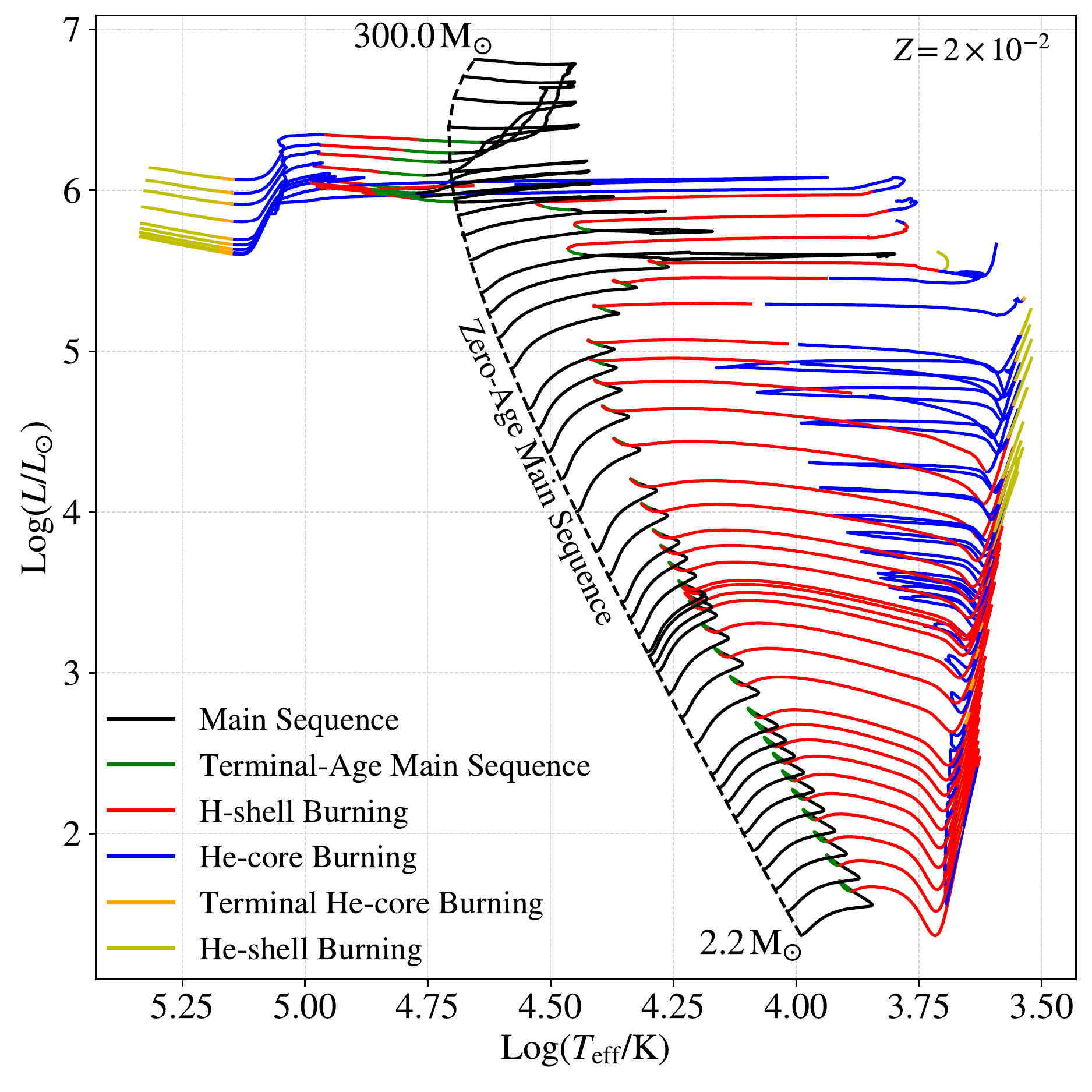}
    \caption{Hertzsprung-Russell diagram (luminosity versus effective temperature) that shows the ZAMS (dashed black line) of stars with masses in the range $2.2\, \msun{}$--$300\, \msun{}$, at metallicity $Z = 2\times 10^{-2}$. The plot also shows the stars' main evolutionary stages (solid lines) until carbon ignition, and it has been obtained through the SEVN population-synthesis code {\citep{spera2017}} coupled with the look-up tables for single-star evolution from the PARSEC code \citep{bressan2012,chen2015}.}
    \label{fig:parseczams}
\end{figure}

Stars spend most of their life on the main sequence, that is transforming hydrogen into helium in their innermost regions. \rev{The moment when a star ignites hydrogen in its core defines the zero age main sequence (ZAMS). The ZAMS line appears as a quasi-diagonal line in the Hertzsprung-Russell diagram (Figure~\ref{fig:parseczams}), which is a standard tool for representing the evolutionary stage of stellar populations. While the bulk of the properties of a star is determined only by its mass at the ZAMS ($\mzams$), the mass of the final compact remnant crucially depends also on parameters like the chemical composition and stellar rotation. These parameters control the efficiency of the processes that affect the final mass of the remnant, i.e., they determine how much mass is lost through stellar winds, how much the stellar core can grow, and how much mass is lost in the supernova (SN) explosion.}

However, as a first approximation, $\mzams$ is indicative of which remnant the star will leave at the end of its life.
Very low-mass stars ($\mzams \lesssim 0.26 \,\msun$) do not reach the threshold temperature for helium ignition, and after their long ($\gtrsim$$70$ Gyr) main sequence phase, they become helium WDs (e.g., \cite{eddington1924}).
Stars with $0.26\,\msun \lesssim \mzams \lesssim 8\,\msun$ ignite helium and form a carbon-oxygen (CO) core but do not reach temperatures high enough to ignite CO. After the formation of a CO-core, nuclear reactions in the core stop. At this point, the star is supported only by electron degeneracy pressure. At the end of its life, the star will eject most of its outer shells creating a planetary nebula. What is left at the center is a CO WD (e.g., \citep{eddington1924}).
Stars with ZAMS masses above ${\sim} 8\,\msun$ can reach iron elements, and their life will end with a SN, possibly leaving behind a NS ($8\, \msun{} \lesssim \mzams{} \lesssim 20\,\msun{}$) or a BH ($\mzams{} \gtrsim 20\,\msun{}$).

The limits of these mass ranges are quite uncertain. We might say that these limits are the \textbf{{zero-level uncertainty}} to our understanding of the mass spectrum of compact objects. On the one hand, the uncertainty stems from theoretical modeling of detailed stellar evolution processes, such as convection, dredge up, wind mass loss, and nuclear reaction rates \cite{woosley2002,smartt2009,horiuchi2011,smartt2015,aguirre2017,vink2021}. On the other hand, the limits also depend on other stellar parameters, such as rotation and chemical composition. 
These uncertainties affect the masses of stellar cores, which, in turn, have an impact on the nature and abundance of compact remnants that stars may form.

\subsection{The Chandrasekhar Limit}

The maximum mass of a WD is well constrained through theoretical arguments, which were firstly outlined by \citet{chandrasekhar1935}.
This limit exists because WDs are sustained against gravity by the pressure of electron degeneracy, which can be either non-relativistic or relativistic. Simple stellar polytropic models show that a star supported by a non-relativistic degenerate electron gas has a radius that is inversely proportional to the cube root of its mass, $R \propto M^{-1/3}$ \cite{chandrasekhar1935}. By looking at the scaling relation, one would expect that the WD radius becomes exceedingly small for exceedingly large masses. However, as the density increases, the electrons become relativistic, and the WD becomes supported by a relativistic degenerate electron gas. In such a state, the corresponding equation of state predicts the existence of a maximum sustainable mass: this is the maximum mass of a WD, also referred to as the \textbf{Chandrasekhar mass limit}. Its precise value depends on the average molecular weight per electron, which, in turn, depends on the chemical composition of the WD. For a typical CO or helium WD, the Chandrasekhar limit is $M_{\rm c} \simeq 1.44 \, \msun$.

\subsection{The Tolman-Oppenheimer-Volkoff Limit}

The maximum mass value for a NS, analogue to the Chandrasekhar limit, is \textbf{the
Tolman–Oppenheimer–Volkoff (TOV) limit} \citep{oppenheimer1939,tolman1939}. In this case, support against gravity is provided by the degenerate pressure of a neutron gas. However, unlike in the case of the degenerate electron gas in WDs, neutron-neutron interactions become a crucial (but very uncertain) ingredient to include in the equation of state. Thus, the TOV limit reflects our uncertainties on the NS equation of state. In principle, depending on the adopted equation of state, the TOV limit can be anywhere from $0.5\,\msun{}$ to $3 \,\msun$ \cite{rhoades1974,kalogera1996,srinivasan2002,ozel2010,margalit2017,shibata2017,linares2018,alsing2018,rezzolla2018,ruiz2018,cromartie2020,li2021,angli2021}. The observations of NS masses $\gtrsim$$1\,\msun$ (e.g., those in binary pulsars, such as PSR B1913~+~16) ruled out the softest equations of state, placing the TOV limit at $M_{\rm TOV}\simeq1.5$--$3\,\msun$.
The detection of the GW signal from merging NSs can also be used to constrain the maximum NS mass. In fact, tidal deformations during the last phase of the inspiral affect the properties of the gravitational waveform, which can then be linked to the NS equation of state. The analysis of GW170817 data constrains $M_{\rm TOV} \lesssim 2.3 \,\msun$ \cite{shibata2019,kashyap2021}. Stellar rotation may also play a role, with rigidly rotating NSs having about 25\% larger allowed masses \cite{friedman1987,lasota1996,lattimer2005,breu2016}. 

The secondary compact object of GW190814 (${\sim} 2.6\, \msun$), if a NS, might challenge our understanding of the maximum mass of NSs. Because of the lack of a clear signature of tidal deformations in the GW190814 signal, and the poor constraints on the secondary's spin, no definitive conclusions on the nature of the less massive component of GW190814 exist to date (see also Section~\ref{subsection:gw190814}).

\subsection{The Role of Stellar Winds}\label{sec:singlewinds}

The nature and final mass of a stellar remnant depends crucially on the final properties of the stellar core, which, in turn, depend on the amount of mass a star has lost during its life. Stellar winds have a central role in this picture since they \textit{drive} mass loss over the lifetime of a star. \textbf{Stellar winds}, especially for massive stars, are uncertain, and even a factor of 2 uncertainty (typical for state-of-the-art models, e.g., \citep{vink2021}) might have important consequences on the nature and mass spectrum of compact objects.

Wind mass loss originates from the complex \textbf{interaction between radiation and matter in stellar atmospheres}. The idea that the outer layers of stars could expand was introduced already at the beginning of the 
\rev{XX century by \citet{saha1919}}. \citet{saha1919} suggested that radiation could be absorbed by matter in the solar atmosphere through an inelastic impact, with a resulting forward velocity of $h\nu/mc$, where $h$ is the Plank length, $\nu$ is the frequency of the photon and $m$ the rest mass of matter. We now know that the strongly anisotropic and continuous component of photons from the innermost layers constantly exchanges energy and momentum with free electrons, ions, atoms and dust grains in stellar atmospheres. The momentum equation, considering only a radial direction of the radiation (1D problem), reads
\begin{equation}
    v\frac{dv}{dr} = - \frac{GM_*}{r^2} - \frac{1}{\rho}\frac{dp}{dr} + g_{\rm rad}
\end{equation}
where $M_*$ is the mass of the star, $r$ is the distance from the center of the star, $v\left(r\right)$ is the local wind velocity, $\rho\left(r\right)$ is the local density, $p\left(r\right)$ is the local pressure of gas, and $g_{\rm rad}$ the radiative acceleration.
For the atmospheres of hot massive stars ($T_{\rm eff}\gtrsim 10^4\, K$), $g_{\rm rad} = g_{\rm con} + g_{\rm line}$, where $g_{\rm con}$ is the electron scattering acceleration, and $g_{\rm line}$ is the \textit{selective} acceleration caused by spectral line opacity. While $g_{\rm con}$ is quite well established, the challenge is to estimate $g_{\rm line}$\endnote{In cool supergiants ($T_{\rm eff}< 10^4\, K$) the mechanism responsible for winds is the absorption of photons by dust grains, i.e., dust- (or continuum-) driven winds.}. The latter gained increasing importance over the years, especially after the discovery of blue-shifted resonance lines of carbon IV, silicon IV, and nitrogen V, in the OB supergiants $\delta$, $\epsilon$, and $\zeta$ Orionis, which suggested expansion velocities ${\sim} 1400 \,\mathrm{km\,s}^{-1}$ and mass outflows of ${\sim} 10^{-6}\, \msun{}\, \mathrm{yr}^{-1}$ \citep{morton1967_A, morton1967_B}.

\citet{castor1975} introduced the idea to express $g_{\rm line}$ through a force multiplier (\textbf{CAK theory}), that is
\begin{equation}
    v\frac{dv}{dr} = - \frac{GM_*}{r^2} - \frac{1}{\rho}\frac{dp}{dr} + g_{\rm con}\left[1 + M\left(\tau\right)\right]
\end{equation}
where $\tau$ is an optical-depth scale for the wind, assumed to depend \textit{only} on the local conditions where the absorption occurs, including the wind velocity gradient \citep{sobolev1960}, and $M\left(\tau\right)=k\tau^{-\alpha}$. \citet{castor1975} showed that all spectral lines should be included in the calculation of $M\left(\tau\right)$, not only the resonance lines, and that $M\left(\tau\right) \gg 1$, that is \textbf{stellar winds in hot massive stars are \textit{line driven}.} 

\textls[-15]{\rev{More and more spectral lines were included in the CAK theory over the years, and the contribution of \textbf{metals}---elements heavier than helium---to stellar winds became increasingly important.} Indeed, hydrogen and helium have very few spectral lines in the UV (i.e., the radiation peak frequency of hot massive stars), thus their contribution is expected to be minimal compared to that coming from metals, which have crowded line spectra in the UV band. \rev{From CAK theory, it was already clear that \textbf{stellar winds are quenched at low metallicity}, that is the mass fraction of metals in a stellar layer. Denoting the mass loss by winds as $\dot{M}_*$ and the metallicity as $Z$, $\dot{M}_*\propto Z^{x}$ with $x$ ranging from ${\sim} 0.5$ \citep{kudritzki1987} to ${\sim} 0.9$ \citep{abbott1982}.}}

\rev{
It is nowadays understood that the amount and type of metals in the stellar atmosphere affect greatly the mass of the star prior to the SN explosion. Figure~\ref{fig:winds} shows the typical impact that different values of metallicity have on the final mass of the stars. It is apparent that stars at low $Z$ retain \textit{significantly} more mass than stars at higher $Z$, thus the former can collapse to significantly heavier BHs.

\begin{figure}[H]
	
	\includegraphics[width=0.7\columnwidth]{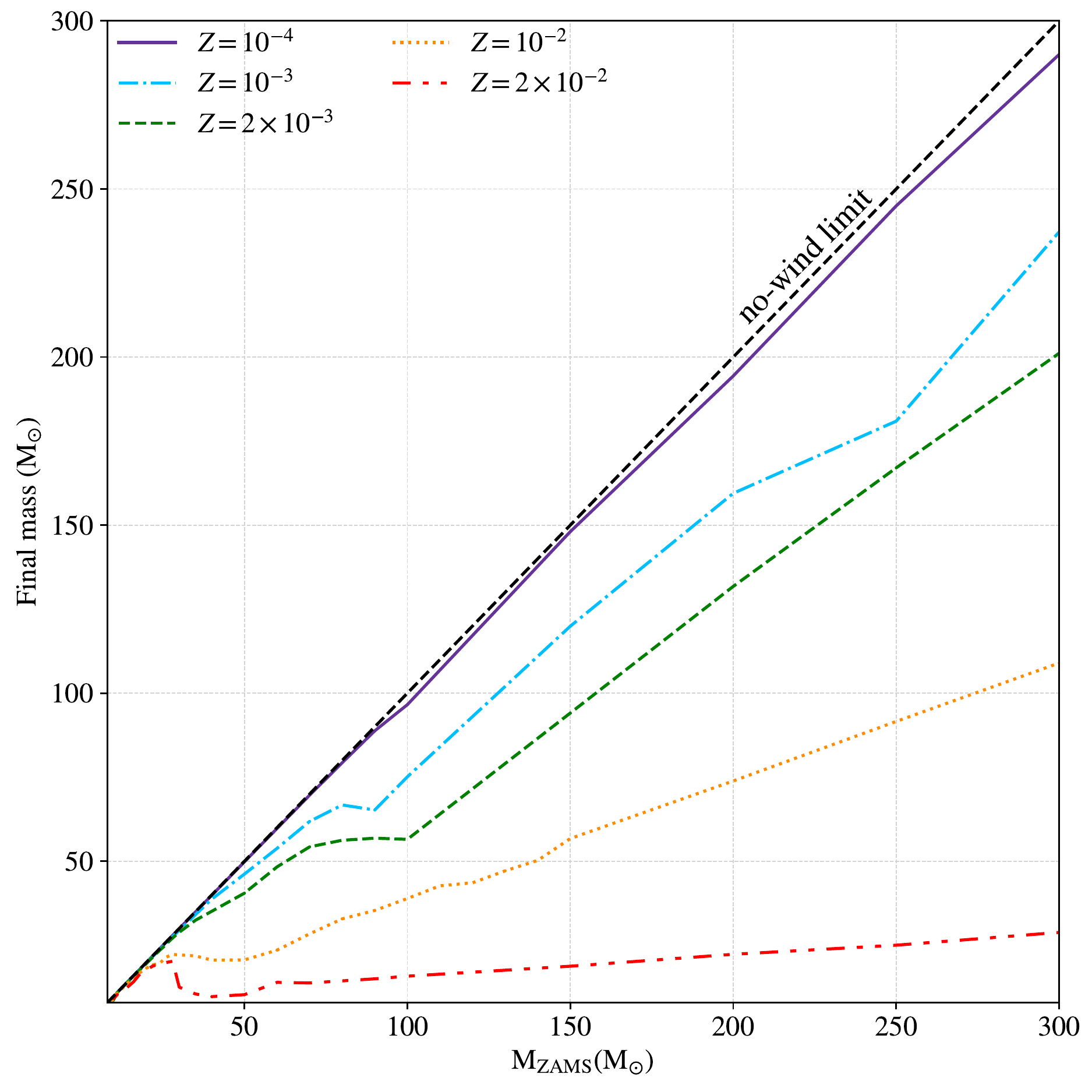}
	\caption{Final mass of the stars as a function of their initial mass, for different values of metallicity. The dashed line at 45 degrees corresponds to the no-wind limit (i.e., final mass = initial mass). Plot obtained with the SEVN population-synthesis code {\citep{spera2017}} with look-up tables for single-star evolution from the PARSEC code \citep{bressan2012}.}
	\label{fig:winds}
\end{figure}
}

\rev{The precise dependency of winds strength on the amount and kind of metals is still matter of debate}. \citet{abbott1985} introduced an alternative approach to CAK based on a Monte Carlo method capable of tracking photon paths and includes the possibility of multiple photon scatterings. 
\citet{vink2001} used this improved approach to re-investigate the $\dot{M}_* - Z$ relation and found $\dot{M}_*\propto Z^{0.69}$ ($\dot{M}_*\propto Z^{0.64}$) for O (B) stars\endnote{It is worth noting that wind mass loss does not depend only on metallicity, but also on luminosity, effective temperature, stellar mass, and the velocity of wind at infinity.}. \citet{vink2001} pointed out that the dominant contribution to the inner (subsonic) stellar wind for O stars at high metallicity ($Z$$\sim$$Z_{\odot}$) comes from the \textbf{Fe-group elements}, which are extremely efficient absorbers because their complex atomic structure allows for millions of different lines. At lower metallicity and in the outer (supersonic) wind, the main contribution comes from CNO elements. Furthermore, the recombination of Fe IV to Fe III for $T_{\rm eff}$ going from ${\sim}\text{27,500}\, {\rm K}$ to ${\sim}\text{22,500}\, {\rm K}$ gives a significant boost to mass loss, despite $\dot{M}_*\propto T_{\rm eff}$ in this temperature range (bi-stability jump).

Mass loss is mainly driven by the \textbf{Fe-group elements even for Wolf–Rayet (WR) stars}, though the dependence on $Z$ cannot be described as a single power-law. The atmospheres of WR stars are \textit{self-enriched} with metals, (e.g., carbon), so the latter can sustain the mass loss of WR stars for $Z\lesssim 10^{-3}\, \zsun{}$, where, indeed, $\dot{M}_*$ becomes insensitive to metallicity~\citep{vink2005, sander2020}. 
The mass loss prescriptions developed by \citet{vink2001} and \citet{vink2005} are the ones adopted by most state-of-the-art stellar evolution codes. A summary of the prescriptions is given in {Section 4 }of \citet{vink2021}. 

Both the CAK theory and the Monte Carlo approach rely on the assumption that a photon with a specific frequency can only be absorbed in an infinitely narrow region of the stellar atmosphere \citep{sobolev1960}. This approximation breaks down if the wind is not supersonic (e.g., for the inner wind, where $\dot{M}_*$ is set) and for \textbf{clumpy winds}. Recent works that do not adopt this assumption predict mass loss rates lower than \citep{vink2001}, though the metallicity dependence is remarkably similar (e.g., \citep{bjorklund2021}).

Another important aspect to consider is that the stars that \textbf{approach the Eddington limit} during their evolution might experience enhanced mass loss, which may even become insensitive to metallicity and occur in the form of pulsations \citep{vink2011_massive, vink2002, yoon2010, chen2015, tang2014, grafener2011, grafner2008}. Our knowledge of such continuum-driven winds is hampered by the uncertainties in modeling the interaction between winds and radiation-dominated envelopes.

Another main source of uncertainty is about the \textbf{homogeneity} of stellar winds. Several observations seem to suggest that winds are clumpy, though the clumps' formation mechanism and evolution is still under debate \citep{davies2005,puls2006}. The geometry, clumpiness level, and nature of clumps (i.e., optically thin or think) are also uncertain, but they might have a significant impact on stellar winds (e.g., \citep{muijres2011}).

Finally, magnetic fields can quench winds and allow the formation of quite massive compact objects even at high metallicities (e.g., \citep{petit2017}), while stellar rotation affects the evolution of stars in many ways, but, overall, it tends to increase mass loss and to allow for the formation of larger stellar cores through enhanced mixing (e.g., \citep{heger2000_rot, maeder2000}).

\subsection{Core-Collapse Supernovae}\label{sec:corecollapsesne}

As briefly described in Section~\ref{sec:overview}, stars with mass $\gtrsim$$8\, \msun{}$ end their life with a SN explosion, ejecting their outer layers in the interstellar medium and leaving a compact remnant behind (either a NS or a BH, depending on the progenitor's mass and structure).
The SN process starts with the collapse of the stellar structure, which, after the formation of the Fe-group elements, is not sustained anymore by either the core's nuclear reactions or electron degeneracy pressure. Electron captures on nuclei accelerate core collapse, and when the temperature reaches ${\sim} 10^{10}\, {\rm K}$ the photodisintegration of the Fe-group elements becomes the dominant interaction mechanism ($^{56}{\rm Fe} + \gamma \rightarrow 13 ^{4}{\rm He} + 4{\rm n}$). This process requires very high energies (${\sim}2\, {\rm MeV}$ per nucleon), thus the core-collapse accelerates and photons reach enough energies to even photodisintegrate $\alpha$ particles ($^{4}{\rm He} + \gamma \rightarrow 2{\rm p} + 2{\rm n}$). At this stage, electrons have high enough energies ($> m_{\rm e}c^2$) to collide with protons and forming neutrons and electron neutrinos (${\rm p} + {\rm e}^{-} \rightarrow {\rm n} + \nu_{\rm e}$). Thus, it is apparent that there is a significant \textbf{enrichment of neutrons}, which eventually form a very compact degenerate structure that can halt the collapse. 

The entire collapse process proceeds typically on the dynamical timescale which, for densities $\gtrsim 10^{10} \mathrm{g\,cm}^{-3}$, might last a few milliseconds. The typical gravitational binding energy of the collapsed core is ${\sim} 10^{53} \mathrm{erg}$, more than enough to power a typical SN explosion, as long as an effective mechanism that transfers this energy to stellar layers exists.

The mechanism that triggers the actual explosion and, consequently, the link between progenitor stars and their compact remnants, are still matter of debate.

A possible scenario is the \textbf{bounce-shock mechanism} (e.g., \citep{colgate1966}). During the collapse phase, the core contraction is not self-similar: only the innermost part of the core contracts all together (\textbf{homologous core}, ${\sim} 0.5\text{--}1\,\msun{}$). When the density in the homologous core rises to ${\sim} 2.7\times 10^{14} \mathrm{g\,cm}^{-3}$, the neutron degeneracy pressure would be high enough to sustain the structure against collapse, though the core overshoots its equilibrium state and when $\rho\simeq 5\times 10^{14} \mathrm{g\,cm}^{-3}$ the repulsive nuclear force makes the core bounce back, creating a \textbf{shock wave}. The latter might carry enough energy to eject the stellar envelope and power a \textit{prompt explosion}. However, the shock dissipates most of its energy while travelling outwards, through the infalling material, until it stalls at about hundreds of kilometers from the center, well within the Fe core, failing to produce a successful SN. 

Neutrinos play a crucial role in reviving the shock through the delayed \textbf{neutrino-driven mechanism} (e.g., \citep{bethe1985}).
At central densities ${\sim} 5 \times 10^{9}\, \mathrm{g\,cm}^{-3}$, the mean free path of neutrinos is comparable to the dimension of the homologous core. At higher densities (${\sim} 10^{11} \mathrm{g\,cm}^{-3}$) neutrinos are basically trapped in the core and they start a congestion that results in the stall of the neutronization process at ${\sim} 10^{12}\, \mathrm{g\,cm}^{-3}$.
The latter completes only seconds after the collapse, when most of the very high-energy neutrinos have had time to escape the core and to deposit part of their energy in the material behind the former shock wave. The rise in pressure in the layer between the proto-NS and the shock wave might revive the latter and power a successful explosion.

One-dimensional simulations of neutrino-driven explosions obtain successful SNe only for low-mass stars with naked O-Ne-Mg cores (ZAMS masses from ${\sim} 7\,\msun{}$ to ${\sim} 10\,\msun{}$, electron-capture SNe---see Section~\ref{sec:ecsne}).
Two-dimensional simulations revealed that the layer where neutrinos deposit their energy experiences non-radial hydrodynamic instabilities, which (i) generate asymmetric explosions, and (ii) convert thermal energy into kinetic energy, further fueling the explosion (\textbf{convection-enhanced neutrino-driven mechanism}, e.g., \citep{janka1995, janka1996, burrows1995}).

\rev{State-of-the-art, three-dimensional hydrodynamical simulations of neutrino-driven SNe predict successful explosions for stars up to ${\sim} 25\,\text{--}\,30\,\msun{}$. Such sophisticated multi-dimensional simulations are subject to major uncertainties and they are computationally intensive, but most state-of-the-art models seem to agree that blowing up massive stars ($\gtrsim$$25~\msun{}$) is quite challenging (see \citep{janka2017_nde, janka2012, mezzacappa2020} and references therein).
For this reason, other explosion mechanisms have been proposed so far (e.g., magnetorotational-driven explosion explosions, \citep{leblanc1970,bisnovati1970,meier1976,mueller1979,fryer2004,kotake2006,nakamura2014,takiwaki2016,summa2018}}).

This does not necessarily undermine the foundations of the neutrino-driven mechanism. The detection of heavy stellar-mass BHs through GWs, the lack of observed SNe with massive progenitor stars ($\gtrsim$$20\,\msun{}$), and the fact that the observed SN energies are small compared to the energy reservoir of neutrinos ($\lesssim$$1\%$), provide clues towards an intrinsically inefficient SN mechanism.

\subsection{Electron-Capture SNe} \label{sec:ecsne}

Stars with masses in the range $8\,\msun{}\lesssim \mzams\lesssim 10\,\msun{}$ ignite carbon and leave Oxygen-Neon-Magnesium cores. The central temperature and density in the core increase enough to reach electron degeneracy, but never enough to ignite Ne. The increasing electron degeneracy favors electron capture on $^{24}{\rm Mg}$ and $^{20}{\rm Ne}$, which lowers pressure support, and initiates the core-collapse phase, which proceeds until the formation of a NS (e.g., \citep{miyaji1980,nomoto1982}, but see also \citep{jones2016}).

\rev{
The steep density profile at the edge of O-Ne-Mg cores favors the propagation of the core-bounce shock, preventing its stagnation and the development of significant non-radial hydrodynamical asymmetries in the neutrino-heated layer. This explains why even one-dimensional neutrino-driven simulations produce successful explosions for low-mass progenitors (e.g., \citep{kitaura2006}). The resulting SN explosion is denoted as an \textbf{electron-capture
supernova (ECSN)}, and its lack of non-radial asymmetries has implication for the strength of SN kicks (see Section~\ref{sec:snkicks}).
}

\subsection{Compact Remnants and the Lower Mass Gap} \label{subsec:compactremnants}
The nature and mass of a compact remnant are close relatives of the SN explosion of the progenitor star. Depending on the energy of the explosion, a fraction $f_{\rm fb}$ of the star's material may fall back onto the proto-compact object, which can eventually exceed the TOV limit and transform into a BH. Failed explosions are associated with the collapse of the entire stellar structure and the formation of a massive BH (i.e., $f_{\rm fb} \simeq 1$, \textbf{direct collapse}). As such, fallback is a key ingredient to understand the mass spectrum of both NSs and BHs, but constraining it is very challenging.

On the one hand, large grids of self-consistent, multi-dimensional SN explosions are currently not feasible since they are computationally expensive and still require improvements in the implemented physics. On the other hand, one-dimensional simulations predict successful explosions only for low-mass progenitors (no convective engine). 
For these reasons, grids of SN explosions are generally constructed via one-dimensional models, where the explosion energy is artificially injected either directly into the convective region (\textbf{energy-driven models}) or modeled through the expansion of a hard surface placed at a specified mass-cut, generally at the outer border of the iron core (\textbf{piston-driven}). In both cases, the convective-enhanced neutrino energy becomes a parameter of the models, and it is generally calibrated using the observed SN luminosities and $^{56}{\rm Ni}$-yields. 
Following this approach, many authors tried to identify the key parameters of the stellar structure that \textbf{drive the explodability of stars} and the amount of fallback.

\textls[-25]{\citet{fryer2012} studied the outcome of several energy-driven SN explosions and introduced a model where the mass of compact remnants and fallback depend mainly on the final mass of the star and on the mass of the carbon-oxygen core ($m_{\rm CO}$). \citet{fryer2012}} considered two cases: (i) the explosion happens in the first ${\sim} 250$ ms and it is driven mainly by the Rayleigh–Taylor instability (\textbf{rapid} model), and (ii) the explosion is delayed (${\sim}$seconds) and its main engine becomes the standing accretion shock instability (\textbf{delayed} model). In both models, fallback has a huge impact on the masses of remnants from stars with  $10\,\msun{}\lesssim \mzams \lesssim 30\,\msun{}$.
Both models predict direct collapse for $m_{\rm CO}\geq 11\,\msun{}$, but the rapid model also for $6\,\msun{}\leq m_{\rm CO}\leq 7\,\msun{}$, which corresponds to $22\,\msun{}\lesssim \mzams\lesssim 26\,\msun{}$ The latter happens because the rapid \rev{mechanism occurs in $\sim$$100$ ms, i.e., when the infalling ram pressure can be still high enough to prevent a successful explosion. Thus, the rapid approach is more prone to a failed explosion than the delayed model, and it is more sensitive to the compactness of the innermost star's regions. Specifically, in \citet{fryer2012}, the stellar models with $22\,\msun{}\lesssim \mzams\lesssim 26\,\msun{}$ develop mixing instabilities and more compact structures during the latest evolutionary stages, causing a failed rapid SN and a \textbf{gap} in the remnants mass spectrum between ${\sim} 3\,\msun{}$ and ${\sim} 5\,\msun{}$, a dearth which seems in agreement with observations (the \textbf{lower mass gap} \cite{ozel2010,ozel2012,farr2011}).} While this argument might suggest a preference for the rapid model, the approach followed by \citet{fryer2012} is simplified and \rev{sensitive to the details of the stellar late-stage burning phases. Thus, the existence of the lower mass gap is still matter of debate and there are no conclusive results on the topic.}

\begin{figure}[H]
    
    \includegraphics[width=0.7\columnwidth]{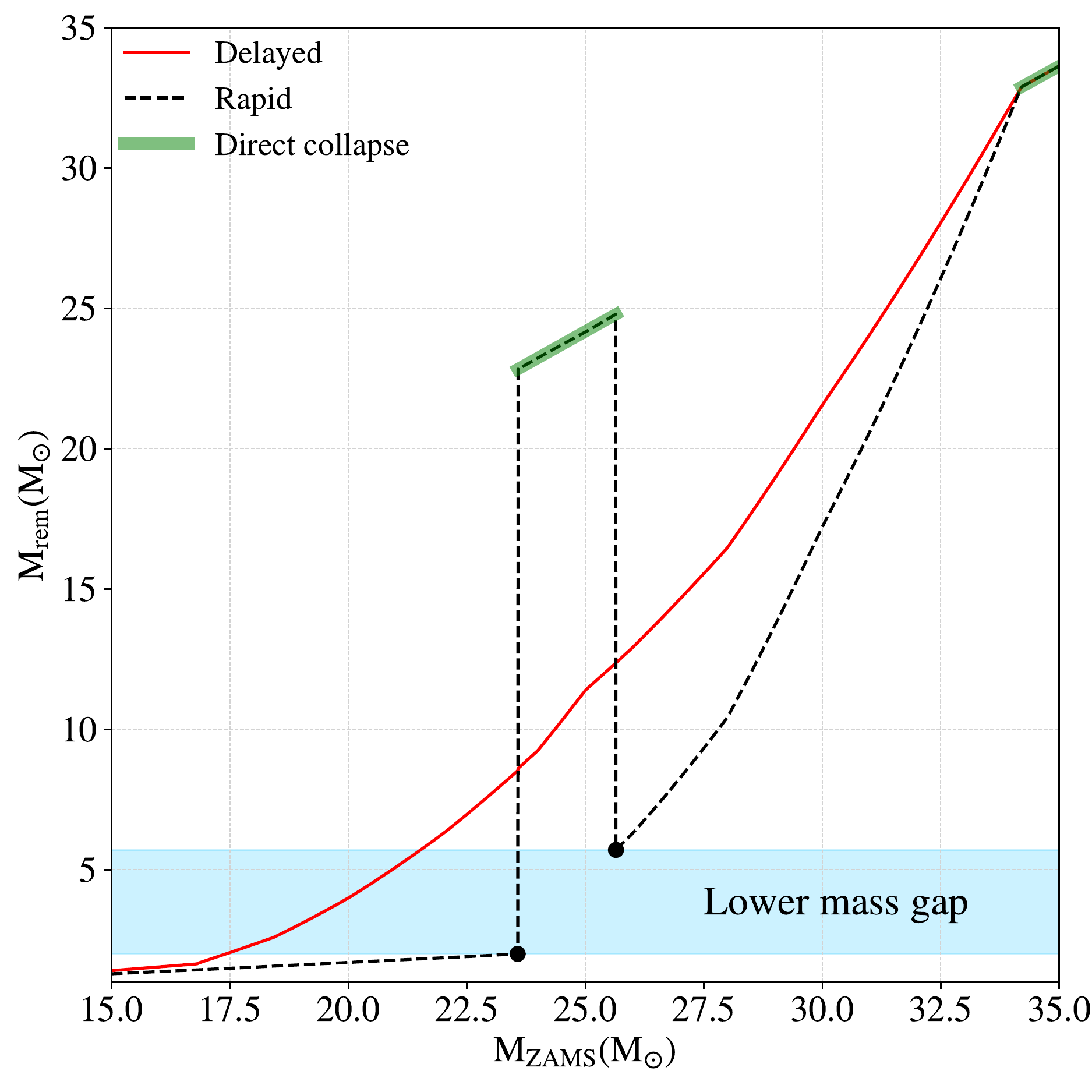}
    \caption{Mass of the compact remnant as a function of the initial mass of its progenitor star obtained with the delayed (solid red line) and the rapid (dashed black line) SN explosion model. The progenitor stars come from the SEVN code {\citep{spera2017}} with look-up tables from PARSEC \citep{bressan2012}. The shaded cyan area shows the location of the lower mass gap, while the semi-transparent green line highlights the region where direct collapse occurs. The two black points define the lower and the upper edge of the mass~gap.}
    \label{fig:lowergapfryer}
\end{figure}

Figure \ref{fig:lowergapfryer} compares the mass spectrum of compact remnants obtained using the rapid and the delayed SN explosion models applied to the progenitor stars of the SEVN code~\citep{spera2017}, at $Z=0.001$ (see also Figure~\ref{fig:winds} for the progenitors). From Figure~\ref{fig:lowergapfryer}, it is apparent that the rapid model creates a gap in the BH mass spectrum between ${\sim} 2\,\msun{}$ and ${\sim} 6\,\msun{}$ (shaded area), because stars with $23\,\msun{}\lesssim \mzams\lesssim 26\,\msun{}$ evolve through direct collapse, while the delayed model does not.

\citet{ugliano2012} adopted a more sophisticated model than \citep{fryer2012}, but still 1D, and showed that the \textbf{compactness parameter}, $\xi_{2.5}$\endnote{$\xi_{m} = \frac{m}{R\left(m\right)}$, where $m$ is a threshold mass in $\msun{}$ and $R\left(m\right)$ is the radius enclosing $m$, in units of 1000 km.} \citep{oconnor2011}, provides better insights than $m_{\rm CO}$ on the explodability of a star. \citet{ugliano2012} simulations revealed that BHs can form via direct collapse for $\mzams \lesssim 20\,\msun{}$ and that successful SNe are possible for $20\,\msun{}\lesssim \mzams\lesssim 40\,\msun{}$. This happens because stellar structure does not vary monotonically with \mzams, and the SN explosion is sensitive to such variations. Rather than a $\xi_{2.5}$-threshold, it is the \textbf{local maxima} in the $\xi_{2.5}\text{--}\mzams$ plane that increase the probability of failed SNe, thus, \textit{islands of explodability} appear for $\mzams\lesssim 40\msun{}$, while direct collapse is dominant for $\mzams \gtrsim 40\,\msun{}$. Specifically, stars with $22\,\msun{}\lesssim \mzams\lesssim 26\,\msun{}$ tend to have higher $\xi_{2.5}$ than neighboring stars and this creates a dearth of remnants with mass between ${\sim} 2\,\msun{}$ and $6.5\,\msun{}$. While this finding qualitatively agrees with the rapid model of \citet{fryer2012}, in \citet{ugliano2012} the gap is naturally produced through a wide range of explosion timescales (from 0.1 s to 1 s) that depend only on the structure of the progenitor at the onset of collapse.

\citet{limongi_windrot2018} showed that there is a tight monotonic correlation between $\xi_{2.5}$ and $m_{\rm CO}$, which can be expressed as $\xi_{2.5}=0.55 -1.1m_{\rm CO}^{-1}$, with $m_{\rm CO}$ given in units of $\msun{}$ \citep{mapelli_rot2020}. This is in qualitative agreement with the simplified approach of \citet{fryer2012}, but different assumptions on carbon nuclear reaction rates might significantly affect the correlation (e.g., \citep{chieffi2021}).

To capture the apparent stochasticity emerging from compactness-based studies, \citet{clausen2015} adopted a probabilistic description to model the NS and BH mass spectrum.

\citet{ertl2016} refined the compactness approach by introducing an even more sophisticated \textbf{two-parameter model} to predict the explodability of stars. The first parameter is the normalized enclosed mass for a dimensionless entropy per nucleon of $s = 4$, $M_4$. This is a good proxy for the proto-NS mass, which corresponds roughly to the iron-core mass at the onset of collapse. The other parameter, $\mu_4$, is the mass derivative at $R\left(M_4\right)$. The advantage of the two-parameter model is that the quantities $x\equiv M_4\mu_4$ and $y\equiv \mu_4$ are strongly connected to the mass accretion rate of the stalling shock and to the neutrino luminosity, respectively, so they are expected to capture the physics of the neutrino-driven explosion better than $\xi_{2.5}$. \citet{ertl2016} predicted successful (failed) SNe for $\mu_4 < y_{\rm sep}\left(x\right)$ ($\mu_4 > y_{\rm sep}\left(x\right)$), where $y_{\rm sep}\left(x\right) = k_1x+k_2$, and $k_1\in\left[0.19;0.28\right]$ and $k_2\in\left[0.043;0.058\right]$, depending on the adopted set of progenitor stars. \citet{ertl2016} (see also \citep{sukhbold2016}) confirmed the presence of islands of explodability, the prevalence of direct collapse for $\mzams \gtrsim 30\,\msun{}$, and that fallback SNe are quite rare (i.e., a gap of compact objects with mass between ${\sim} 2\,\msun{}$ and $5\,\msun{}$).

Recently, \citet{ertl2020} investigated the explodability of a set of evolved naked-helium stars. They confirmed the presence of islands of explodability and the robustness of the two-parameter method to predict progenitors' fate. Furthermore, they showed that the number of fallback SNe that form remnants between ${\sim} 2\,\msun{}$ and $6\,\msun{}$ is larger than that predicted by \citet{ertl2016} and \citet{sukhbold2016}. 

It is worth noting that all the features emerging through parametric (1D-based) approaches to the explodability of stars represent only the first step towards an exhaustive scenario for the SN mechanism and the masses of compact remnants. As such, \textbf{they should be taken with a grain of salt}, as all the features might be either confirmed or gone by the time we will have a realistic framework for 3D explosion models, which still need improvements and should be considered only as provisional (e.g., \citep{burrows2020}).

\subsection{Core-Collapse SNe in Population Synthesis Calculations}
Performing accurate calculations of the internal structure of stars and sophisticated one-dimensional simulations of the SN mechanism is prohibitive for fast population-synthesis simulations of either single or binary stars.

To calculate the mass of compact remnants, the prescriptions of \citet{fryer2012} are easy to implement and do not require accurate calculations of the internal structure of stars. As such, while very simplified, they are still the most commonly used for fast population-synthesis simulations.

\citet{ertl2020} and \citet{woosley2020} provide tables and fitting formulas to calculate the remnant mass as a function of the initial and pre-SN helium core mass of the star, including also the effect of PISNe and pulsational pair-instability SNe (see Section~\ref{sec:uppermassgap}). Such relations are expected to capture the physics behind the SN explosion mechanism better than \citep{fryer2012}.

A similar approach was followed by \citet{patton2020} who evolved a set of naked carbon-oxygen cores until the onset of collapse and provided values of $\xi_{2.5}$ and $M_4$ as a function of the initial $m_{\rm CO}$ and carbon abundance $X_{\rm C}$. The provided tables can be easily implemented in population-synthesis codes using a compactness- or a two-parameter based method for explodability.

\textls[-15]{\citet{patton2022} investigated the mass spectrum of NSs and BHs from population-synthesis simulations comparing the different approaches of \citet{fryer2012}, \citet{patton2020}, and \citet{woosley2020}, for single and binary stars. They found qualitative agreement between the prescriptions of \citet{patton2020} and \citet{woosley2020},} which give results roughly consistent with \citet{ertl2016} and \citet{sukhbold2016}. Significant differences emerge between \citet{patton2020} and \citet{fryer2012} for the mass distribution of NSs, and for the BH mass spectrum at low masses ($10\,\msun{}\text{--}15\,\msun{}$).

\subsection{Pair-Instability SNe and the Upper Mass Gap} \label{sec:uppermassgap}

\textls[-15]{Theoretical models of single-star evolution predict the existence of another gap in the mass spectrum of compact remnant, which extends from ${\sim} 60\,\msun{}$ to ${\sim} 120\,\msun{}$. This is also known as \textbf{the upper mass gap}, as opposed to the lower mass gap which corresponds to a dearth of observations of compact objects with mass between ${\sim} 2.5\,\msun{}$ and ${\sim} 5\,\msun{}$~\cite{ozel2010,ozel2012,farr2011} }(see also Section~\ref{subsec:compactremnants}).
The pulsational pair-instability supernova (PPISN) \cite{woosley2017} and the PISN \cite{heger2002} are the main mechanisms behind the formation of the upper mass gap.

The relation that links the birth mass of a BH ($m_{\rm BH}$) to \mzams{} is complex because it reflects the uncertainties we have on the evolution of massive stars, on stellar winds, and on the SN explosion mechanism (e.g., \cite{heger2003,fryer2012}). Assuming that a progenitor star does not lose mass through stellar winds and that, at the end of its life, the entire stellar structure collapses into a BH without any ejecta, then $m_{\rm BH}$$\sim$$\mzams$ (e.g., \citep{spera2015} and Figure~\ref{fig:winds}). This is a continuous and monotonically increasing function, thus it predicts no gaps in the BH mass spectrum. Such a simplified relation would work reasonably well for massive stars ($\mzams\gtrsim 40\,\msun{}$) at low metallicity ($Z\lesssim 10^{-3}$), as long as PPISNe and PISNe are not effective. However, if stars' core temperatures rise above ${\sim} 7\times 10^8\, {\rm K}$, photons become energetic enough to \textbf{create electron-positron pairs} \cite{fowler1964,barkat1967,rakavy1967,woosley2017}. This process converts energy (gamma photons) into rest mass (electrons and positrons), thus it lowers radiation pressure and it triggers stellar collapse. In stars with helium core masses between ${\sim} 34\,\msun{}$ and ${\sim} 64\,\msun{}$, the collapse is reversed by oxygen- or silicon- core burning, which shows up as a pulse and makes the core expand and cool. The flash is not energetic enough to disrupt the star \rev{and the core begins a series of contractions and expansions (stellar pulsations) that \textbf{significantly enhance mass loss}, especially from the outermost stellar layers, and continue until the entropy becomes low enough to avoid the pair instability and stabilize the core until the core-collapse SN explosion. Such pulsational instabilities are referred to as pulsational pair-instability supernova \cite{heger2002,woosley2007,woosley2017}}.  In contrast, in stars with helium core masses between ${\sim} 64\,\msun{}$ and ${\sim} 130\,\msun{}$, the first pulse is energetic enough to \textbf{completely disrupt the entire star} (i.e., PISN \cite{bond1984,fryer2001,chatzopoulos2012})\endnote{It is worth noting that a PISN is driven by a thermonuclear explosion, i.e., very different from neutrino-driven core-collapse SNe.}. Stars with helium cores above ${\sim} 130\,\msun{}$ experience a rapid pair instability-induced collapse but the energy released by nuclear burning is not enough to reverse the collapse before photodisintegration (endothermic) becomes the dominant photon-interaction mechanism \citep{heger2002}. Thus, the direct collapse to a massive BH (mass $\gtrsim 130\,\msun{}$) becomes unavoidable.

Stars with $Z\gtrsim 5\times 10^{-3}$ ($Z\gtrsim 2\times 10^{-2}$) will not evolve through the PISN (PPISN) phase, since their helium core masses cannot grow above ${\sim} 64\,\msun{}$ (${\sim} 34\,\msun{}$) (e.g., \citep{spera2017}), though the metallicity limits are uncertain and depend on the prescriptions adopted to model several processes, including stellar winds, overshooting and stellar rotation. 

The main consequence of PPISNe and PISNe is that they create a gap in the BH mass spectrum. At $Z\lesssim 10^{-3}$, the $m_{\rm BH}$ versus \mzams{} relation flattens out for stars with $\mzams \gtrsim 60\, \msun{}$ because of the enhanced mass loss caused by PPISNe, which removes all the hydrogen envelopes. Then, the BH mass continues to increase following the growth of the helium core until \mzams{} becomes large enough (${\sim} 150\,\msun{}$) to allow for the formation of helium cores $\gtrsim$$64\,\msun{}$, when stars evolve through the PISN and $m_{\rm BH} = 0$. This means that the $m_{\rm BH}$ versus \mzams{} relation has a \textbf{local maximum} which corresponds to the \textbf{lower edge of the upper mass gap} ($M_{\rm low}$) \cite{belc2016b,fishbach2017, spera2017,stevenson2019,farmer2019,dicarlo2020,farmer2020}. Pair creation triggers direct collapse for stars with helium core masses $\gtrsim$$130\,\msun{}$ (i.e., $\mzams\gtrsim 300\,\msun{}$ for $Z\simeq 10^{-3}$), thus these stars form massive ($\gtrsim$$130\,\msun{}$) BHs. This BH mass corresponds to a \textbf{local minimum} of the $m_{\rm BH}\text{--}\mzams$ curve, for $\mzams\gtrsim 300\,\msun{}$, and it is referred to as \textbf{the upper edge of the upper mass gap} ($M_{\rm high}$) \cite{spera2017,mangiagli2019,fishbach2020,kohler2021}.

\begin{figure}[H]
    
    \includegraphics[width=\columnwidth]{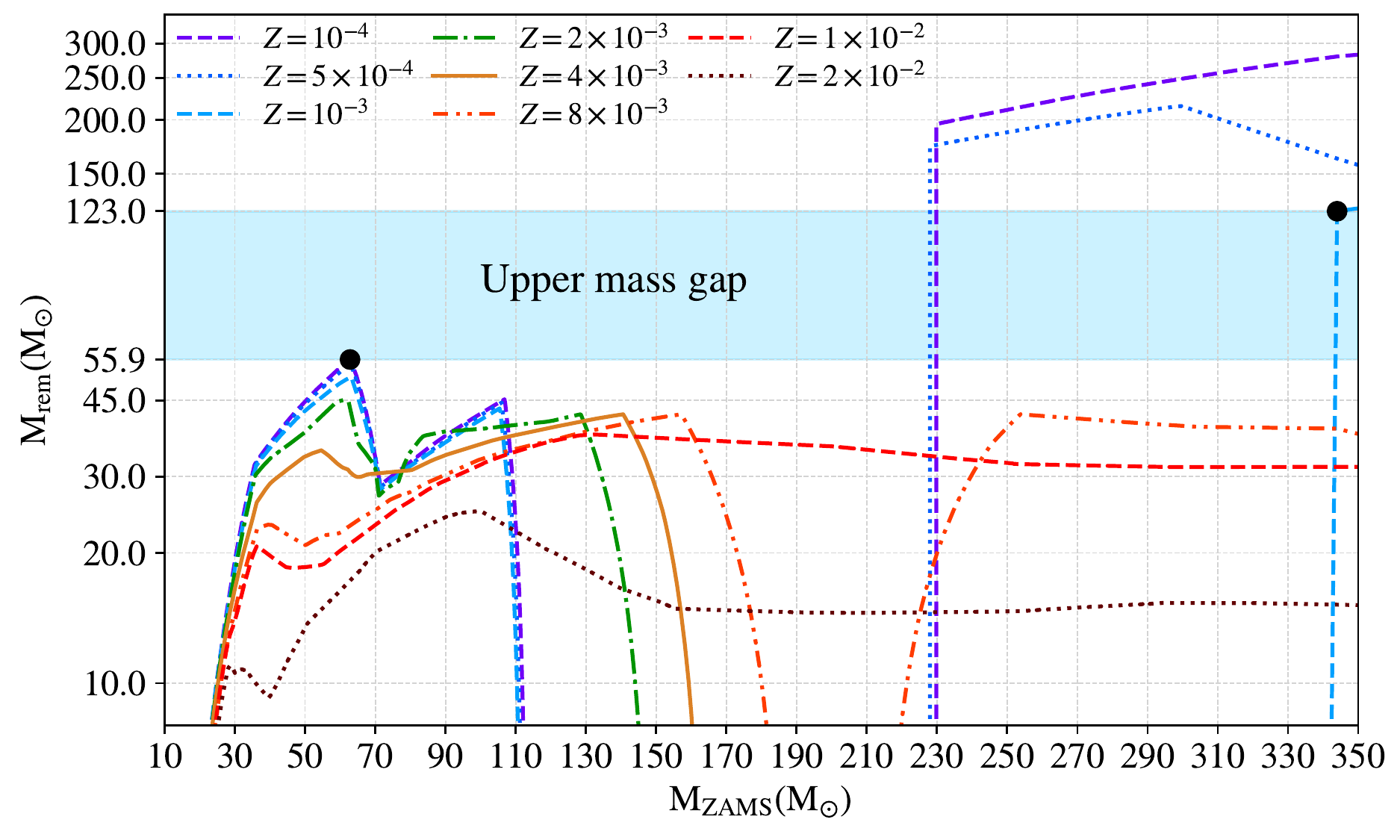}
    \caption{Mass of the BH as a function of the initial mass of its progenitor star, for different values of metallicity $Z\in [10^{-4};2\times 10^{-2}]$. The shaded cyan area shows the location of the upper mass gap. The two black points set the lower edge (${\sim} 55.9\,\msun$) and the upper edge (${\sim} 123\,\msun$) of the gap. The plot has been adapted from {\citet{spera2017}} and it has been obtained through the SEVN code coupled with look-up tables from PARSEC \citep{bressan2012} and the delayed SN explosion model \citep{fryer2012}.}
    \label{fig:spera2017}
\end{figure}

\rev{Figure \ref{fig:spera2017} shows a typical example of a BH mass spectrum with the upper mass gap, obtained from a population of single stars, at various metallicities, through the SEVN population-synthesis code \citep{spera2017}.}
 
The values of $M_{\rm low}$ and $M_{\rm high}$ are highly uncertain because they strongly depend on metallicity, on the adopted stellar-wind models, on the boundaries of helium core masses for the occurrence of PPISNe and PISNe, on nuclear reaction rates, on stellar rotation, on the treatment of convection, and on the SN explosion mechanism.
While most of these processes affect the value of $M_{\rm low}$ and $M_{\rm high}$ by a few percent \citep{marchant2020,mapelli_rot2020,woosley2021,renzo2020}, the two main sources of uncertainty are our understanding of the failed SN mechanism and of the \cago{} reaction rate.

As concerns the failed SN mechanism, most theoretical models seem to agree on the occurrence of direct collapse for massive stars but we still do not know if the \textbf{hydrogen envelope of the progenitor star} is ejected during the collapse, leaving a remnant with the mass of about the helium core, or if it also contributes to the BH's growth \citep{nadezhin1980,lovegrove2013,clausen2015,kochanek2015,sukhbold2016}. The collapse of the hydrogen envelope gives an uncertainty on $M_{\rm low}$ of about $20\,\msun{}$ \citep{mapelli_rot2020}. 

\textls[-20]{The impact on $M_{\rm high}$ is more difficult to quantify. As example, \citet{spera2017} (see also Figure~\ref{fig:spera2017})} show that stellar winds of massive stars with $Z\gtrsim 10^{-3}$ are likely strong enough to remove most of the hydrogen envelope during the main sequence phase, preventing the formation of helium cores with mass $\gtrsim 130\,\msun{}$, even for extremely massive stars ($\gtrsim$$300\,\msun{}$). Very massive stars ($\gtrsim$$350\,\msun{}$) with $Z\simeq 10^{-3}$ might die as Wolf-Rayet stars with mass $\simeq$$130\,\msun{}$, thus they might be the stars with the smallest \mzams{} to form BHs beyond the gap, setting the value of $M_{\rm high}$ to ${\sim} 120\,\msun{}$. At $Z\lesssim 10^{-3}$, stars retain a significant fraction of their hydrogen envelope prior to collapse and the stars with the smallest \mzams{} to reach helium core masses above ${\sim} 130\,\msun{}$ form BHs with masses above the gap. Indeed, at $Z\simeq 10^{-4}$, the stars with the smallest \mzams{} to form BHs beyond the gap have $\mzams \simeq 230\,\msun{}$ and \rev{$m_{\rm BH}\simeq 210\,\msun{}$ when the hydrogen envelope is accreted by the BH and $m_{\rm BH}\simeq 120\,\msun{}\simeq M_{\rm high}$ when it is not.}
However, the uncertainties on the evolution of such extremely massive stars are significant and the discussed limits are uncertain. Following these arguments, \citet{woosley2017} pointed out that the upper gap might be seen more as a cut off of BHs with mass $\gtrsim M_{\rm low}$, because BHs with masses $\gtrsim M_{\rm high}$ can form only from the collapse of extremely massive stars, \rev{which are supposed to be exceedingly rare and live only for a few Myr (e.g., \citep{kroupa2001}).}

Another significant source of uncertainty is given by our knowledge of \textbf{the temperature dependence of the \cago{} reaction rate} (e.g., \citep{deboer2017}). While changing the \cago{} rate has no significant impact on the stellar structure, it governs the relative amount of oxygen with respect to carbon in the core. Low \cago{} rates translate into large carbon reservoirs at the end of helium-core burning and into a prolonged carbon-burning phase, which contributes to suppress pair production, stabilize the oxygen core, and delay the latter ignition. In contrast, high \cago{} rates imply significant carbon depletion in favor of oxygen, which ignites explosively just after the helium burning phase. This has a strong impact on the upper mass gap because low (high) \cago{} rates push $M_{\rm low}$ and $M_{\rm high}$ towards higher (lower) values \citep{takahashi2018,farmer2019,farmer2020,costa2021,woosley2021}. Furthermore, massive stars with low \cago{} rates might experience significant dredge up which tends to stabilize the oxygen core even further. In this scenario, if very low ($-3\sigma$) \cago{} rates are considered together with the collapse of the hydrogen envelope, the upper mass gap might even disappear \citep{costa2021,tanikawa2022_popsyn}. Conservatively, the impact of \cago{} rates on both $M_{\rm low}$ and $M_{\rm high}$ is about $15\, \msun{}$.

It is also worth mentioning that the presence of physics beyond the standard model might also significantly affect the edges of the upper mass gap (e.g., \citep{sakstein2020}).

Overall, considering all the main known uncertainties so far, $M_{\rm low}\in \left[40\,\msun{};75\,\msun{}\right]$. As for $M_{\rm high}$, the scenario is more complex. A left endpoint of ${\sim} 120\,\msun{}$ for the mass interval seems to be a quite robust prediction, while the right endpoint is more uncertain. For instance, considering only stars with $\mzams \lesssim 250\,\msun{}$ and the collapse of the entire hydrogen envelope, $M_{\rm high}$ might rise to ${\sim} 200\,\msun{}$. More conservative limits on \mzams{} (e.g., $\mzams\lesssim 150\,\msun{})$ might even cause the mass gap to be seen as a cut off at $M_{\rm low}$. However, while rare, very massive stars might form from a series of stellar mergers, thus $M_{\rm high}\in \left[120\msun{};200\msun{}\right]$ might be considered as a reasonable range for $M_{\rm high}$.

To model the enhanced mass loss predicted by PPISNe, population-synthesis codes generally adopt fitting formulas to detailed stellar evolution calculations (e.g., \citep{spera2017, marchant2019}).

\subsubsection{Piling-Up BHs}

A direct consequence of the enhanced mass loss caused by PPISNe is that most stars with initial mass $60\,\msun{}\lesssim \mzams \lesssim 150\,\msun{}$ die as naked helium cores, even at very low metallicity.
The final helium cores of these stars have a mass $35\,\msun{} \lesssim m_{\rm HE} \lesssim 65\,\msun{}$, but pair-instability pulses are stronger for heavier cores, thus the helium cores left by PPISNe lie more likely in the range $35\,\msun{} \lesssim m_{\rm HE} \lesssim 45\,\msun{}$. Thus, we might expect \textbf{an excess of BHs (pile-up)} with masses in about the same range \cite{spera2016_memsait,belc2016b,spera2017,woosley2020}.

Constraining the extent of the BH bump and its mass limits is challenging. From a purely single stellar evolution perspective, the significance of the PPISNe bump depends primarily on the slope of the $m_{\rm BH}\text{--}\mzams{}$ theoretical curve for $\mzams \lesssim 60\, \msun{}$ (not affected by PPISNe) compared to that for $60\,\msun{}\lesssim \mzams \lesssim 150\,\msun{}$ (affected by PPISNe). With a standard initial mass function (e.g., \citet{kroupa2001}), if the $m_{\rm BH}\text{--}\mzams{}$ curve at $60\,\msun{}\lesssim \mzams \lesssim 150\,\msun{}$ is significantly flatter than the curve at $\mzams \lesssim 60\, \msun{}$, then the bump in the BH mass distribution will be more relevant and narrower, especially in correspondence of the kink between the two slopes. However, the relative slopes and the position of the \textbf{kink in the $m_{\rm BH}\text{--}\mzams{}$ curve are highly uncertain} since they depend on many ingredients including metallicity, stellar winds, the details of the growth of helium cores inside massive stars, nuclear reaction rates (e.g., \cago{}), and the collapse of the hydrogen envelope.

\begin{figure}[H]
    
    \includegraphics[width=\columnwidth]{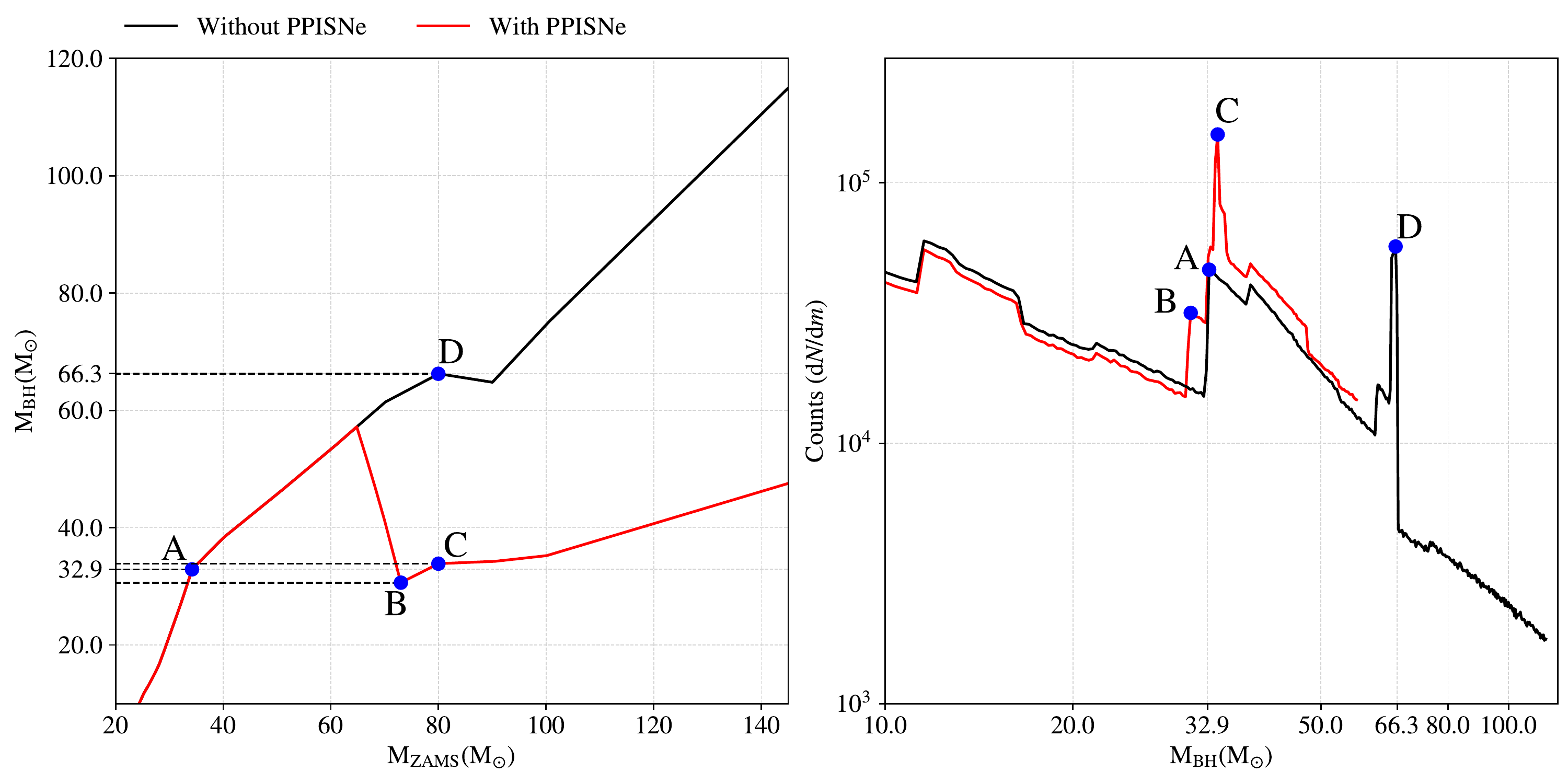}
    \caption{Mass spectrum ($m_{\rm BH}$ vs \mzams, left panel) and mass distribution (${\rm d}N/{\rm d}m$ vs $m_{\rm BH}$, right panel) of BHs obtained evolving a population of $10^7$ single stars. The progenitors follow a \rev{power-law initial mass function that scales as ${\rm d}N/{\rm d}\mzams{} \propto \mzams{}^{-2.3}$ \citep{kroupa2001}} with masses between $25\,\msun$ and $145\,\msun$. The black (red) curve shows the results without (with) PPISNe. The points A, B, C, and D represent the kinks of the $m_{\rm BH}$--\mzams{} curve. The plot has been obtained for $Z=10^{-3}$ through the SEVN population-synthesis code {\citep{spera2015}} with PPISNe fitting formulas from \citep{spera2017}, look-up tables for single stellar evolution from the PARSEC code \citep{bressan2012}, and the delayed SN explosion mechanism \citep{fryer2012}.}
    \label{fig:pilingup}
\end{figure}

Figure \ref{fig:pilingup} shows the BH mass spectrum (left panel) and the BH mass distribution (right panel) obtained from a population of $10^7$ single stars at $Z=10^{-3}$. The progenitors follow a \citet{kroupa2001} initial mass function with masses in the range $[25;145]\msun$. The black curve does not include the contribution of PPISNe, while the red does.
It is apparent that the bumps in the BH mass distribution are close relatives of the kinks of the $m_{\rm BH}$-\mzams{} curve. The pile-up of BHs through PPISNe happens at about $33\,\msun{}$, around the kinks B and C. However, the \textbf{model we adopted} to make Figure~\ref{fig:pilingup} has an additional kink in A, which also piles up BHs at $33\,\msun{}$, even without PPISNe. The pile-up at $66.3\,\msun{}$ (kink D) disappears when PPISNe are considered because the latter force $m_{\rm BH}\lesssim 55\,\msun{}$. Therefore, besides depending on metallicity, the existence and position of the kinks A, B, C, and D is \textit{strongly} model-dependent.

A pile-up of compact objects at mass $30\text{--}40\,\msun{}$ in GW detections might suggest a PPISNe signature in the BH mass spectrum (e.g., \citep{gwtc3_disc_2021,lvcgwtc3_2021}). 
However, the effect of piling up \textit{merging} BHs at specific masses is not only model-dependent (e.g., Figure~\ref{fig:pilingup}) but it might also have \textit{strong} degeneracies with stellar dynamics and binary stellar evolution processes, thus disentangling the various contributions might be challenging.

\subsubsection{Populating the Gap}

\textls[-15]{The upper mass gap is a feature in the BH mass spectrum that comes from our theoretical knowledge of \textit{single stellar evolution} processes. However, we know that stars are not born in isolation and, during their life, they may evolve through a series of events that can change their fate in many ways. Specifically, there are various (not exotic) \textbf{astrophysical processes that can form BHs in the mass gap}, without contradicting its existence (e.g.,~\cite{miller2002,antonini2016a,kimpson2016,rodriguez2018,yang2019,arcasedda2020b,dicarlo2019a,dicarlo2019b,roupas2019,natarajan2020,safarzadeh_accr2020,tanikawa2021b,tanikawa2022_popsyn,farrell2021,liu_popiii2020,spera2019,kremer2020}).}

For instance, a Nth-generation BH, which is a BH coming from a previous merger of two $(N-1)$th-generation BHs, can easily fall into the upper mass gap. 
Furthermore, assuming that the hydrogen envelope participates in the BH formation process, stars with an oversized hydrogen envelope (gained through mass accretion from a companion star or even through a merger), but small enough helium cores to avoid the PISN, might form oversized BHs, with masses in the upper mass gap. The latter scenario is quite rare but possible for the isolated binary evolution channel (e.g., \cite{spera2019}).

Nth-generation BHs and oversized BHs coming from binary evolution can be retained, acquire a companion, and merge within a Hubble time in dense stellar environments. Therefore, they can become loud (and detectable) sources of GWs with at least one member in the upper mass gap (e.g., \citep{dicarlo2019a,dicarlo2019b}).

Testing the presence of the upper mass gap and its underlying stellar astrophysics through GW detections might be challenging. The latest population analysis coming from the GWTC-3 shows a monotonically decreasing BH mass distribution at masses $\gtrsim$$50\,\msun{}$ and an inconclusive evidence for the presence of the upper mass gap \citep{lvcgwtc3_2021}.
On the other hand, observing GW events with members in the upper mass gap might be the distinguishing feature of astrophysical formation channels other than single or isolated binary-star evolution, not necessarily contradicting the existence of the gap itself.

We will explore more scenarios on how to populate the upper mass gap in Section~\ref{sec:lessonslearned}.

\subsection{SNe Asymmetries and Kicks}\label{sec:snkicks}
Galactic pulsars are observed with \textbf{fairly large spatial velocities}, which can be as high as thousands of kilometers per second. Such values are too large to be explained through Blaauw kicks \citep{blaauw1961} from SN explosions in binary systems. Thus, some, if not all, compact objects should receive quite high \textbf{kicks at birth} \citep{gunn1970}. Kicks have a huge impact on merging compact objects if they are members of isolated binaries (e.g., change of orbital parameters, unbinding the binary) or if they reside in dense stellar environments (e.g., ejections).

From the observational point of view, \citet{hobbs2005} presented an up-to-date catalogue of Galactic pulsars and showed that the mean three-dimensional velocity of young (age $\lesssim 3\, \mathrm{Myr}$) pulsars is $400\pm 40 \,\mathrm{km\,s}^{-1}$. The three-dimensional speeds are well fit by a Maxwell-Boltzmann distribution with $\sigma_{\rm 1D}=265\, \mathrm{km\,s}^{-1}$.

A more recent work \citep{verbunt2017} found that a double-Maxwell-Boltzmann distribution $f\left(v\right) = wf_1\left(v;\sigma_1\right) + \left(1-w\right)f_2\left(v;\sigma_2\right)$ with $\sigma_1 \simeq 80\, \mathrm{km\,s}^{-1}$, $\sigma_2 \simeq 320\, \mathrm{km\,s}^{-1}$, and $w\simeq 0.4$ significantly improve the theoretical description of the pulsar data.

Analyses of a new dataset of proper motions and parallaxes \citep{igoshev2020} confirm the bi-modality feature of the velocity distribution with $w=\valpm{0.2}{0.11}{0.10}$, $\sigma_1=\valpm{56}{25}{15}\,\mathrm{km\,s}^{-1}$, and $\sigma_2=\valpm{336}{45}{45}\,\mathrm{km\,s}^{-1}$.

From the theoretical point of view, \textbf{asymmetries in the SN ejecta} can impart high kicks to newly-born compact objects (e.g., \citep{wongwathanarat2013}). The kicks can vary from ${\sim} 10\, \mathrm{km\,s}^{-1}$ to ${\sim} 1000\, \mathrm{km\,s}^{-1}$, depending mainly on the steepness of the density profile at the outer edge of the stellar core (i.e., compactness), and on the stochastic variations of non-radial instabilities associated with the SN engine. Shallow (steeper) density profiles are more (less) prone to SN shock stalling, thus neutrinos will be able to interact with more (less) material and produce more- (less-)asymmetric ejecta (e.g., \citep{janka2017}). As already discussed in Section~\ref{sec:ecsne}, the progenitors of \textbf{electron-capture SNe} have very steep density profiles, so the explosion is expected to impart \textbf{low kicks} to NSs (tens of $\mathrm{km\,s}^{-1}$, e.g., \citep{gessner2018}). Similar results have been obtained for \textbf{ultra-stripped stars}, which might form during mass transfer in close binaries~\citep{tauris2015b,muller2018,suwa2015}.

Assuming that BHs share the same formation mechanism as NSs and that kicks are driven by the asymmetries in the SN ejecta, BH kicks are expected to be smaller than NSs', with differences coming mainly from the heavier mass of BHs and the smaller amounts of ejecta.
This theoretical argument cannot be supported by observations, because BH kicks lack strong observational constraints. \citet{repetto2012} and \citet{repetto2015} showed that the large distances from the Galactic plane of some BH X-ray binaries can only be explained if BHs acquire high kicks at birth (as high as those of NSs), even though they cannot completely rule out smaller BH kicks (see also \citep{mandel2016b}).

For rapid population-synthesis calculations, most codes assign birth kicks assuming two velocity distributions that distinguish between electron-capture and ultra-stripped SNe from all the other core-collapse SNe. Following \citep{fryer2012}, compact-object kicks ($v_{\rm k}$) are generally assumed to be the same as those coming from observations of pulsars ($v_{\rm NS}$), but modulated by the fraction of fallback ($f_{\rm fb}$), so that BHs that form via direct collapse (i.e., no SN ejecta) do not receive kicks:
\begin{equation}\label{eq:kicksfryer}
    v_{\rm k}=(1-f_{\rm fb})v_{\rm NS}=\frac{m_{\rm ej}}{m_{\rm ej}+m_{\rm rem}-m_{\rm proto}}v_{\rm NS},
\end{equation}
where $m_{\rm rem}$ is the mass of the compact object, $m_{\rm proto}$ is the mass of the proto-compact object, and $m_{\rm ej}$ is the mass of the SN ejecta. A threshold of $m_{\rm rem}=2\text{--}3\,\msun{}$ is assumed for distinguishing BHs from NSs.

Adopting small kicks for ECSNe and SNe from ultra-stripped progenitors is crucial to match the rates of merging NSs inferred by the LVK collaboration (e.g., \citep{giacobbo2019,bray2018}), but the simple model by \citet{fryer2012} fails to produce low kicks for compact objects from such channels, because the kick prescription is not very sensitive to $m_{\rm ej}$ (it appears in both the numerator and the denominator of Equation~\eqref{eq:kicksfryer}).
Here is an instructive example. Case~1: a progenitor star with $m_1 = 10\,\msun{}$, $m_{\rm He,1}=6\,\msun{}$, and $m_{\rm CO,1}=4\,\msun{}$. The delayed SN prescription in \citet{fryer2012} predicts $m_{\rm rem,1}\simeq 2.5\,\msun{}$, $m_{\rm ej,1}\simeq 7.5\,\msun{}$, and $f_{\rm fb,1}\simeq 0.14$. Case 2: same progenitor that becomes ultra-stripped via binary evolutionary processes. We will have $m_2\simeq m_{\rm He,2}\simeq m_{\rm CO,2}\simeq 4\,\msun{}$, and the delayed SN prescription predicts $m_{\rm rem,2}\simeq 2.1\, \msun{}$, $m_{\rm ej,2}\simeq 1.9\,\msun{}$, and $f_{\rm fb,2}\simeq 0.31$. This means that the birth-kick ratio between the two cases would be
\begin{equation}
    \frac{v_{{\rm k},1}}{v_{{\rm k},2}} = \frac{1-f_{\rm fb,1}}{1-f_{\rm fb,2}} \simeq 1.2
\end{equation}
that is, mostly the same kick for case 1 and case 2, despite the factor of $\simeq$$4$ difference in $m_{\rm ej}$. This is in contrast with both momentum-conserving arguments and with the low kicks obtained through multi-dimensional SN explosions from ultra-stripped progenitors.

Recently, \citet{giacobbo2020} proposed a \textbf{unified approach} derived from momentum-conserving arguments, inspired by \citet{bray2016,bray2018}. Independent of the progenitor, the nature of compact remnant, and the SN explosion engine, the birth kick ($v_{\rm k}$) is expressed as
\begin{equation}
    v_{\rm k}=v_{\rm NS} \frac{m_{\rm ej}}{m_{\rm rem}}\frac{\left<m_{\rm NS}\right>}{\left<m_{\rm ej}\right>}
\end{equation}
where $\left<m_{\rm NS}\right>$ ($\left<m_{\rm ej}\right>$) is the average NS (ejecta) mass obtained from a population of isolated stars. This method naturally reproduces the bi-modal kick distribution obtained from observations: it predicts small kicks for compact objects from ECSNe and ultra-stripped progenitors (for which $m_{\rm ej}\ll \left<m_{\rm ej}\right>$), and the normalization $\left<m_{\rm NS}\right>/\left<m_{\rm ej}\right>$ ensures $v_{\rm k}\simeq v_{\rm NS}$ for isolated progenitors, so that kicks match those observed for Galactic pulsars.

Constraints on the kick magnitudes can be obtained through multi-dimensional simulations of SN explosions, though, as already discussed in Section~\ref{sec:corecollapsesne}, these sophisticated simulations are complex and highly uncertain, thus their results cannot be considered as~conclusive.

\subsection{Spins} \label{subsection:spins}
The topic of spins of compact objects would deserve a review in its own right. Here, we discuss the main aspects that are relevant for the current catalog of GW detections, thus we focus mainly on BHs.

\textls[-15]{The spin rate of compact objects at birth is very uncertain, because it depends on the angular momentum transport mechanisms in the stellar interior, the efficiency of which is still matter of debate.
Asteroseismic observations (\cite{beck2012,mosser2012,deheuvels2014,trianna2017,hermes2017,gehan2018}) have shown that the spin rates of red giant cores and WDs are slower than theoretically predicted by nearly all angular momentum transport mechanisms, such as meridional circulation, shear instabilities or propagation of gravity waves \cite{eggenberger2012,marquees2013,ceillier2013,tayar2013,cantiello2014,fuller2014,Belkacem2015,fuller2015,spada2016,eggenberger2017,ouazzani2019}. The magnetohydrodynamical instability known as the Tayler–Spruit dynamo \cite{tayler1973,spruit2002} can provide more efficient angular momentum transport than other mechanisms, albeit early works predicted spin rates roughly an order of magnitude too large \cite{cantiello2014}. \rev{The typical BH natal spins predicted by the Tayler-Spruit dynamo are $\chi$$\sim$$0.05$--$0.15$}. \citet{fuller2019_A} and \citet{fuller2019} (see also \citet{heger2005}) }show that the Tayler–Spruit instability can persist in red giant branch stars despite the existence of strong composition gradients, and they argue that its growth will saturate in a different manner than proposed by \citet{tayler1973} and \citet{spruit2002}. \rev{In their formulation, sometimes referred to as the modiﬁed Tayler--Spruit dynamo, the instability can grow to larger amplitudes and produce stronger magnetic torques, which lead to an even more efficient angular momentum transport and lower natal spins ($\chi$$\sim$$0.01$). The modified Tayler–Spruit instability proposed by \citet{fuller2019_A} may correctly explain the angular momentum contained in the core of stars, and hence the spin of the BH that is formed upon the collapse of massive stars.} \textls[-15]{\citet{eggenberger2019} shows that the revised prescription for the transport by the Tayler–Spruit instability does not provide a complete solution to the missing angular-momentum transport revealed by asteroseismology of evolved stars.} However, asteroseismic observations and new detailed theoretical models show that the angular momentum transport from the cores to the envelopes of massive stars may be very efficient, thus the first-born BH in a binary may form with low spin (e.g., \citep{qin2018, fuller2019_A, fuller2019, eggenberger2019, belc2020b}).
The GW observations of low effective spins in most BBHs also hint at an efficient angular momentum transport in their progenitors~\cite{miller2020,abbott2021,lvcgwtc3_2021,zevin2021}.

\textls[-15]{Stars in binaries may spin up via tides, while compact objects in binaries may spin up though accretion.
For instance, first-generation, highly spinning BHs can form if their progenitor stars become chemically homogeneous due to tides in close binaries, but BBHs originating from this channel have large and nearly equal masses \citep{demink2016, mandel2016a, marchant2016}.
BHs in X-ray binaries tend to spin faster than BHs observed through GWs, but whether these systems are distinct populations or two sides of the same coin is still matter of debate \citep{belc2021,fishbach2021}.
Spinning up first-born BHs in binaries would require a significant amount of accretion, which is unlikely to occur over the short evolutionary timescale of the massive, non-degenerate companion star or during a common-envelope phase (e.g., \citep{macleod2014}). 
\rev{However, \citet{olejak2021} find out that evolutionary sequences that do not involve a CE phase can still lead to an appreciable fraction (${\sim}10\%$) of systems that are spun up via tides, reaching $\chi \gtrsim 0.4$, considering the classic Tayler-Spruit angular momentum transport. \citet{bavera2021}} consider highly efficient angular momentum transport with the modified Tayler-Spruit dynamo, and obtain high spins for the second-born BH at low metallicity due to tidal spin-up. Conversely, they find that at high metallicity the second-born BH has a negligible spin because of wind mass-loss that spins down the progenitor and widens the binaries, weakening tidal interactions.}

Finally, the merger remnants of coalescing BBHs will be spinning even if the progenitor BHs were non-spinning. As the two BHs merge, part of the angular momentum of the binary is converted into the spin of the final BH. This has important implications for the signatures of \textbf{hierarchical mergers}, wherein GW events are produced by second or higher-generation BHs formed from the coalescence of BBHs, rather than from SN explosions of their progenitor stars. 
\rev{Repeated mergers of BHs can thus produce higher and higher spinning remnants, and naively one might expect to achieve maximally spinning BHs, i.e., BHs with dimensionless spin $\chi \simeq 1$. However, this is not the case, because the spin of the final remnant depends also also on the spin of the progenitor BHs and their relative orientation with respect to the binary angular momentum vector. Spins anti-aligned with the binary angular momentum will subtract from the total angular momentum budget of the final BH. Therefore, we expect hierarchical mergers to produce, after several generations of mergers, BHs with an average of $\chi \simeq 0.7$ \citep{berti2008,gerosa2017,rodriguez2018,rodriguez2019,fishbach2017,mapelli_hier2021,tagawa2021_hier}. The latter is true for nearly-equal mass mergers of higher generation BHs. On the other hand, if many first generation BHs coalesce into a single, massive merger product (as massive BH runaway formation scenarios, e.g., see \cite{ebisuzaki2001}), the final BH spin will decrease on average. This is because, at next-to-leading order, the decrease in BH spin is proportional to the mass ratio of the binary, and thus on average the spin distribution of merger products will decrease after asymmetric mergers (e.g., \citep{hughes2003}).
Any hierarchical merger scenario, requires mechanisms to assemble higher-generation BHs into merging binaries, which will be described later in Section~\ref{sec:dynchannel}.}

The spins of merging BBHs can be probed via GW observations.
GWs carry information about the spin magnitude and orientation of the merging BHs into two phenomenological parameters: the effective inspiral spin parameter $\chi_\mathrm{eff}$ and the effective precession parameter $\chi_\mathrm{p}$ \citep{damour2001,racine2008,schmidt2012,schmidt2015}. 
The effective inspiral spin parameter relates to the component of the BHs' spins aligned to angular momentum of the binary orbit, and can be expressed as:
\begin{equation}
\chi_\mathrm{eff} = \frac{m_1 \chi_1 \cos{\theta_1} + m_2 \chi_2 \cos{\theta_2} }{ m_1 + m_2}
\end{equation}
where $m_1$ and $m_2$ are the masses of the two BHs, $\chi_1$ and $\chi_2$ are the two dimensionless spins, and $\theta_1$ and $\theta_2$ are the obliquities, i.e., the angle between the spin direction and the normal to the plane of the binary.
The effective spin parameter remains approximatively constant during the inspiral, and it can be used to constrain the individual spins of the BHs. Note that the effective spin parameter is the mass-weighted average of the spin component parallel to the angular momentum vector. Consequently, high-mass-ratio inspirals will mostly carry information about the primary's spin (e.g., \rev{GW190412}, GW190814, \rev{Sections~\ref{subsection:gw190814} and \ref{subsection:gw190412}}).

The second phenomenological parameter, $\chi_\mathrm{p}$, relates to the orbital precession caused by the in-plane spin component, and is expressed as:
\begin{equation}
    \chi_\mathrm{p} = \max{\left[ \chi_1 \sin{\theta_1}, \chi_2 \sin{\theta_2}\frac{ 4q + 3}{4 + 3q}q \right]}
\end{equation}
where the indices $_1$ and $_2$ refer to the primary and secondary BHs, so that $m_1 > m_2$, and $q = m_2 / m_1 \leq 1 $ is the mass ratio.

Correlations between $\chi_{\rm eff}$ and other parameters can be used to disentangle the formation scenarios of GW sources. For example, \citet{zevin2022} find that isolated binary evolution cannot produce BBHs with significant mass asymmetry and high spins in the primary BHs, unless either inefficient angular momentum transport or super-Eddington accretion are assumed \cite{bavera2021}.
The active galactic nuclei (AGN) scenario instead predicts values of $\chi_{\rm eff}$ that are anti-correlated with the mass ratio $q$, which is roughly consistent with the observed distribution \citep{tagawa2020a,mckernan2021}. \rev{We caution that these studies focus on a small region of the available parameter space of BBH mergers, and  more systematic investigations will be required for a better understanding of the correlations between $\chi_{\rm eff}$ and the other observables.}

\section{Binary Stars}\label{sec: binarystars}

As briefly discussed in Section~\ref{sec:introduction}, two main mechanisms to pair compact objects have been proposed. The isolated binary evolution pathway, in which the GW progenitors evolve through various binary evolution processes in isolation, and the dynamical scenario, where the evolution of binaries is mediated by gravitational interactions with other bodies. Here we focus on the isolated binary mechanism, and leave the discussion of the dynamical scenarios for Section~\ref{sec:dynchannel}. We note, however, that some of the formation pathways that have been proposed so far blur the lines between these two categories, requiring elements of both binary stellar evolution and gravitational dynamics.

Many stars, especially the more massive ones, are born in binaries or higher multiple stellar systems. \citet{moe2017} showed that the multiplicity of stellar systems increases with the stellar mass. This crucial result suggests that \textbf{most BH and NS progenitors are not isolated}, but members of binaries, triples, and even quadruple stellar systems. Studying the interactions between close stars is crucial to understand the evolutionary histories of GW mergers.

\textls[-20]{The timescales of GW coalescence were derived analytically by \citet{peters1963},} which showed that a circular BBH with masses $m_1$, $m_2$ will coalesce in a time $t_{\rm GW}$ given by:
\begin{equation}\label{eq:peters}
	t_{\rm GW} = \frac{5}{256} \frac{c^5 a^4}{G^3 m_1 m_2 (m_1 + m_2)}	
\end{equation}
where $a$ is the semimajor axis of the binary. Considering two BHs of $10\, \msun$, it follows that, in order to merge within ${\sim} 13 \gyr$, the separation must be smaller than ${\sim} 0.1 \au$. At solar metallicity, the stellar progenitors of $10\, \msun$ BHs have about $30\, \msun$ ZAMS mass and a radius of ${\sim}20\, \rsun \simeq 0.18 \au$. It is apparent that such a binary could not have been born at a separation of ${\lesssim}0.1 \au$, because the stars would have collided during the main-sequence phase, even without considering the following giant phase. 
Therefore, the progenitors of merging BBHs from isolated stellar evolution must have been born at wider separations, and subsequently brought to smaller separations by various mechanisms. Here, we begin by describing the evolutionary processes that can affect binary stars.

\subsection{Stellar Tides}

When a stellar binary is very tight, \textbf{the point mass approximation is not enough} to describe its motion, because finite-size effects (i.e., \textbf{tidal forces}) become significant. An elegant derivation of the equations of motion for a binary affected by tides can be found in~\citep{eggleton1998, hut1981}. The main idea behind these equations is that the star is deformed by its companion, generating a gravitational quadrupole moment. Due to dissipation sources in the stellar interior, the response of the quadrupole moment is not instantaneous with respect to the tidal field. 
This delay, called time-lag, allows the coupling between the rotational and orbital angular momenta, in addition to the dissipation of orbital energy in the stellar interior.

While the equations of \cite{hut1981} have been used to model a variety of different tidal dissipation mechanisms, the precise source of tidal dissipation depends on the stellar structure \citep{zahn1977}.
Tidal dissipation in convective layers occurs via the convective motion of large eddies. The convective flows counteracts the tidal flow, which gives rise to dissipation and the lag of the tidal bulges. This kind of tide is referred to as \textbf{the equilibrium tide}.
For this reason, to quantify the tidal dissipation of evolved stars, we need to characterize the timescale of the convective motion, which is the eddie turnover timescale $\tau_\mathrm{conv}$. This time scale can be calculated in several ways, either from the bulk properties of the star e.g.,~\cite{rasio1996,hurley2002} or from the mixing length parameters adopted in the stellar models \cite{brandenburg2017}.

In stars with a radiative envelope, the source of tidal dissipation is the damping of low-frequency gravity waves near the surface of the star \citep{zahn1975}. This kind of tide, called \textbf{dynamical tide}, is generally modeled following \citet{hut1981}, who relies on the ideas of quadrupole deformation and time lag, which are more suitable to describe the equilibrium tide. Nonetheless, just like the equilibrium tide, the tidal dissipation constants of the dynamical tide depend on the details of the stellar structure. The dissipation rate of the dynamical tide scales linearly with a dimensionless tidal torque constant, named $E_2$, which must be calculated from the stellar density profile e.g., \cite{claret1997,siess2013}.
Tabulated values for $E_2$ were provided by \citet{zahn1975}, and were later fitted as a function of stellar mass by \citet{hurley2002}, to use in population-synthesis codes. More recent fitting formulae can be found in \citet{qin2018}, which, in turn, are based on the ones of \citet{yoon2010}. An alternative formulation for the dynamical tide, which avoids entirely the tidal torque constant $E_2$, was proposed by \citet{kushnir2017}. 

Compact stars (WDs, NSs) also experience tides, although their rates of tidal dissipation are poorly constrained \cite{hurley2002,piran1996,Thorne1977,Fryer1996}.

The main effects of tides are the following. First, they tend to \textbf{circularize eccentric binaries}, shrinking their semimajor axis. Second, they tend to \textbf{spin-up stars in close binaries}, synchronizing their rotation period to the orbital period, and aligning the spin directions with the angular momentum vector of the binary.  Both effects are especially important in the context of GWs. Specifically, tidal spin-up can change both magnitude and orientation of the spins of compact objects with respect to the orbital angular momentum vector, and GW observations may give us insights into  these two parameters (see also Section~\ref{subsection:spins}). 

Finally, another crucial consequence of tides is that they can \textbf{radically change the structure and evolution of a star}. Tidal spin-up in a close binary introduces rotational mixing of the stellar interior, which tends to flatten its chemical composition gradient. For very close massive binaries, rotational mixing drives large-scale Eddington–Sweet circulations \cite{endal1978,zahn1992}, so that the entire star is fully mixed. These stars undergo \textbf{chemical
homogeneous evolution (CHE)}, which has been proposed as a formation pathway for BBHs \cite{demink2008a,demink2008b,demink2009,demink2016,mandel2016a,marchant2016}. Chemically-homogeneous stars skip entirely the evolved giant phase because they do not develop a core-envelope boundary. Since such stars remain compact even during the post-MS phases, they can evolve very close to each other without merging via unstable mass transfer (Section~\ref{sec:masstrans}). Therefore, CHE can produce BBHs that merge within the age of the Universe.
Because this scenario involves tight binaries with synchronized spins, it predicts BH mergers with large aligned spins. It also favors high BH masses  ($>$20 M$_\odot$) and nearly equal mass ratios ($q \simeq 1$).

\subsection{Mass Loss, Mass Transfer and Accretion}\label{sec:masstrans}

As detailed in Section~\ref{sec:singlewinds}, stars lose mass through stellar winds. In binaries, such mass loss leads to changes in the orbit of the binary. If the mass loss by winds is isotropic (i.e., no net change in momentum) and adiabatic (slow with respect to the orbital period), the rate of change in orbital semimajor axis $a$ is:
\begin{equation}\label{eq:semimajormloss}
	\dot{a} = - a \, \frac{\dot{m}_1 + \dot{m}_2}{m_1 + m_2}
\end{equation}
where $\dot{m}_1$ and $\dot{m}_2$ are the (negative) mass change rates of the two stars. If the mass loss is slow, the only effect of mass loss is the increase in size of the binary, while its eccentricity remains constant \citep{hadjidemetriou1963,dosopoulou2016}.

However, the wind lost by one star may be partially accreted by its companion, which introduces a positive rate of mass change. In addition, part of the accreted material may carry linear and angular momentum, further affecting the binary orbit.
\textbf{The wind accretion rate} can be calculated using the \citet{bondi1944} accretion model. Given a binary with eccentricity $e$, donor wind speed $v_{\rm w}$, and mean orbital velocity $v_{\rm circ}=\sqrt{G\,(m_1+m_2)/a}$, \citet{hurley2002} approximate the mass accretion rate as: 
\begin{equation}
	\dot{m}_2=\frac{1}{\sqrt{1-e^2}} \left(\frac{G\,{}m_2}{v_{\rm w}^2}\right)^2\,{}\frac{\alpha_{\rm w}}{2\,{}a^2}\,{}\frac{\left|\dot{m}_{\rm 1}\right|}{[1+(v_{\rm circ}/v_{\rm w})^2]^{3/2}}
\end{equation}
where $G$ is the gravitational constant, $\alpha_{\rm w}\sim{}3/2$ is an efficiency constant,  and $\dot{m}_1$ is the donor mass loss rate due to stellar winds. Understanding how the orbit responds to mass transfer is complex, because it depends not only on the amount of mass transferred or lost, but also on the linear and angular momentum that is carried out or accreted. Detailed equations for the orbital response including various mass transfer models were recently developed by \citet{dosopoulou2016,dosopoulou2016b} and \citet{hamers2019_masstrans}.

Another way to transfer mass from a star to its companion is via \textbf{Roche lobe overflow}. If the stellar radius is relatively large compared to the size of the binary, the external layers of the star may be stripped out by the gravity of the companion star and the centrifugal force of the binary motion. The region in space where this occurs is approximated by the \textbf{Roche lobe}, the equipotential surface shaped like two tear-drops that surround both stars, with the two lobes connected by a saddle point at the center (also known as the first Lagrangian point, $L_1$) \cite{sepinsky2007,sepinsky2009,sepinsky2010}.
In general, Roche-lobe overflow can be caused by either the primary star entering the giant phase and increasing in radius, or by the shrinking of the binary orbit due to tides.

\textls[-25]{Commonly, the Roche lobe is approximated as an equal-volume sphere of effective size $R_{\rm L}$, the \textbf{Roche radius}. A convenient analytic approximation for $R_{\rm L}$ was given by~\citet{eggleton1983}:}
\begin{equation}\label{eq:rocherad}
	R_{\rm L,1}=a\,\frac{0.49\,q^{2/3}}{0.6\,q^{2/3}+\ln{\left(1+q^{1/3}\right)}},
\end{equation}
where $q=m_1/m_2$ is the mass ratio, and $R_{\rm L,1}$ corresponds to the Roche radius of the star of mass $m_1$.
If the radius of one of the stars exceeds its Roche radius, some material will flow through the saddle point. Part of the material will be accreted by the companion star, while some material will be dispersed in a circumbinary disk.
If all the mass lost by one star is accreted by the other and no mass is dispersed, we are in the case of \textbf{conservative mass transfer}. The material that is lost during \textbf{non-conservative mass transfer} will carry out not only mass but also angular momentum from the binary.

The typical rate at which mass transfer proceeds through $L_1$ can be estimated, at \textit{order of magnitude}, through Bernoulli's equation, assuming an isentropic, adiabatic, and irrotational fluid, and that the velocity of the flow is parallel to the axis connecting the centers of the two stars. Under these assumptions, the mass transfer rate can be expressed as (e.g., \citet{ge2010})
\begin{equation}\label{eq:fillingroche}
    \dot{m}_1\simeq -\frac{m_1}{P_{\rm orb}}\left(\frac{\Delta R}{R_{\rm L,1}}\right)^{n+\frac{3}{2}},
\end{equation}
where $P_{\rm orb}$ is the binary orbital period, $\Delta R = R_1 - R_{\rm L,1}$ is the Roche-lobe filling factor, and $n$ and $R_1$ are the envelope's polytropic index and the radius of the donor star, respectively.

Besides changing the binary orbit, mass loss and accretion via Roche lobe overflow will change the structure of both the donor and the accretor. Modeling \textbf{the response of the donor star to mass transfer is crucial} to predict the evolution of binary stars but it is also very challenging since it requires an in-depth knowledge of the internal structure of the star and possibly non-equilibrium solutions for it.

\rev{
To assess the response of the star, we must carefully consider the relevant timescales of stellar evolution. The shortest characteristic time is the dynamical timescale, which is the time on which a star reacts to a perturbation to the hydrostatic equilibrium: 
\begin{equation}\label{eq:dyntimescale}
	t_\mathrm{dyn}  \simeq \sqrt{\frac{R_1}{g}} \simeq \sqrt{\frac{R_1^3}{G m_1}} \,,
\end{equation}
where $g\approx G m_1 /R_1^2$ is the gravitational acceleration at the stellar surface. The dynamical timescale can be also interpreted as the timescale for the star to collapse (or `free-fall') if the gas pressure would suddenly disappear. For the Sun, this timescale is about $\tau_\mathrm{dyn} \approx 0.5\rm\,h$, which means that main sequence stars are extremely close to hydrostatic equilibrium.

The thermal timescale, also known as the Kelvin-Helmholtz timescale, describes how fast the changes in the thermal structure of a star can be. A star without a nuclear energy source contracts by radiating away its internal energy content. The Kelvin-Helmholtz timescale is the timescale at which this gravitational contraction occurs:
\begin{equation}\label{eq:thermtimescale}
	t_\mathrm{KH} \simeq \frac{E_{\rm int}}{L} \simeq \frac{G m_1^2}{R_1 L} \,,
\end{equation}
where $E_{\rm int} \approx G m_1^2 / R_1$ is the gravitational energy budget of the star, and $L$ is the stellar luminosity. For the Sun, the Kelvin-Helmholtz timescale turns out to be $t_\mathrm{KH} \approx 15 \rm\,Myr$.

Finally, the longest stellar timescale is the nuclear timescale, which is the timescale for the nuclear fuel to be exhausted. The energy source of nuclear fusion is the direct conversion of a small fraction $\phi$ of the rest mass of the reacting nuclei into energy. For hydrogen fusion, $\phi \approx 0.007$, while for helium fusion it is about 10 times smaller. The total nuclear energy supply can be then written as $E_{\rm nuc} = \phi f_{\rm nuc} m_1 c^2$, where $f_{\rm nuc}$ is the fraction of stellar mass that can serve as nuclear fuel (${\sim}$10\%). The nuclear timescale can be written~as
\begin{equation}\label{eq:nuctimescale}
	t_{\rm MS} \simeq \frac{E_{\rm nuc}}{L} = \phi f_{\rm nuc} \frac{m_1 c^2}{L}.
\end{equation}
{Using} solar values, we obtain $t_{\rm MS} \approx 10 \rm\, Gyr$, which is consistent with the time the Sun will spend on the main~sequence.
}

During mass transfer, stars continue their evolution in thermal equilibrium if the mass loss rate remains above the \textbf{nuclear timescale} but well below the \textbf{thermal timescale} for mass transfer, that is
\begin{equation}
   t_{\rm MS} \simeq \frac{\phi X m_{1,\rm c}c^2}{L} < \frac{m_1}{\left|\dot{m}_1\right|} < \frac{Gm_1^2}{R_1L} \simeq t_{\rm KH},
\end{equation}
where $X$ is the initial hydrogen fraction, $m_{1,\rm c}$ the mass of the stellar core, $L$ the luminosity of the star, and $c$ the speed of light.
However, if the mass loss timescale becomes comparable or larger than the \textbf{thermal timescale}, but remains below the \textbf{dynamical timescale}, that is 

\begin{equation}
    t_{\rm KH} < \frac{m_1}{\left|\dot{m}_1\right|} < \frac{1}{\sqrt{G\rho}} \simeq t_{\rm dyn},
\end{equation}
the star cannot be considered in thermal equilibrium anymore and its response in terms of luminosity and radius variation can be significantly different.
In the most extreme cases, mass transfer can reach values as high as the dynamical timescale,
crucially altering the response of the donor.

The mass-loss rate depends on the Roche-filling factor (see Equation~\eqref{eq:fillingroche}), which, in turn, depends crucially on how the Roche-lobe radius (Equation~\eqref{eq:rocherad}) and the donor's radius respond to mass transfer/loss. Generally speaking, if during mass transfer the Roche lobe contracts more than the star does, then the Roche lobe filling factor increases and mass transfer accelerates. Such relative contractions/expansions are evaluated through the \textbf{$\zeta$ coefficients}:
\begin{equation}
    \zeta_*\equiv\dlog{R_1}{m_1}, \,\,\,\,\, \zeta_{\rm L}\equiv \dlog{R_{\rm L}}{m_1}
\end{equation}
where $\zeta_*=\zeta_{\rm eq},\zeta_{\rm th}, \zeta_{\rm ad}$, depending on whether the star is in equilibrium, out of thermal equilibrium, or out of both thermal and hydrodynamical equilibrium, respectively.

\subsubsection*{Approximate Solutions for Population Synthesis Simulations}
The $\zeta_*$ coefficient is a complex function that depends on many physical stellar parameters, including the star's evolutionary phase. An on-the-fly, self-consistent calculation of such coefficients in population-synthesis simulations is prohibitive, considering also that $\zeta_{\rm th}$ and $\zeta_{\rm ad}$ refer to non-equilibrium states. 

The response of the star in population-synthesis codes is approximated considering the star always in both thermal and hydrodynamical equilibrium, thus the same fitting formulas or look-up tables for single stars apply to stripped/oversized stars in binaries. Furthermore, to predict whether the mass transfer will remain stable (say, at $\sim$ constant $\Delta R$) or not (e.g., runaway Roche-lobe filling), such codes adopt a criterion based on a \textbf{critical mass ratio, $q_{\rm crit}$.} Indeed, if we assume that mass transfer is conservative, i.e., $M = m_1 + m_2 = {\rm const}.$, and that the total angular momentum of the binary ($J$) is conserved as well, then
\begin{equation}\label{eq:atilde}
    a = \widetilde{a}\frac{\left(1+q\right)^4}{q^2} \,\,\,\, \mathrm{and}\,\,\,\, \zeta_{\rm L}\left(q\right) = \left(1 + q\right) \dlog{R_{\rm L}}{q}\,\,\, \mathrm{with}\,\, \widetilde{a}=\frac{J^2}{GM^3}
\end{equation}
and thus, the condition for unstable (hydrodynamical) mass transfer can be thought of as a condition that depends \textit{only} on the mass ratio, that is

\begin{equation}
    \zeta_{\rm L}\left(q_{\rm crit}\right) > \zeta_{\rm ad}.
\end{equation}

The $\zeta_{\rm ad}$ values are generally approximated with simplified fitting formulas or constant values that depend on the stellar evolution phase, and they are generally calibrated to match the merger rate of compact objects inferred by the LVK and/or on other Galactic observations (e.g., \citet{hurley2002,leiner2021,neijssel2019}).

The values of $q_{\rm crit}$ have a dramatic impact on merging compact-object binaries especially because, if the mass transfer is unstable, the binary orbit shrinks on a dynamical timescale, and the stars are destined to merge. However, if the donor is an evolved star, constituted by a massive core surrounded by a low-density envelope, the binary enters a further phase of binary evolution that has crucial consequences for compact objects: the CE~phase.

\begin{figure}[H]
    
    \includegraphics[width=\columnwidth]{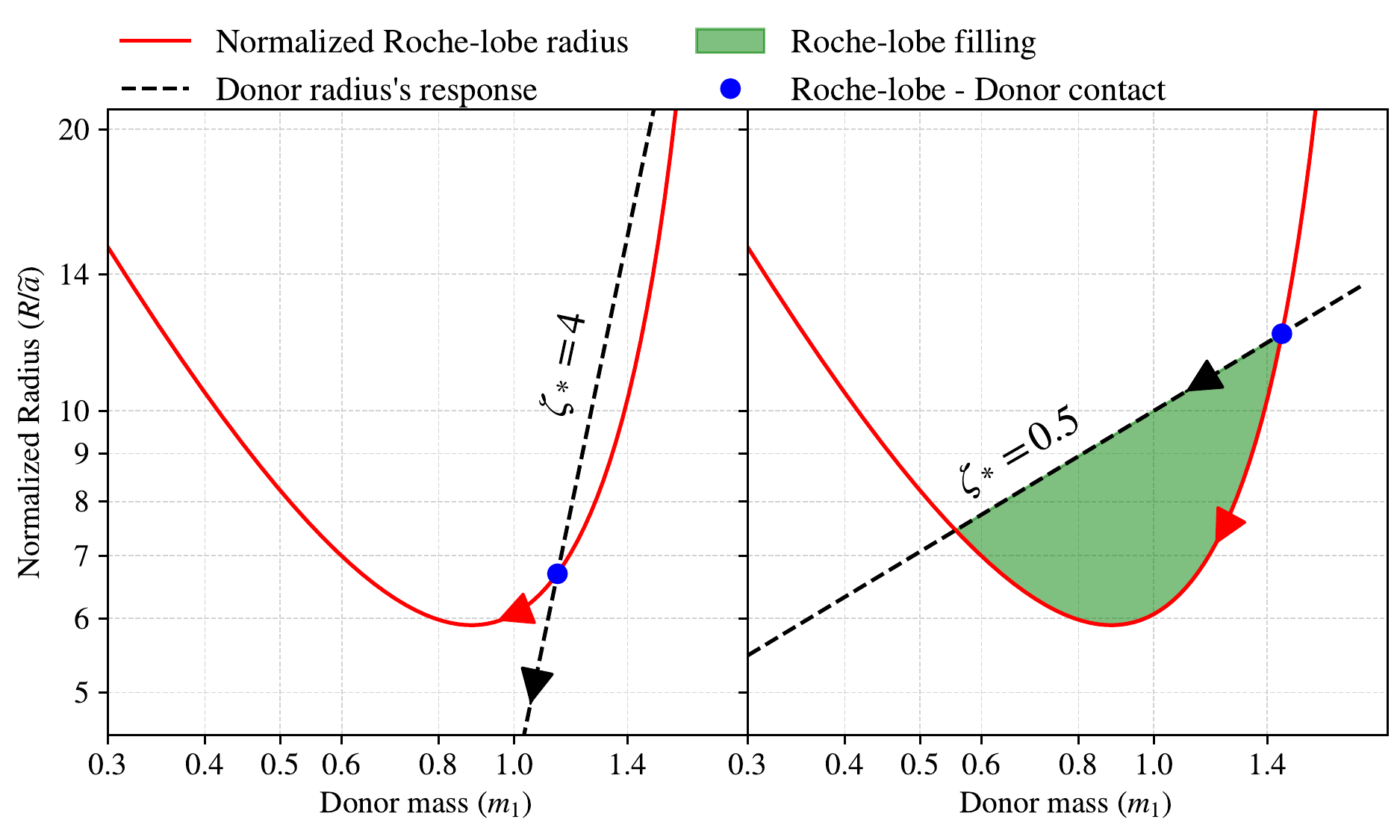}
    \caption{Normalized Roche-lobe radius (solid red line) and donor radius's response (dashed black line) as a function of the donor mass \rev{in arbitrary units}. Left (right) panel: donor star with steep (shallow) response to mass stripping, $\zeta_*=4$ ($\zeta_*=0.5$), and mass $m_1\simeq 1.14$ ($m_1\simeq1.5$). The blue dot is where the donor fills its Roche lobe. The red (black) arrow shows the direction in which the Roche-lobe (donor) radius evolves after mass stripping. The shaded green area represents the amount of Roche-lobe filling. The total angular momentum and the total mass of the system are taken as constants.}
    \label{fig:RLfilling}
\end{figure}

\rev{
Figure \ref{fig:RLfilling} is a cartoon that shows the normalized Roche-lobe radius ($R_{\rm L}/\widetilde{a}$, see Equation~\eqref{eq:atilde}) and the donor's radius response as a function of the mass of the donor star, under the assumptions $J=\mathrm{constant}$ and $m_1 + m_2 = 2 =\mathrm{constant}$, with masses in arbitrary units. In the left panel, a donor with a very steep response to mass stripping ($\zeta_*=4$) and mass $m_1\simeq 1.14$ fills its Roche lobe in correspondence to the blue point in the figure. As soon as mass transfer initiates, the star contracts significantly more (black arrow) than $R_{\rm L}$ does (red arrow), i.e., $\zeta_{\rm L} < \zeta_*$. Thus, the star remains in thermal equilibrium, and mass transfer stops until nuclear reactions cause a further expansion of the donor. This is a typical case of (nuclear) stable mass transfer. In the right panel, a star with a shallower response ($\zeta_*=0.5$) and $m_1\simeq 1.5$ is considered. In this case, $R_{\rm L}$ shrinks more than $R_*$ (i.e., $\zeta_{\rm L}>\zeta_*$), therefore increasing the Roche-lobe filling factor and the mass loss rate. This is a typical case of the beginning of a thermally unstable mass transfer, for which $\zeta_*=\zeta_{\rm th}$. By looking at the right panel of Figure~\ref{fig:RLfilling}, this configuration might last until $m_1\simeq 0.55 \,\msun$, though, depending on the $R_L$-filling factor, mass loss can become fast enough to drive the donor out of dynamical equilibrium, so that $\zeta_* = \zeta_{\rm ad}$. For typical giant stars, mass loss easily evolves towards a very high rate. This happens because $\zeta_{\rm ad}\simeq -0.3$ (e.g., \citep{hurley2002}), thus, such donors easily initiate a runaway $R_{\rm L}$-filling process that likely leads to CE evolution.
}
\vspace{-6pt}

\subsection{Common Envelope}\label{sec:CE}

CE is the process by which one component of a binary gets engulfed in the envelope of its companion \cite{webbink1984,vignagomez2018,postnov2014,dominik2012, ivanova2013}. The gaseous envelope becomes gravitationally focused and exerts a drag force onto the secondary, which can be considered as a second core within the envelope of the primary. Because of the drag, the two cores begin an inspiral phase, during which orbital energy and angular momentum are transferred to the envelope, which heats up and expands. 
After the inspiral sets, only two outcomes are possible. If the envelope is too tightly bound, the inspiral continues until the two cores are tidally disrupted and the binary merges into a single star. 
If, instead, the envelope is ejected, a short-period binary forms.

\textbf{The CE is a key process in the formation of GW events} from isolated binary stars, because it can shrink binary separations by a factor of hundreds, decreasing the coalescence time of compact-object binaries \cite{dominik2015,rodriguez2016,giacobbo2018a,mapelli2018,belc2020b,tanikawa2021a,zevin2021}. \rev{In particular, \citet{dominik2012} found that the coalescence time distribution of post-CE compact object binaries approximately follows a log-uniform distribution, $n(t_{\rm GW}) \propto t^{-1}_{\rm GW}$. The reason for this peculiar scaling comes from Equation~\eqref{eq:peters}, which imposes $t_{\rm GW} \propto a^4$. If we assume that the distribution of semimajor axis of post-CE binaries follows a power-law as in 
\begin{equation}
    \frac{dN}{da} \propto a^{-\gamma},
\end{equation}
then, from Equation~\eqref{eq:peters}, the distribution of coalescence times can be written as
\begin{equation}
    \frac{dN~~~}{dt_{\rm GW}} \propto \frac{dN}{da} \frac{da~~~}{dt_{\rm GW}} \propto t_{\rm GW}^{-(3+\gamma)/4}.
\end{equation}
{Therefore, }even a moderately steep ($\gamma \simeq 1$--2) distribution of post-CE semimajor axes will result in a coalescence time distribution very close to a log-uniform distribution.
}

CE evolution is also important for explaining other astrophysical phenomena related to short period binaries, such as supernovae Ia, X-ray binaries and double neutron stars. In fact, the CE phase was first introduced to explain the existence of short-period dwarf binaries \cite{paczynski1976}.

Due to the complexity of CE evolution, we are still lacking a complete theoretical picture to distinguish between its two possible outcomes---close binary formation or binary merger. Attempts at modeling the CE phase have been mostly limited to two approaches: hydrodynamic simulations \cite{sandquist1998,ricker2012,passy2012,ivanova2016,ohlmann2017,iaconi2018,reichardt2019,reichardt2020,glanz2021a} or analytic prescriptions \cite{vanDenHeuvel1976,paczynski1976,webbink1984,livio1988,dekool1990,nelemans2000,nelemans2005}. The latter are particularly convenient because they can be applied to a wide sample of binaries with minimal computing effort.

Hydrodynamic simulations may appear as the best way to model CE evolution, however they come with their limitations. First, hydrodynamic codes can follow only the initial plunge-in stage of CE evolution, which advances on a dynamical timescale. However, the later stages of the CE phase, where the envelope is heated up and expands, proceed on a thermal timescale. In these later stages, the energy deposited by the inspiral is transported to the surface of the star and radiated away. Therefore, modeling this stage requires proper treatment of radiative and convective energy transfer  and equation of state for the stellar layers. The pre-CE phase, wherein the binary undergoes unstable mass transfer or other structural instabilities, is also difficult to model with hydrodynamic codes. In fact, many hydrodynamic simulations place the companion stars on a grazing orbit, artificially triggering the plunge-in stage. This approach, however, misses important physics of the early CE stage, such as tidal effects and mass transfer, which can significantly change the stellar structure of the stars prior to the inspiral. Furthermore, hydrodynamic simulations are numerically expensive, which makes them impractical to use for fast population-synthesis calculations.

Analytic models of CE evolution are numerically inexpensive and easy to implement, but they also come with their drawbacks.
The most commonly adopted formalism in fast population-synthesis codes is the  \textbf{$\alpha$--$\lambda$ model} \cite{vanDenHeuvel1976,webbink1984,livio1988}, which is based on energy balance considerations. The main idea of this approach is to compare the orbital energy of the binary at the onset of CE with the binding energy of the envelope. By comparing these two energies, it is possible to determine whether or not the binary will survive the CE and to estimate the final size of the binary.

The model depends on two unknown parameters, $\alpha$ and $\lambda$, which parametrize the CE efficiency (i.e., how efficienty orbital energy is used to unbind the envelope) and envelope binding energy, respectively.
\rev{The binding energy of the envelope is parametrized through the $\lambda$ parameter as \cite{dekool1990}
\begin{equation}\label{eq:bindenergylambda}
	E_{\rm bind} = -\frac{G m_{1} m_{1\rm, env}}{\lambda R},
\end{equation}
where $R$ is the stellar radius, and $m_{1\rm, env}$ is the mass of the stellar envelope.} When both stars are giants at the onset of CE, the binding energy of both envelopes is included in Equation~\eqref{eq:bindenergylambda}. The $\lambda$ parameter can be estimated from polytropic models or fit to detailed 1D stellar evolution models \cite{hurley2002,dewi2000,xu2010,claeys2014,kruckow2018}.

The envelope binding energy is then compared to the difference in orbital energy:
\begin{equation}\label{eq:orbenergy}
	\Delta E_{\rm orb} = -\frac{G m_{1\rm, c} m_2}{2a_{\rm f}} + \frac{G m_1 m_2}{2a_{\rm i}},
\end{equation}
where $a_{\rm i}$ and $a_{\rm f}$ are the pre- and post-CE semimajor axes. $\alpha{}$ is a dimensionless parameter that measures the efficiency of the CE phase, i.e., how much orbital energy is transferred to the envelope. By equating Equations~\eqref{eq:bindenergylambda} and \eqref{eq:orbenergy} we can estimate the size of the final orbit~$a_{\rm f}$:

\rev{
\begin{equation}\label{eq:alphalamb}
	\frac{G m_{1} m_{1\rm, env}}{R} = \lambda\alpha\left(\frac{G m_{1\rm, c} m_2}{2a_{\rm f}} - \frac{G m_1 m_2}{2a_{\rm i}}\right).
\end{equation}

In the above equation, the $\alpha$ parameter appears as a fudge factor that determines how much gravitational binding energy is used up to unbind the binary. For $\alpha=1$, all the binary orbital energy is used to unbind the envelope, while for $\alpha<1$ only a fraction of the orbital energy is transferred to the envelope. In other words, the $\alpha$ parameter controls the efficiency of common envelope inspiral.
}

Because the  $\alpha$ and $\lambda$ parameters appear multiplied together, and are both of order unity, they can be sometime dumped together into a single, unknown parameter. This however neglects differences in envelope binding energies between different stars, which can be instead measured from detailed stellar evolution models \cite{hurley2002,dewi2000,xu2010,claeys2014,kruckow2018}. The $\alpha$ parameter is mostly unconstrained. Some clues on its values can come from observations of post-CE binaries \cite{zorotovic2010,davis2010,davis2012,demarco2011}, and from 3D hydrodynamical simulations of low-mass giant donors (e.g., \citep{nandez2016}). However, large values of the $\alpha$ parameter in population-synthesis simulations ($\gtrsim$3) seem to be necessary to match the merger rate of BNSs inferred by LVK (e.g., \citep{giacobbo2018a}). Some recent one-dimensional hydrodynamic simulations also obtain high values for $\alpha$~\citep{fragos2019}. \rev{
Values of $\alpha$ greater than one deserve a particular attention, because according to Equation~\eqref{eq:alphalamb}, they imply that more energy than available was used up during the inspiral, or in other words, energy was \textit{generated} during CE phase. One important source of energy is the recombination energy, i.e., energy released during the recombination of the hydrogen plasma into atoms and molecules. Recombination energy has been suggested as a potential driving mechanism for the ejection of the envelope, once it has been expanded by viscous and gravitational drags. Other potential energy sources include nuclear and accretion energies.
}

The $\alpha$--$\lambda$ model is a very simplistic approach, because it hides behind two parameters the complex physics processes that happen during CE evolution. One glaring issue is that it assumes that the entire envelope is ejected, while hydrodynamic simulations have shown that partial envelope ejection is possible. The envelope ejection is also fine-tuned to carry away zero kinetic energy at the infinity, which is not necessarily true. Finally, the $\alpha$--$\lambda$ formalism neglects angular momentum transfer, tidal heating, and recombination energy of envelope material \cite{ivanova2013}, \rev{all processes that can affect the outcome of CE evolution. Another shortcoming of the $\alpha$--$\lambda$ is that it cannot reproduce the eccentricities of observed post-CE binaries, which were shown to deviate from a perfect circular orbit \citep{lynch2012,kawka2015,kruckow2021}. Finally, the $\alpha$--$\lambda$ model does not follow the binary inspiral, but prescribes an instantaneous change of the binary orbital parameters. Therefore, it is difficult to incorporate such formalism in numerical models that follow the continuous time-evolution of multiple stellar systems. Recent works have been tackling some of these issues. Progress has been done to include the recombination energy into the $\lambda$ description e.g., \citep{xu2010,claeys2014,kruckow2018}, and alternative models of CE inspiral that can follow the binary angular momentum are being investigated \citep{trani22}.
}

\subsection{Supernovae: Blaauw and Velocity Kicks}
SNe can change the binary orbit because they can impart significant natal kicks to compact objects. \rev{The natal kicks are caused by two main physical mechanisms. 

The first is the impulsive mass loss caused by the (possible) ejection of mass during the SN explosion. The sudden loss of mass can change the orbit and even unbind the binary, and this process is sometimes referred to as the \textbf{Blaauw kick}. This unbinding mechanism was originally proposed to explain runaway O- and B-type stars, because the stars inherit a fraction of the binary orbital velocity following the breakup \cite{blaauw1961}.}

Unlike adiabatic mass loss (Section~\ref{sec:masstrans}), the mass ejection during a SN explosion is an impulsive event. Mass loss decreases the gravitational potential, and can therefore even unbind a tight binary without the need of a velocity kick. A circular binary needs to lose half of its total mass during a SN explosion to become unbound, while for an eccentric binary this will depend on the binary phase at the explosion time \cite{hills1983}. If a binary of total mass $M$ and semimajor axis $a_0$, loses instantaneously a mass $\Delta m$, the semimajor axis $a_1$ after the explosion is:  
\begin{equation}\label{eq:snmassloss_4}
	\frac{a_1}{a_0} = \frac{M-\Delta m }{M - 2 \Delta m \frac{a_0}{r}},
\end{equation}
where $r$ is the distance between the two bodies at the moment of the explosion ($r \equiv a_0$ for a circular binary).

\rev{
The second physical mechanism is the velocity kick caused by \textbf{the asymmetries in the SN ejecta} (see Section~\ref{sec:snkicks}). The asymmetries can also affect the spin of the newly-born compact remnant \cite{janka2012}, by changing its magnitude and orientation.

Unlike the Blaauw kick, the velocity kicks can either tighten or unbind binaries, depending on the relative orientation of the velocity kick with respect to the orbital velocity. Besides changing the orbital energy, velocity kicks also alter the eccentricity of the binary, which can then trigger mass transfer episodes or tides.

Velocity kicks have multiple consequences for the formation of merging compact objects. Besides breaking binaries, they can re-align the binary orbital plane, or misalign it with respect to the stellar spin vectors. 
However, the velocity magnitude of velocity kicks is poorly constrained, especially those of BHs (see Section~\ref{sec:snkicks}).

Strong velocity kicks are the only mechanism that can form merging BBHs with anti-aligned spins in isolated binary evolution \cite{belc2020a,callister2021,steinle2021}. In fact, SN kicks can even flip the orbital plane of the binary, resulting in anti-aligned spin-orbit vectors.
On the other hand, extremely strong kicks are required to produce significantly misaligned and anti-aligned spins in isolated binary evolution. The reason is that BBHs that merge within a Hubble time have very short separations (${\ll}0.1 \,\au$) and therefore high orbital velocities (${\gg}400 \,\rm km\, s^{-1}$). In order to significantly tilt the binary plane, SN kicks need to be comparable to the orbital velocity, but such extremely high BH natal kicks are in tension with the estimates obtained from the proper motion of X-ray binaries with BH companions \cite{repetto2012,repetto2015,mandel2016a} and GW observations (e.g., \cite{wysocki2018}).
}

An important consequence of natal kicks is that they can eject compact objects and binaries from their birth star clusters. This prevents them to pair with other compact objects through dynamical interactions, which are described in the next section.

\begin{figure}[H]
    
    \includegraphics[width=\columnwidth]{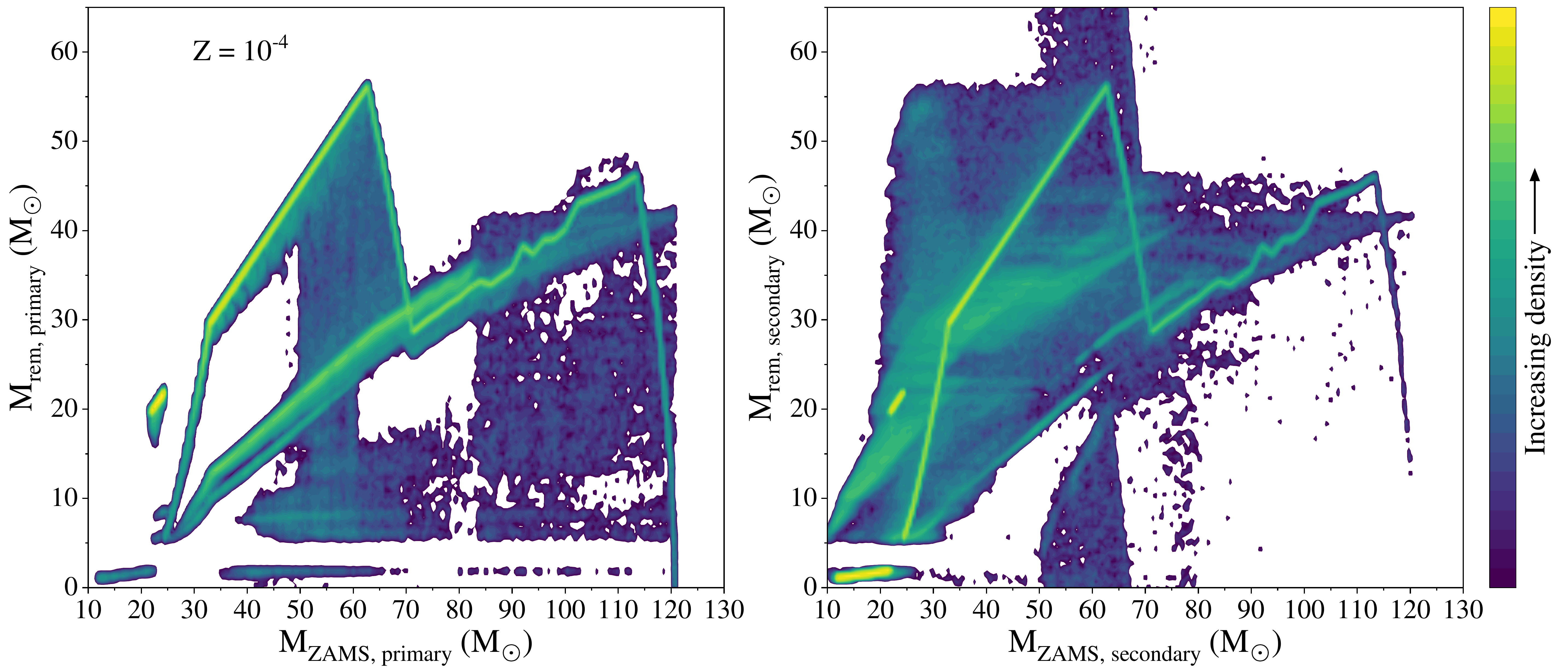}
    \caption{Mass of the compact remnant as a function of the ZAMS mass of its progenitor. The figure has been obtained evolving a population of $10^7$ binary stars at $Z=10^{-4}$ with the SEVN code, as in {\citet{spera2019}.} \rev{The masses of the primary stars are drawn from a Kroupa initial mass function~\citep{kroupa2001} while the secondary masses, initial orbital periods and eccentricities are distributed according to \citep{sana2012}}.  The left (right) panel refers to primary (secondary) stars, i.e., those that have the largest (smallest) $\mzams$ in each binary. The color map represents the number of compact objects at each location of the plot, with counts that increase going from dark to light colors. }
    \label{fig:binarymessup}
\end{figure}

We conclude Section~\ref{sec: binarystars} with Figure \ref{fig:binarymessup}, which shows an example on how binary evolution processes reshape the mass spectrum of compact remnants compared to that obtained for isolated stars. The figure has been obtained by evolving a population of binary stars with the SEVN code, and it has been adapted from \citet{spera2019}. By looking at the left panel of Figure~\ref{fig:binarymessup}, we see that most primaries lose their envelopes through Roche-lobe overflow or CE evolution, thus they form much lighter remnants than those they would have formed as single stars (cfr. curve at $Z=10^{-4}$ in Figure~\ref{fig:spera2017}). In contrast, the secondaries can acquire a significant amount of mass from the primary and they can either retain it and form a heavier compact remnant, or lose it through Roche-lobe mass transfer or CE evolution. The lower the metallicity, the larger the deviations from the mass spectrum from single stars: stellar winds are quenched at low $Z$, thus stars can donate/accrete more mass.

\section{Stellar Dynamics}\label{sec:dynchannel}

Another way to form merging compact-object binaries is through gravitational interactions. Usually, the \textbf{dynamical scenario} involves the formation and hardening of binaries through few-body encounters in stellar clusters. However, in recent years, other forms of dynamical scenarios have been proposed, which involve not only massive star clusters but also small multiple systems like triples and quadruples. 

\subsection{Dense Stellar Environments}

Most, if not all, stars are born in stellar clusters: relatively dense ensembles of stars bound together by gravity.
The evolution of star clusters is mainly driven by long-range gravitational interactions between individual stars, through a process called \textbf{two-body relaxation}. Two-body relaxation leads star clusters to develop a high density core surrounded by a low-density stellar halo.
Given their relatively high central density ($\gtrsim10^3\,\msun$ pc$^{-3}$)~\cite{ladalada2003,portegieszwart2010,ward2020,weidner2006,weidner2009}, the cores of star clusters are the ideal environments for extreme dynamical interactions between stars and binary stars, including stellar collisions, which are unlikely to occur in the galactic field \citep{portegieszwart2010,mapelli2016}. 

Star clusters are distinguished according to their age, density and mass. \textbf{globular
clusters (GCs)} are typically old systems ($\sim$Universe's age, $\sim$12~Gyr), very massive ($\geq$10$^4\,\msun$) and dense (central density $\rho_{\rm c} \geq 10^4\,\msun$). They are evolved systems which do not contain gas, dust or young stars.
Because of their mass and high central density, many studies for the dynamical formation of compact-object binaries focus on GCs \cite{sigurdsson1993,portegieszwart2000,mapelli2005,downing2010,tanikawa2013,breen2013a,samsing2014,rodriguez2016,rodriguez2015,askar2017,samsing2018,askar2018,arcasedda2018,kremer2019,rodriguez2019,longwang2021}.
\textbf{young dense star clusters (YDSCs)} are relatively young ($<$100 Myr) systems, thought to be the most common birthplace of massive stars \cite{ladalada2003, portegieszwart2010}. The central density of YDSCs can be as high as that of GCs, although YDSCs have smaller sizes. Some YDSCs can have comparable masses to present-days GCs and, for this reason, they are thought to be close relatives of the ancient progenitors of GCs. However, because of the stellar mass loss during their evolution, YDSCs are not massive enough to evolve into present-day GCs.
\textbf{open clusters (OCs)} are irregular star clusters composed of 10 --- a~few~$10^3$ stars. They are generally younger than GCs and they may still contain gas from the molecular cloud from which they formed. Studies of compact binary mergers in young and open clusters include \cite{banerjee2010,banerjee2017,banerjee2018,banerjee2021,dicarlo2019a,dicarlo2020,fujii2017,kimpson2016,kumamoto2019,kumamoto2020,mapelli2016,portegieszwart2000,ziosi2014}.
Finally, \textbf{nuclear star clusters (NSCs)} reside in the nuclei of galaxies. Nuclear star clusters are rather common in galaxies, including our own \cite{boker2002,graham2009}. NSCs are usually more massive and denser than GCs, and may host a super-massive black hole (SMBH) at their center. 
Regardless of the presence of a NSC, the environment close to SMBHs can also be a site for formation of GW progenitors. Cusps or disks of BHs may form around SMBHs, where they can interact with other compact objects \cite{bahcall1976,szolgyen2018,gondan2018,leigh2018,trani2019b}. 
In AGN, BHs can be trapped by the SMBH accretion disk, wherein they can migrate and merge~\cite{mckernan2018,tagawa2020a,mckernan2020a,mckernan2020b,ford2021,mckernan2012,mckernan2014,bellovary2016,bartos2017,stone2017,secunda2019,yang2019,leigh2018,tagawa2020b}.


YDSCs and OCs are not long-lived because they inevitably disrupt into the tidal field of their host galaxy as they lose mass during their evolution. Processes that contribute to their disruption include gas expulsion and stellar evaporation. The first may happen during the early life of gas-rich star clusters. Stellar winds and SNe can eject the remaining gas from the cluster, decreasing its gravitational binding and potentially causing its disruption, a process referred to as infant mortality. \textbf{Stellar evaporation} is instead the inevitable consequence of two-body relaxation.

\subsection{Two-Body Relaxation and the Gravothermal Instability}

Two-body relaxation is the consequence of small, but frequent deflections in stellar velocities due to long-range gravitational interactions. Over time, two-body relaxation causes the kinetic energy of stars to be redistributed across the clusters. This process occurs over the timescale \cite{binney2008}:
\begin{equation}\label{eq:twobodyrel}
 t_{\rm rl}=\frac{\sigma^3}{15.4\,G^2\,\langle m\rangle\,\rho\,\ln{\Lambda}} \, ,
\end{equation}
where $\sigma$ is the velocity dispersion, $G$ is the gravitational constant, $\langle m\rangle$ the mean stellar mass, $\rho$ the stellar mass density, and $\ln{\Lambda}$ is the Coulomb logarithm. 
This last term arises from the uncertain range in impact parameters for the interactions that drive the two-body relaxation, and it is generally assumed to be $\ln{\Lambda}$$\sim$$O(10)$. In simple terms, the two-body relaxation timescale is the time necessary for the system to ``forget'' its initial conditions through two-body gravitational interactions. The two-body relaxation of star clusters can range from 200 Myr to few Gyr, depending on clusters' size and mass. For reference, the typical timescale for a star to traverse the cluster, called \textbf{crossing time}, is:
\begin{equation}\label{eq:crosstime}
	t_{\rm cross} = \frac{R}{\sigma} \simeq \frac{8 \ln{N}}{N} \, t_{\rm rl}
\end{equation}
where $R$ is the cluster size, and $N$ is the number of stars. The crossing time is much shorter than the two-body relaxation timescale. For example, for a cluster of $10^5$ stars, the relaxation timescale is about $10^3$ times longer than the crossing time.

Two-body relaxation drives a flux of kinetic energy, carried by stars, from the center of the cluster to its outskirts. When stars move from the core of the cluster to the halo, they carry \textit{heat} with them, i.e., kinetic energy. As the core loses energy to support against its gravity, it collapses, thus it increases its velocity dispersion. This causes even more stars to flow into the halo, which expands. As the stellar halo expands, some stars reach high enough velocities to escape the cluster, in a process called \textbf{evaporation}. In OCs and YDSCs, the expansion of the cluster accelerates the disruption of the cluster due to the tidal field of the galaxy.
This runaway process is called \textbf{gravothermal catastrophe}, and in physical terms it is a consequence of the negative heat capacity typical of every self-gravitating system \citep{lyndenbell1968}. A system with negative heat capacity loses energy and becomes hotter. To become hotter, a self-gravitating system contracts so that its velocity dispersion (i.e., the average speed of the stars) increases.

In summary, the gravothermal catastrophe is a runaway process that triggers the collapse of the core and the expansion of the halo. This process is accelerated if the stars have a mass spectrum. In fact, massive stars are affected by two more physical processes: \textbf{dynamical friction and energy equipartition}.

\subsection{Dynamical Friction, Energy Equipartition and Mass Segregation}

\textbf{Dynamical friction is a drag force} that acts on massive bodies that travel through a medium of less massive objects. The gravity of the massive body attracts the lighter ones, which form a wake behind it. The overdensity of light bodies tends to decelerate the motion of the massive one via a gravitational drag. The massive body decelerates until it is finally at rest with respect to the lighter bodies. The timescale of dynamical friction for a body of mass $M$ is \cite{chandrasekhar1943a,chandrasekhar1943b,chandrasekhar1943c}:
\begin{equation}\label{eq:dynfric}
	t_{\rm df} = \frac{3}{4\,\left(2\,\pi\right)^{1/2}\,G^2\,\ln{\Lambda}}\,\frac{\sigma^3}{M\,\rho}
\end{equation}
where $\rho$ is the mass density of the lighter bodies. The dynamical friction timescale is much shorter than the two-body relaxation time. In star clusters with a mass spectrum, the core collapse can occur on the dynamical friction timescale, rather than on the two-body relaxation one. The steeper the mass spectrum, the faster the core collapse \cite{fujii2014}.
The consequence of dynamical friction is that massive stars become slower and sink to the center of the cluster. This phenomenon, called \textbf{mass segregation}, is characterized by a varying mass spectrum across the cluster radius, with heavier stars sinking to the cluster's core and the lighter ones crowding the halo. A more dramatic rearrangement of the massive stars in the cluster can be caused by energy equipartition, or rather, the lack of it.

\textbf{Energy equipartition} is the tendency for stars to equalize their average kinetic energy
\begin{equation}\label{eq:avekinstar}
	E_{\rm k} = \frac{1}{2} m \sigma^2 
\end{equation}
where $\sigma^2 \equiv \langle v^2 \rangle$ is the velocity dispersion of the stars of mass $m$. Therefore, two populations of masses will have a different velocity dispersion, in order to have the same kinetic energy
\begin{equation}\label{eq:equipartition}
	\frac{1}{2} m_i \sigma^2_i = \frac{1}{2} m_j \sigma^2_j \Rightarrow \sigma_i = \sigma_j\sqrt{\frac{m_j}{m_i}}.
\end{equation}
{For} $m_i$ > $m_j$, $\sigma^2_i < \sigma^2_j$, i.e., more massive stars will be on average slower. However, for typical initial mass functions (e.g., $N(m) \propto m^{-2.35}$ \cite{salpeter1955}), the pathway towards equipartition breaks, and the population of BHs (i.e., the most massive objects one can find in a cluster) decouples from the lightest objects. The BHs interact only among themselves and form an independent sub-cluster, which starts acting as an additional internal energy source for the whole system \cite{spitzer1969,merritt1981,inagaki1984,breen2013a,bianchini2016,spera2016,webb2017,kremer2018}. 

\subsection{Halting Core Collapse with Binaries}

The core collapse cannot proceed indefinitely. As the density of star increases, so does the chance that stars and binaries have a close gravitational interaction. The rate of encounter between a binary and a single of masses $m_{\rm bin}$ and $m$ is \cite{binney2008}:
\begin{equation}\label{eq:encrate}
	\Gamma_{\rm 1+2\: enc} = 4 \sqrt{\pi} n \sigma  \,r^2_{\rm enc} \left(1 + \frac{G (m_{\rm bin}+m) }{2 r_{\rm enc} \sigma^2}\right)
\end{equation}
where $n$ is the number density of single stars, $m$ is their average mass, $\sigma$ the velocity dispersion, and $r_{\rm enc}$ is the maximal closest approach distance. For the encounter to result in energy exchange between the single and the binary, the encounter distance $r_{\rm enc}$ needs to be of the order of the binary semimajor axis, e.g., $r_{\rm enc} \simeq 2a$.
If in the core of the cluster there are no binaries, they can be formed via three-body encounters of single stars. These occur on a timescale given by \cite{lee1995}:
\begin{equation}\label{eq:3bb}
	\Gamma_{\rm 1+1+1\: enc} = 126 \frac{G^5 m^5 n^3}{\sigma^9}
\end{equation}

During a three-body encounter between a binary and a single star, a fraction of the internal energy of the binary can be redistributed as translational energy among the interacting bodies. This means that \textbf{Binaries can considered as a reservoir of kinetic energy}. The kinetic energy released through three-body encounters can be used to \textit{reverse} core collapse.

Statistically, three-body encounters can have different outcomes depending on the kinetic energy of the single,
\begin{equation}\label{eq:singlekin}
	E_{\rm k} = \frac{1}{2} m_{\rm sin} v^2_{\infty}
\end{equation} 
and the internal energy of the binary
\begin{equation}\label{eq:binenergy}
	E_{\rm bin} = \frac{G m_{1} m_{2}}{2a}
\end{equation}
where, $m_1$ $m_2$ are the masses of the binary members and $a$ the binary semimajor axis. In the context of stellar clusters, a binary is considered as \textbf{hard} if its internal energy $E_{\rm bin}$ is greater than the average kinetic energy $E_{\rm k}$ of neighboring stars, while it is \textbf{soft} in the opposite case. On average, subsequent encounters make hard binaries harder (i.e., their semimajor axis shrinks), while soft binaries tend to become softer (i.e., wider semimajor axis) until they break up \citep{heggie1975}.
It is worth noting that hardness is a property of the binary relative to its environment. Due to the higher velocity dispersion, the same binary in the core of a cluster might be soft, whereas in the halo it would be hard.

\subsection{Forming Merging Compact-Object Binaries}

\begin{figure}[H]
	
	\includegraphics[width=0.98\columnwidth]{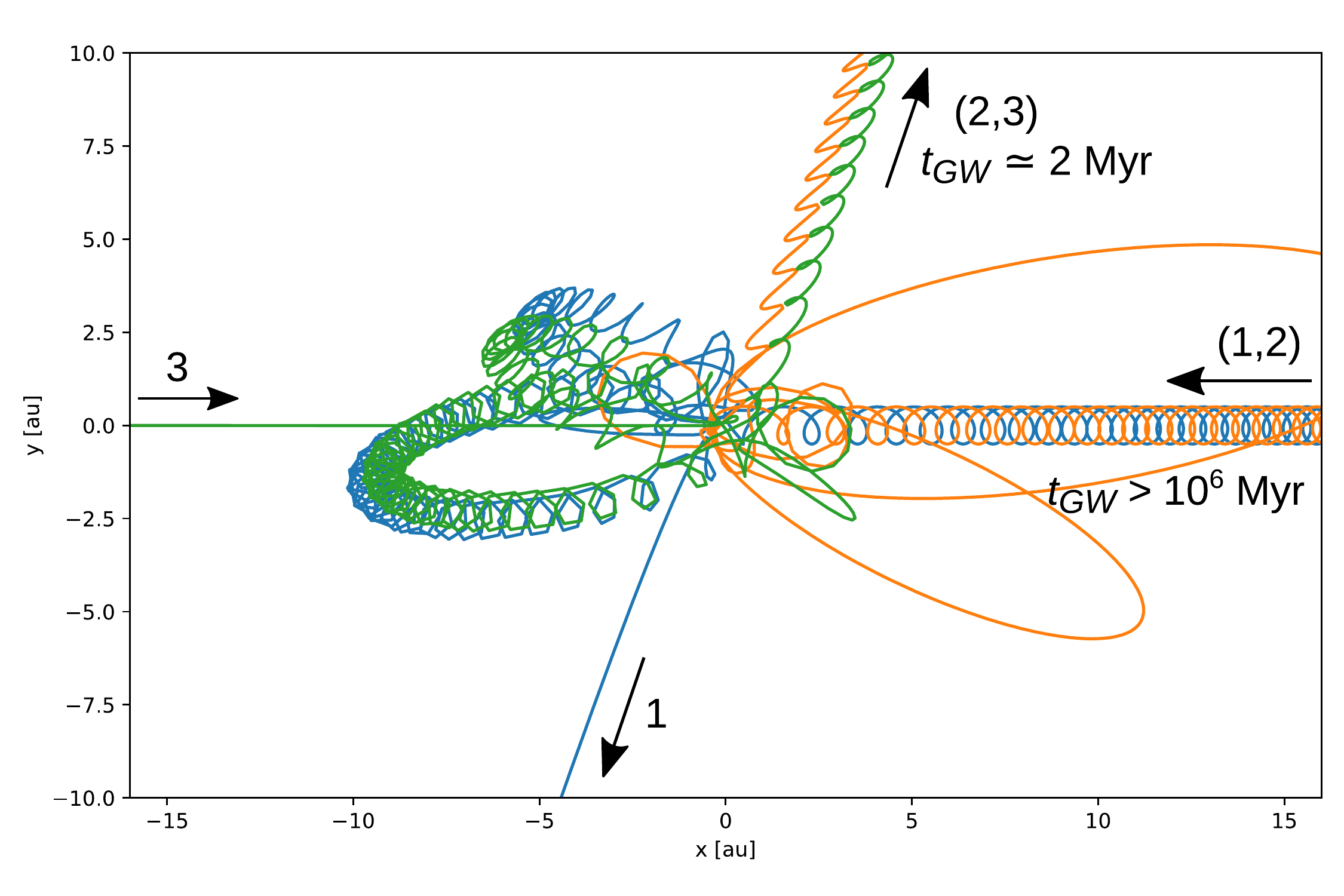}
	\caption{{Trajectories} during a three-body encounter between a binary (1,2) and a single (3). After a brief chaotic interaction, the binary and the single undergo an exchange. The encounter is concluded by the ejection of the new binary (2,3) and the new single (1). The initial coalescence time of the binary is $t_{\rm GW} > 10^6\,\rm Myr$, much greater than a Hubble time. After the interaction, the coalescence time is $t_{\rm GW} \simeq 2\,\rm Myr$. The initial approach of the binary-single is hyperbolic, with an impact parameter of $b=0.6 \,\au$ and velocity at infinity of $v_\infty = 0.1 \,\rm km/s$. \rev{All the bodies are $50\,\msun$ BHs integrated with \textsc{tsunami} ({Trani et al.,} in prep.), including post-Newtonian corrections of order 1PN, 2PN and 2.5PN to the equations of motion}.
		\label{fig:threebody}}
\end{figure}

The process of \textbf{binary hardening} in stellar clusters is a direct consequence of the core collapse and the gravothermal catastrophe, and it is argued to be one of the most critical processes for the formation of BBHs. 
Shrinking the semimajor axis of compact object binaries can dramatically shorten the coalescence time of binaries, because the GW coalescence timescale scales as $t_{\rm GW} \propto a^4$ (Equation~\eqref{eq:peters}). Another important consequence of three-body encounters is that they tend to excite the orbital eccentricity of binaries. In fact, the probability distribution of binary eccentricities after a three-body encounter is close to a thermal distribution ($N(e) \propto e$), and can be even super-thermal in the case of low angular momentum encounters \citep{jeans1919,heggie1975,monaghan76a,monaghan76b,valtonen2005,stone2019,kol2021}. The orbital eccentricity has an even greater impact on the GW coalescence timescale, because, for eccentricities close to 1, the coalescence timescale shortens as $t_{\rm GW} \propto (1-e^2)^{7/2}$ \citep{peters1964}. 
An example of the effects of three-body encounters on binary coalescence times is given in Figure~\ref{fig:threebody}. In this simulated three-body encounter, a binary and a single meet on a hyperbolic orbit with a small impact parameter and a velocity at infinity of $0.1 \,\rm km\,s^{-1}$, which is typical of small OCs. The initial binary is not close enough to merge within the age of the Universe. However, the outgoing binary has a shorter separation and a much higher eccentricity, which will cause the binary to coalesce in about 2 Myr after the encounter.

Due to dynamical friction and mass segregation, BHs sink to the core of star clusters~\cite{breen2013a}, where they undergo three-body encounters during the core-collapse phase.
A binary will keep undergoing three-body encounters in the core until (i) core collapse is reversed, i.e., the cluster's core ``bounces back'' and the density in the core decreases, quenching the encounter rate, (ii) the recoil velocity of the binary becomes so high that it is ejected from the core, or (iii) the binary merges due to GW radiation.

The second outcome can happen because when the binary becomes harder, the difference between the initial and final internal binary energies is redistributed as kinetic energy $\Delta E = E_{\rm bin, f} - E_{\rm bin, i}$. This kinetic energy is redistributed between the single and the binary following momentum conservation, i.e., a fraction $m_{\rm bin}/ (m_{\rm sin} + m_{\rm bin})$ is gained by the single, and a fraction $m_{\rm sin}/ (m_{\rm sin} + m_{\rm bin})$ is gained by the binary. In general, the binary recoil velocity is less than the ejection velocity of the single. However, if the injected kinetic energy is sufficiently high, the binary can be ejected from the core and even from the cluster itself. Early studies found that the \textbf{dynamical formation of binaries occurs in star clusters}, but most of the merger events happen from binaries that were hardened in the core and then \textbf{ejected} \cite{portegieszwart2000,downing2010,bae2014,ziosi2014,rodriguez2015,askar2017}.

Only recently, attention has been brought to \textbf{in-cluster mergers that can occur during few-body encounters} in the core (e.g., \cite{samsing2018,rodriguez2018,zevin2019}). These mergers result from the short pericenter passages that can happen during chaotic three-body encounters, which can trigger rapid gravitational wave coalescence. These kind of mergers can be extremely rapid, due to the initial short separation between the compact objects. For this reason, binaries formed through this scenario can retain detectable eccentricities in the LVK band~\citep{lower2018,huerta2018}.
Another possible scenario for producing eccentric mergers in the LIGO band are GW captures during hyperbolic single-single interactions \cite{gondan2018,samsing2020,hoang2020}. Because the cross section for single-single captures is extremely small compared to binaries, hyperbolic captures are likely to happen only in the most dense environments, like NSCs and GCs. 

\begin{figure}[H]
	\centering
	\includegraphics[width=\columnwidth]{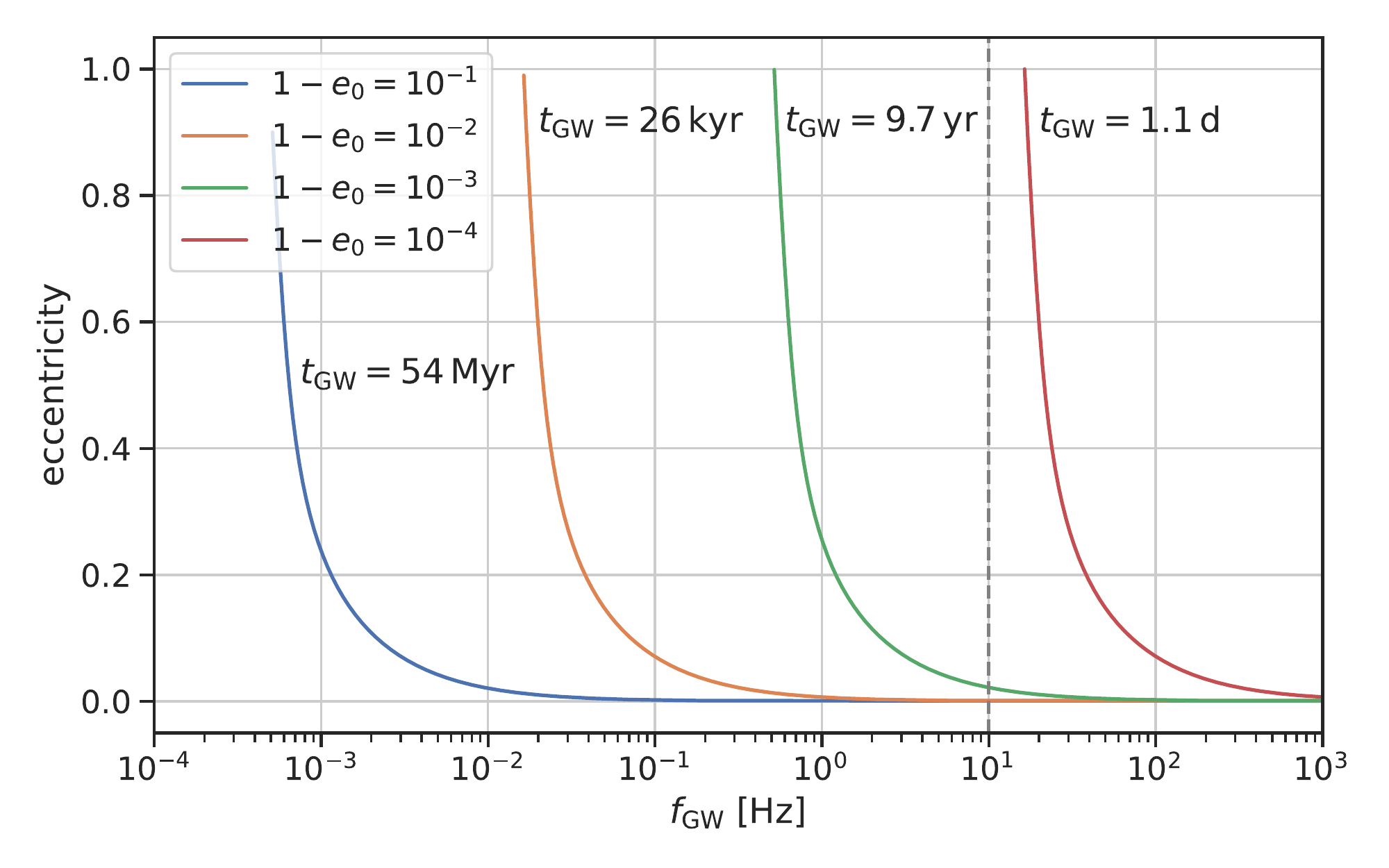}
	\caption{Eccentricity as a function of peak GW frequency for inspiralling binaries. Each curve indicates a binary with the same semimajor axis ($1\,\au$), but different initial eccentricity, as indicated in the legend. Coalescence timescales for each initial eccentricity are written next to the respective curves. The curves were calculated using the equations in {\citet{peters1964}} and the fitting formula for the peak harmonic GW frequency in \citet{hamers2021_peakgw}.{)}.
		\label{fig:eccentmerg}}
\end{figure}

During three-body encounters, one of the members of the binary might be swapped with the initially single body. Binaries formed through this process are referred to as \textbf{exchanged binaries}. Statistically, the lightest body has greater chances to be ejected. For this reason, binaries formed through three-body encounters tend to have higher masses and equal mass ratios. 
Furthermore, even if binaries formed through dynamical interactions tend to have high eccentricities \cite{jeans1919,heggie1975}, by the time they reach the LVK band, GW emission circularizes them. Therefore, most ejected binaries are not expected to have any residual eccentricity at ${>}10$ Hz \cite{abbott_150914_astro}. Figure~\ref{fig:eccentmerg} illustrates the impact of circularization of GW radiation on merging binaries. In the example, a binary with an initial eccentricity of $e_0 = 0.99$ will be completely circularized by the time it reaches a peak GW frequency of 10~Hz. Only binaries with an extreme eccentricity ($e_0 > 0.999$) can retain some eccentricity at 10 Hz, but their coalescence time will be extremely small (${\sim}${days). }

Furthermore, dynamically assembled binaries should not have any correlation between the orientation of BHs spins. Consequently, the predicted $\chi_{\rm eff}$ distribution of dynamically assembled binaries is symmetric and centered around $\chi_{\rm eff}$$\sim$$0$ \cite{rodriguez2018,safarzadeh_chieff2020, tagawa2020a}.

Finally, dense environments have an important consequence for hierarchical mergers, i.e., the GW coalescence of BHs formed from a prior BH coalescence. Asymmetric dissipation of linear momentum during the \textbf{GW coalescence imparts a recoil kick} to the final remnant. Depending on the mass ratio and spins of the merging BHs, these GW recoil kicks can be of the order of 100 km s$^{-1}$, which is much higher then the escape velocity of most clusters \citep{campanelli2007a,campanelli2007b,boyle2008,lousto2011,lousto2013,lousto2014}. Therefore, it is expected that hierarchical mergers only occur in massive stellar environments such as GCs, NSCs, and close to SMBHs \cite{rodriguez2019}. Runaway hierarchical mergers are also a proposed pathway to form intermediate mass BHs from stellar mass BHs \cite{miller2002,giersz2015,antonini2019,fragione2020,mapelli_hier2021,mapelli2020c}).

\subsection{Small-N Systems} \label{sec:smallN}
GW mergers can also be mediated by gravitational interactions in small-N systems, namely \textbf{triples, quadruples and higher hierarchical systems}.
Hierarchical triple systems are constituted by a binary orbited by an outer object, in a stable configuration. Many stellar triple systems have been observeed so far, and it is reasonable to expect that in many triples the inner binary is composed of compact objects. In fact, massive stars are especially found in triples and higher multiples \cite{tokovinin2008,tokovinin2014a,tokovinin2014b,tokovinin2018}. The fraction of stars found in multiples increases for more massive stars, up to ${\sim}50$\% for B-type stars \cite{duchene2013,sana2014,moe2017,toonen2016}. These multiple systems may be formed primordially as a natural outcome of star formation, or may also form dynamically from few-body encounters in stellar clusters.

\begin{figure}[H]
	
	\includegraphics[width=\columnwidth]{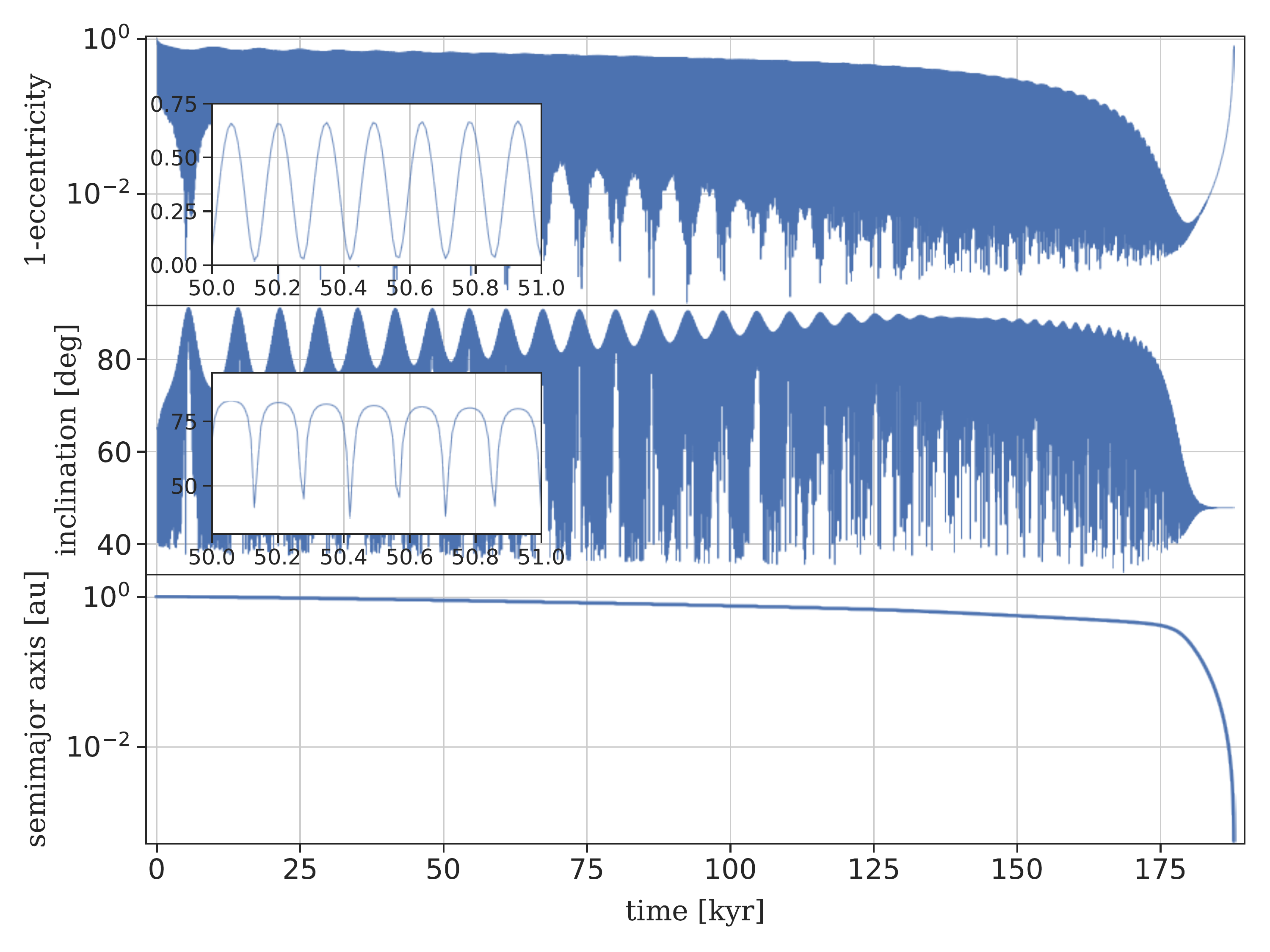}
	\caption{\rev{Gravitational-wave merger induced by von~Zeipel-Kozai-Lidov evolution in a triple system. Orbital parameters as a function of time for a triple system. Top panel: eccentricity of the inner orbit. Middle panel: inclination of the outer body's orbit with respect to the inner binary. Bottom panel: semimajor axis of the inner binary. The insets in the top and middle panels highlight the von~Zeipel-Kozai-Lidov oscillations in eccentricity and inclination. The initial orbital parameters of the triples are: mass of the outer body $m_3 = 40\,\msun$, mass of the inner bodies $m_1 = 10 \,\msun$, $m_2=20\,\msun$, inner semimajor axis $a_1 = 1 \,\au$, outer semimajor axis $a_2 = 16 \,\au$, mutual inclination $i_{\rm mut} = 65^\circ$. Evolved with \textsc{okinami}, which solves the orbit-averaged equations of motion in Delaunay coordinates  ({Trani et al.,} in prep.).}
	\label{fig:triplemerge}}
\end{figure}

Gravitational interactions can strongly affect the evolution of triple systems.
If the outer object is sufficiently inclined with respect to the inner binary, the latter can exchange angular momentum with the outer orbit on a secular timescale, which is longer than the orbital period of the inner binary. These secular exchanges of angular momentum manifest themselves as the \textbf{von Zeipel-Kozai-Lidov (ZKL)} \cite{lidov1962,kozai1962,zeipel1910,Naoz2016,Shevchenko2017,ito2019}. During such oscillations, the eccentricity of the inner binary can reach values very close to 1, inducing very close pericenter passages. If the inner binary is composed of compact objects, GW radiation can be efficiently emitted during these short pericenter passages, leading to the rapid coalescence of the binary. Figure~\ref{fig:triplemerge} shows the typical evolution of a GW coalescence triggered by ZKL oscillations. There are two main effects of ZKL oscillations. First, they can accelerate the merging of compact object binaries, allowing them to merge within the age of the Universe. Secondly, they excite very high eccentricity in the inner compact binary, which can affect the GW emission waveform, and therefore can be potentially inferred by GW observations at different frequencies \cite{hoang2019,gupta2020,arcasedda2021}.

The evolution in triple systems can also be affected by all the processes that occur in binary stars, and more. Mass can be lost by the tertiary star and transferred to the inner binary. The outer star could also be affected by tertiary tides \citep{gao2020} or become a giant and undergo a phase of triple CE \citep{glanz2021b}. In general, modeling the evolution of a triple system is quite complex, because of the strong interplay between dynamics and stellar evolution~\citep{toonen2020}. Due to the relative lower abundance of triples, the predicted merger rate of BBHs from triples is generally smaller by a factor of ${\sim}100$ than that from dynamically assembled binaries \citep{thompson2011,antonini2017,antonini2016b,grishin2018,martinez2020,martinez2021,arcasedda2021,silsbee2017,toonen2018,rodriguez2018,vignagomez2021,trani2021}.

The ZKL effect has applications also to triples in which the outer body is a massive BH. This may happen in NSCs hosting a SMBH, which can be orbited by a stellar BBH~\cite{antonini2012,vanLandingham2016,fragione2019_nuclei,hamers2018_smbh,hoang18,trani2020}. 
Just like in the case of a tertiary star, the massive BH can excite the eccentricity of the inner binary, causing the binary to merge.

Other forms of dynamical scenarios include dynamical perturbations of multiple stellar systems in the field \citep{michaely2019,michaely2020,grishin2022}, or interactions of binaries with the tidal fields of star clusters \citep{hamilton2019}.

A few authors have recently started a systematic investigation of the evolution of hierarchical systems with $N>3$ \citep{vynatheya2022,hamers2021_quad}. Specifically, \citet{vynatheya2022} investigated compact-object mergers in quadruples, finding that the expected rates for BBHs are of the order of $10 \,\gpcyr$ (see also \citep{liu2019}). Other studies on quadruples focused on specific GW events will be discussed in Section~\ref{sec:lessonslearned}.

\subsection{Hybrid Scenarios} \label{sec:hybridscenarios}
Distinguishing between the isolated binary channel and the dynamical channel might not be straightforward. \textbf{Stellar evolution and stellar dynamics are inseparable processes} that are active at the same time. Indeed, besides evolving in complete isolation, stellar binaries can be found in star clusters as well, where they can be affected by close encounters just like compact-object binaries. These stellar binaries can therefore experience processes from both the binary evolution pathway (e.g., mass transfers, CEs) and the dynamical pathway (e.g., exchanges, excitation of eccentricity). 
For instance, in young star clusters, some BBHs are formed via CE evolution of dynamically-assembled \textit{main sequence} binaries that, at some point of their life, are ejected from the cluster, and merge in the field, appearing as if they had evolved in complete isolation. They can contribute to the merger rate more than dynamically assembled BBHs \cite{kumamoto2019,dicarlo2020,rastello2020,dicarlo2019a}. These binaries undergo a CE phase, like in the isolated channel, but they also undergo dynamical interactions before and after collapsing to BHs. The CE phase might even be triggered by such dynamical interactions, so that the same binaries would not have merged without the crucial contribution of stellar dynamics. \rev{
Some specific scenarios require elements from both channels. For example, three-body encounters of tidally spun-up, post-common-envelope binaries have been proposed as a mechanism to produce BBH with misaligned spins \cite{trani2021}.
}

\section{Astrophysical Insights from Exceptional Gravitational-Wave Events}

\label{sec:lessonslearned}
In the previous sections, we presented the main astrophysical processes that can drive the formation of merging compact-object binaries. Despite the uncertainties and degeneracies in the astrophysical models, in this section we investigate whether some of the exceptional GW events have distinguishing features that may point us towards a specific formation scenario. \rev{An exhaustive analysis of all the exceptional GW events is beyond the scope of this review, which focuses mainly on the astrophysical processes behind merging compact-object binaries. Thus, in this section, we decided to restrict our analysis only to the events containing at least one BH and with physical properties that can be the smoking gun of a specific formation pathway: heavy BHs, as in GW190521, very asymmetric masses, as in GW190814 and GW190412, and mixed components, as in GW200105\_162426 and GW200115\_042309.}

We stress that it is \textit{currently premature to make any kind of conclusive statements}, especially on the origin of \textit{single} GW events, though these exceptional events carry a number of astrophysical traces that are worth exploring here in some detail.

\subsection{GW190814} 
\label{subsection:gw190814}
GW190814 is a special LIGO-Virgo Collaboration (LVC) event detected during the first part of the third observing run and presented in \citet{LVK190814_2020}. It is a coalescence with significantly unequal masses (mass ratio $q=0.112_{-0.009}^{+0.008}$). Furthermore, the secondary component of GW190814 is either the lightest BH or the heaviest NS ever discovered in a compact-object binary system, (secondary mass $m_2=2.59_{-0.09}^{+0.08}$). The mass of GW190814's secondary falls into the range of the hypothesized lower mass gap ($2.5$--$5\,\msun{}$), questioning the existence of this gap. GW190814 is also the GW event with the strongest constraint on the spin precession parameter ($\chi_{\rm p}=0.04_{-0.03}^{+0.04}$) and on the spin of the primary component ($\chi_1\leq 0.07$)\rev{, though both of these constraints are consistent with non-spinning components.}

\textls[-15]{The secondary of GW190814 might be either a BH or a NS. \rev{The non-detection of any electromagnetic counterparts \citep{dobie2019, gomez2019, ackley2020, andreoni2020, antier2020, gompertz2020, morgan2020, page2020, thakur2020, vieira2020, watson2020}, the fact that there were neither clear signatures of tides or spin-induced quadrupole effects in the waveform \citep{essick2020,Lackey2015},} and the uncertainties on both the theoretical estimates of the maximum NS mass and the NS equation of state \citep{malik2022,Antoniadis2013,Fonseca2021,Romani2021},} prevent us to determine the nature of the compact object. Post-merger electromagnetic studies suggest that the merger product of GW170817 likely collapsed into a (highly-spinning) BH with a mass comparable to GW190814's secondary (${\sim} 2.6\,\msun{}$)~\citep{annala2021}. The latter result has been used by various authors as a starting point to constrain the NS equation of state and the maximum mass of a non-rotating NS, suggesting that the secondary of GW190814 is likely a BH \citep{tews2021,essick2020,fattoyev2020}. However, the hypothesis of a rapidly rotating NS cannot be completely excluded \citep{dexheimer2021,tsokaros2020,nathanail2021,most2020,biswas2021,zhang2020,tan2020,godzieba2021,wu2021,huang2020,pegios2021}, even in the context of extended theories of gravity (e.g., \citet{asta2020}).

The population analysis of GWTC-3 \citep{lvcgwtc3_2021} \rev{shows that (i) the secondary of GW190814 is an outlier from the population of NSs of BNS and of the secondaries of NSBH systems detected} so far through GWs, and that (ii) GW190814 is an outlier from the population of observed BBHs. These findings support the idea that GW190814 belongs to a distinct population of merging compact-object binaries.

Understanding the formation and evolutionary history of GW190814 is challenging for all current astrophysical models. The challenge is to find a theoretical model that can accommodate, at the same time, the GW190814 mass ratio, masses, and the LVC inferred merger rate.
As already discussed in Section~\ref{sec:singlestars}, most theoretical models do not have predictive power on the nature of compact objects and only a mass-threshold criterion is used to distinguish between NSs and BHs. Thus, depending on the adopted threshold, GW190814-like events might be marked as either NSBH or BBH mergers.

GW190814-like systems are outliers in terms of rates, masses, and mass ratios in most of the distributions of merging NSBH and merging BBH systems predicted by isolated binary evolution models \citep{zevin2020,broekgaarden2021,safarzadeh_accr2020,dominik2012,giacobbo2018a,kruckow2018,spera2019,mapelli2018,neijssel2019,olejak2020,mandel2021,dominik2015}. \citet{zevin2020} identified two possible formation channels for GW190814-like systems, but (i) they form only if the delayed supernova explosion mechanism is adopted (see also Section~\ref{sec:corecollapsesne}), and (ii) their merger rates are at least one order of magnitude below the inferred LVC rates. Furthermore, \citet{zevin2020} showed that the mass of GW190814's secondary is likely its birth mass, therefore GW190814 might give insights on the existence of the lower mass gap and, possibly, on the supernova explosion engine. It is worth mentioning that, in the isolated binary scenario, unlimited super-Eddington accretion onto compact objects might be the key to obtain significantly higher merger rates for GW190814-like systems, within the LVC inferred rate (see e.g., \citep{eldridge2016, eldridge2017}).

GW190814 might be a second-generation merger event from a hierarchical triple system \citep{lu2021,gupta2020,samsing2019,martinez2021,toonen2016,cholis2021, trani2022}. This is not an exotic formation pathway for GW190814, considering that most massive B-type stars, $\gtrsim$$50\%$, are in triples and the percentage tends to increase in low-metallicity environments (see also Section~\ref{sec:smallN}) \citep{jimenez2019,duquennoy1991, sana2017, duchene2013, moe2017, sana2014, gao2014, yuan2015, moe2019}. Compared to the population of BH mergers from the isolated binary channel, mergers in triples tend to enhance the number of mergers with more unequal masses, resulting in a flatter mass-ratio distribution down to $q\simeq 0.3$ \citep{martinez2021, trani2022}. In this context, a possible formation pathway for GW190814 is that the members of the inner binary are two stars that evolve through the CE process and form a tight BNS system. The latter merges within a Hubble time, leaving a low-mass ($\sim$$2.6\msun{}$), highly-spinning BH. The tertiary body is a star that turns into a $\sim$$26\msun{}-$BH at its death. \citet{lu2021} show that there is a non-negligible chance that, at its formation, the low-mass BH is kicked into a low angular momentum orbit so that it can merge with the tertiary compact object within a Hubble time. The merger rate of GW190814-like sources obtained through this channel is consistent with the one inferred in \citet{LVK190814_2020}, provided that $\gtrsim$$10\%$ of BNS mergers occur in triples and that the tertiary orbital semi-major axis is less than a few au. A high spin of the low-mass BH would be a distinguishing feature of this scenario, but it cannot be corroborated through the analysis of the GW190814 signal because \rev{of the uninformative spin posterior of the secondary}.
\citet{cholis2021} investigated the same formation scenario and obtained similar results as \citet{lu2021}, but they considered the possibility that the secondary of GW190814 is a Thorne--Zytkow object, i.e., a metastable object formed from the collision of a NS with a red giant star, possibly turning into a low-mass BH at the end if its life \citep{thorne1975}. The local merger rate density from the latter channel can be as high as a few $\gpcyr{}$, which is within the LVC inferred rate.

Other authors investigated similar dynamical formation pathways for GW190814 involving systems with higher multiplicity, e.g., hierarchical quadruples \citep{safarzadeh_quad2020,fragione_quad2020,fragione_quadr2019,liu2021}, finding similar results.

GW190814-like systems might form in dense stellar environments through binary-single dynamical exchanges. However, the dynamical scenario favors the formation of compact-object binaries with similar masses ($q \gtrsim 0.5$), pairing up the most massive objects (e.g., BHs and their stellar progenitors) in the dense cores of clusters, while low-mass objects (e.g., NSs) are likely ejected from the cluster during close three-body gravitational interactions (see also Section~\ref{sec:dynchannel}) \citep{sigurdsson1993,sigurdsson1995,portegieszwart2000,rodriguez2016,rodriguez2015,rodriguez2019,fragione_dyngen2018,ziosi2014,mapelli2016,banerjee2018}. In contrast, direct GW captures coming from single-single gravitational interactions seem to be exceedingly rare events \citep{samsing2020}. Numerical simulations of GCs \citep{clausen2013,clausen2014,devecchi2007,ye2020} and binary-single scattering experiments in GC-like environments \citep{kritos2021,arcasedda190814_2021, arcasedda2020a} show that the local merger-rate density of GW190814-like systems is $\lesssim$$0.1\gpcyr{}$, well below the single-event rate inferred by the LVC ($\valpm{7}{16}{6}\gpcyr{}$). YDSCs might also have an impact in the formation of GW190814-like systems. In contrast to GCs, the dynamical evolution of YDSCs proceeds on much shorter timescales, and they are thought to be the nurseries of massive stars and the building blocks of galaxies. Furthermore, YDSCs form continuously in the Universe, at all redshifts, thus their contribution to the local merger-rate densities of merging compact-object binaries may potentially be higher than that of GCs \citep{rastello2020, santoliquido2020,arcasedda2020a,dicarlo2019a,ziosi2014}. Specifically, the N-body simulations of YDSCs presented in \citet{rastello2020} support a dynamical formation scenario for GW190814, with an estimated local merger-rate density of GW190814-like systems of \valpm{8}{4}{4}\gpcyr{}. Most of the merging NSBH systems presented in \citet{rastello2020} come from dynamically-perturbed original binaries (i.e., from progenitor stars already bound in the initial conditions, but that would not have formed a merging NSBH if evolved in isolation). Furthermore, all mergers happened in the field because all the progenitor binaries were ejected from the star cluster before the binary members reach coalescence (i.e., hybrid scenario, see Sections~\ref{sec:introduction} and \ref{sec:hybridscenarios}). In contrast, \citet{fragione2020} performed direct N-body simulations of 65 YDSCs and estimated the local NSBH merger-rate density for these environments to be $\lesssim 3\times10^{-2}\gpcyr{}$. However, \citet{fragione2020} and \citet{rastello2020} consider different types of clusters with different initial masses and concentrations, different NS birth kick models, and different fractions of the cosmic star-formation rate that goes into clusters.

GW190814-like systems may also form in AGN. In such crowded environments, the gas-driven formation mechanisms and the hardening process of binaries are effective and gas accretion on compact objects with birth mass $\lesssim$$2\,\msun{}$ might be high enough to bring them in (or even beyond) the low-mass gap, before they merge with another compact objects \citep{hiromichi2021, mckernan2020a, mckernan2020b}.
Furthermore, the secondary of GW190814 might also be a compact object coming from a previous merger event in an AGN disk. This is not an exotic scenario because merger products are easily retained in AGN thanks to high escape speeds, thus next-generation compact objects can participate in additional mergers \citep{hiromichi2021,yang2020}. Specifically, \citet{mckernan2020a,mckernan2020b} show that the \rev{typical} value for the mass ratio of the population of NSBH mergers in AGN disks is $q\simeq 0.1$, with a corresponding local merger rate of about a few $\gpcyr{}$, both compatible with GW190814.

\citet{bombaci2021} discussed the hypothesis that the secondary of GW190814 is a strange quark star, which is a non-ordinary NS composed of a deconfined mixture of up, down and strange quarks. The authors showed that quark stars can reach masses comparable to the secondary of GW190814 while using physically motivated equations of state for hadrons and quarks, and without assuming an exceedingly large speed of sound.

Various authors investigated the hypothesis of the secondary component of GW190814 being a primordial BH, i.e., a BH formed from density fluctuations in the early Universe, a fraction of second after the big bang \citep{jedamzik2021, vattis2020}. \citet{vattis2020} show that the ${\sim} 2.6\,\msun{}$ member is rather unlikely to be a primordial BH because the typical time needed for a primordial BH to pair and then merge with a $\sim 26\,\msun{}$, stellar-origin BH is very close to (or even larger than) the age of the Universe, contradicting the GW190814 detection (but see also \citep{carr2021,sebastien2021}). 

It is apparent that the parameter space relevant for GW190814-like systems still needs to be fully explored and future GW detections will be crucial to shed light on the nature and formation channels of GW190814-like systems.

\subsection{GW190521} \label{subsection:gw190521}
\rev{GW190521 is currently the GW event with the heaviest BHs in the catalog \citep{lvcgwtc3_2021}}. The masses of the two BHs are $m_1=\valpm{85}{21}{14}\,\msun{}$ and $m_2=\valpm{66}{17}{18}\,\msun{}$, respectively, and the mass of the merger product is $M=\valpm{142}{28}{16}\,\msun{}$ \citep{lvk_190521_astro2020, lvk_190521_2020}. The mass of the primary BH falls  confidently into the hypothesized upper mass gap ($\sim 60\,\msun{}$--$120\,\msun{}$) and the merger product is the first strong observational evidence for the existence of intermediate-mass BHs ($\sim$$10^2\msun{}$--$10^4\, \msun{}$).

Being the shortest-duration signal among the GW detections, GW190521 is a quite difficult event to analyze and measurements of spins and their in-plane components are quite uncertain. The data analysis presented in \citet{lvk_190521_2020} supports mild evidences for in-plane spin components with high spin magnitudes for both the BHs, but the values of the BH dimensionless spins remain uninformative at 90\% credible intervals (i.e., $\chi_{1,2}$$\sim$ 0.1--0.9). The hypotheses of a non-precessing signal with orbital eccentricity $e\geq 0.1$ at 10 Hz \citep{romero2020} and of a precessing signal with orbital eccentricity $e\simeq 0.7$ \citep{gayathri2020} cannot be excluded, even though various authors suggest that GW190521-like binaries likely enter the 10 Hz-band with $e \ll 0.7$ (e.g., \citep{holgado2021}). Furthermore, subsequent analysis of the LVC data brought out the intriguing possibility of GW190521 being an \rev{intermediate-mass ratio inspiral}, with $m_1=\valpm{168}{15}{61}\,\msun{}$ and $m_2=\valpm{66}{33}{3}\,\msun{}$ \citep{nitz2021, fishbach2020}.

The formation channel of GW190521 is uncertain. The physical properties of the event seem to favor the dynamical formation pathway, while an explanation through the isolated binary channel is challenging. 

Population-synthesis simulations of binary stars show that the most massive BBHs that can form at a given metallicity are unlikely to merge within a Hubble time, making it hard to even explain the formation of systems with lower masses, such as GW170729 \citep{spera2019}. In contrast, \citet{belc2020a} shows that if heavy ($\gtrsim 180~\msun{}$) progenitor stars are considered and a very low \cago{} rate is adopted ($-2.5\sigma$ with respect to the fiducial value of \citet{farmer2020}), the isolated-binary channel becomes a plausible formation pathway for GW190521. \cite{tanikawa2021b} found that GW190521-like systems can be formed from population III
(Pop III) binaries, but only for stellar evolution models with a small convective overshooting parameter.

The hypothesis of repeated mergers in dense stellar systems seems to be a promising scenario to explain GW190521 \citep{liu2021, fragione_repeated2020, fragione_nsc2021, mapelli_hier2021, mapelli_symm2021, baibhav2021, anagnostou2020, kimball2021}. 

The escape speed of GCs may be higher than the GW recoil kick imparted to some second-generation BHs, especially if BHs are born with low spins. Thus, second-generation BHs can be retained in GCs and \citet{rodriguez2019} show that the fraction of merging BBHs that have components formed from previous mergers in GCs can be $\gtrsim$$10\%$ \rev{$\gtrsim$$20\%$ for the detectable population}. \citet{kimball2021} created a phenomenological population model for merging BBHs derived from \citep{rodriguez2019}, and they use that to infer the population properties of BH mergers in the second LVC catalog, finding that the members of GW190521 are likely second-generation (2~g) BHs. Triple systems in GCs may further enhance the retention probability of 2~g BHs, strengthening the \citet{kimball2021} and \citet{rodriguez2019} findings (e.g., \citep{martinez2020}).
Repeated minor mergers (>2~g) might also have formed the BHs of GW190521 but only in environments with escape velocities $\gtrsim$$200\, \mathrm{km\,s}^{-1}$, such as the most massive GCs or nuclear star clusters (e.g., \citep{fragione_repeated2020, arcasedda2021}). 

The escape speed of YDSCs is significantly smaller than globular and nuclear star clusters, thus second-generation BH mergers are expected to be exceedingly rare in such environments \citep{fragione_nohighmass2021, fragione_nsc2021}. However,
\citep{dicarlo2020,dicarlo2021,dicarlo2019b, dallamico2021} show that GW190521-like systems can still form in YDSCs via three-body encounters or multiple stellar mergers. In the latter scenario, a heavy BH can originate from the merger of an evolved star with a main-sequence star. If the helium core of the primary star is small enough to avoid PISN and the secondary star is still on the main-sequence, the merger product might be an evolved star with an oversized hydrogen envelope. If a significant fraction of the hydrogen envelope is retained and participate in the final collapse, a BH in the PISN mass gap can form \citep{spera2019}. Such heavy, single BHs can acquire a companion in dense stellar environments and explain the formation of GW190521-like systems. Furthermore, BHs born via this mechanism can preserve most of the spin of the progenitor star, leading to a high-spin BH even if they are 1~g BHs \citep{mapelli_rot2020}, in agreement with the properties of GW190521, even though the efficiency of angular momentum transport inside stars is matter of debate \citep{marchant2020}. The same formation pathway has been identified also in GCs \citep{kremer2020, gonzalez2021}.
It is worth noting that, as in GCs, triple systems in YDSCs may enhance the overall retention probability of 2~g BHs, making the 2~g-channel for GW190521 a viable possibility also for such environments.

\citet{palmese2021} investigated the hypothesis that the two BHs of GW190521 were at the centers of two ultra-dwarf galaxies, assuming an extrapolation of the low-end of the central BH-galaxy mass relation (e.g., \citep{reines2015}). This scenario assumes that, after the merger of the two galaxies, the central BHs can merge, and the merger rates for GW190521-like systems from this channel match well those inferred by the LVC.

In AGN, \citet{hiromichi2021} show that GW190521-like systems can form either via repeated mergers (>2~g at high metallicity, 2~g at low metallicity), or via mergers of 1g BHs that have grown via super-Eddington accretion. In such gas-rich environments, the interaction between the GW-recoiled merger product and the accretion disk of the active galactic nucleus might generate a delayed ($\sim$days after the merger), off-center UV flare, potentially detectable as an electromagnetic transient counterpart of the GW event \citep{mckernan_flare2019}. About 26 days after the merger of GW190521, the Zwicky transient facility \cite{Ashton2021} identified a flare from the AGN J124942.3~+~344929, the latter located at the 78\% spatial contour and within $1.6\sigma$ from the peak marginal luminosity distance of GW190521.
\citet{graham2020} identified the flare as a plausible electromagnetic counterpart to the BBH merger GW190521 (ZTF19abanrhr). While the possibility is intriguing and it is a possible distinguishing feature of the AGN formation scenario, the 90\% localization area of GW190521 contains thousands of AGN and, currently, it is not possible to establish whether GW190521-ZTF19abanrhr is a real association or a chance coincidence \citep{Ashton2021, palmese_agn2021}. Future high-mass detections and follow-up investigations will be crucial to shed light on this interesting possibility.

Various authors suggest Pop III stars as promising progenitors of the BHs members of GW190521 \citep{liu_popiii2020,farrell2021,kinugawa2021,tanikawa2021a,tanikawa2022_popsyn}. The key idea is that Pop III stars might retain most of their hydrogen envelope until the end of their life and might form heavy BHs via direct collapse, with masses up to $\sim$$85~\msun{}$.
Specifically, \citet{liu_popiii2020} investigated a simplified binary evolution scenario, with an initial binary population taken from N-body simulations of Pop III star clusters \citep{liu_nbody2021}, and the dynamical hardening process of binaries in high-redshift ($z\simeq 10$) nuclear star clusters. The authors show that both the scenarios can explain the masses and merger rates of GW190521-like systems. In the same context, \citet{safarzadeh_accr2020} developed an illustrative toy model showing that the collapse of relatively massive Pop III stars in high-redshift minihalos can form BH seeds that, in $O\left(10\text{--}10^2\right) {\rm Myr}$, can double their mass via gas accretion, reach the PISN mass gap, and then merge within a Hubble time, explaining the formation of GW190521. Similar results were obtained by \citet{rice2021}, who investigated the growth of stellar BH seeds in dense molecular clouds.

\citet{deluca2021} studied the hypothesis of a primordial BH origin for GW190521. Their in-depth analysis shows that the primordial BH scenario is unlikely for GW190521 only if primordial BHs do not accrete mass during their cosmological evolution. In contrast, if accretion is efficient (see also \citep{yang2021}), the scenario can explain the properties of GW190521, including the mild evidence for high BH spins with non-zero in-plane components (see also~\citep{cruzosorio2021}). A mixed scenario for GW190521-like systems, where primordial BHs coexist, interact and possibly merge with astrophysical (stellar-origin) BHs in dense stellar environments, is also a viable formation scenario \citep{kritos_pbh2021}.

More exotic explanations for the massive BHs of GW190521 are conceivable, and they have been investigated by various authors. They include horizonless vector boson stars, beyond-standard-model physics affecting the edges of the upper mass gap, and dark matter
annihilation within stars to avoid PISNe (e.g., \citep{bustillo2021,sakstein2020,ziegler2021}).

\subsection{GW190412} \label{subsection:gw190412}
At the time of its detection, GW190412 was the first GW event with support for asymmetric masses, and the one with the strongest constraint on the individual spin magnitude of the primary BH \citep{lvk_190412_2020}. The masses of the BHs are $m_1=\valpm{30.1}{4.6}{5.3}\,\msun{}$ and $m_2=\valpm{8.3}{1.6}{0.9}\,\msun{}$, and the mass ratio is $q=\valpm{0.28}{0.12}{0.07}$. The signal shows a mild evidence of precession coming from non-zero, in-plane spin components, with $0.15\leq\chi_{\rm p}\leq0.50$ at 90\% credibility. The LVC analysis reports a dimensionless spin of the primary BH of $\chi_1=\valpm{0.44}{0.16}{0.22}$, at 90\% credibility, obtained using uninformative priors for the spins of both the compact objects ($\chi_{1,2}\in \left[0, 0.99\right]$).

In contrast, \citet{mandel2020} chose an informative spin prior, assuming that the event formed via the isolated binary channel, and they suggested an alternative interpretation of GW190412 as a merging BBH with a non-spinning primary and a rapidly spinning (tidally spun-up) secondary. \citet{zevin2020_priors} studied the impact of different spin prior assumptions on the analysis of GW190412 and they found that the uninformative prior with both BHs spinning is preferred over other configurations.

Multiple scenarios can explain the formation of GW190412. As already discussed for GW190814, most of the merging BHs that originate from the isolated binary channel have mass ratios $q\gtrsim 0.5$, but the tail of the mass-ratio distributions extends down to $q$$\sim$$0.2$, especially in low-metallicity environments \citep{stevenson2017,olejak2020,dominik2012,giacobbo2018a,kruckow2018,neijssel2019,spera2019}. Furthermore, the assumption of super-Eddington accretion onto compact objects might significantly increase the number of merging systems with $q\lesssim 0.5$ \citep{eldridge2016,eldridge2017}. Thus, GW190412 cannot be considered as an outlier for the isolated binary scenario in terms of masses and mass ratio. The moderately high spin of the primary BH may be difficult to reconcile with the isolated channel \citep{safarzadeh_accr2020,zevin2022}, but the uncertainties on the rotation speed and on the dominant mechanisms for angular momentum transport inside massive stars hamper our knowledge of the birth-spin distribution of stellar BHs.

First-generation, highly spinning BHs can form if their progenitor stars undergo CHE, but most of the BHs originating from this channel have large and nearly equal masses, making this scenario unlikely to explain GW190412 \citep{demink2016, mandel2016a, marchant2016, riley2021}.

Repeated mergers of BHs in dense stellar environments might explain the mass ratio and the spin of the primary BH of GW190412. Assuming that BHs have negligible spins at birth, \citet{rodriguez2020_ssc} show that GW190412-like systems can form in super star clusters with central escape speeds of, at least, $90\,\mathrm{km\,s}^{-1}$ and that such events are consistent with a first-generation BH of $\sim$$10~\msun{}$ merging with a BH formed from the subsequent merger of \rev{three} $\sim$$10~\msun{}$ BHs (i.e., a third-generation BH). \citet{gerosa2020} also support the hierarchical merger scenario in dynamical environments with central escape speeds $\gtrsim$$150\,\mathrm{km\,s}^{-1}$, but they discussed the possibility of the primary of GW190412 being a second-generation BH. 

The higher central escape speeds of NSCs and AGN can significantly enhance the retention fraction of >1-generation BH mergers, thus naturally producing hierarchical-merger events compatible with GW190412 \citep{miller2009, antonini2016a, antonini2019, tagawa2020a}.

On the other hand, \citet{dicarlo2020} and \citet{rastello2020} show that GW190412-like systems can also form in metal-poor YDSCs, \rev{but} from merging first-generation BHs, \rev{because} the central escape speed of such clusters is too low to allow for any retention of >1-generation BHs and most binaries are ejected from the cluster before they reach coalescence.

A first-generation origin for GW190412 is also supported by the phenomenological approach to mergers in dense stellar environments presented in \citet{kimball2020} and by von Zeipel-Kozai-Lidov induced mergers in isolated triples \citep{martinez2021, trani2022}. The moderately high spin of the primary of GW190412 may be difficult to reconcile with first-generation BHs, even though, as already discussed, the birth spin of BHs is uncertain (see also Section~\ref{subsection:spins}).

The high spin of the primary BH, the mass ratio, and the masses of GW190412 can also be explained through hierarchical mergers in isolated quadruples \citep{hamers2020, fragione_quad2020}.

\subsection{GW200105\_162426 and GW200115\_042309} \label{subsection:gw200105_gw200115}
GW200105\_162426 and GW200115\_042309 (hereafter GW200105 and GW200115) are the first two GW events \rev{that are consistent with} merging NSBH systems. During the first part of the third observing run, GW190426\_152155 and GW190814 were also reported as possible merging NSBH candidates. However, the false alarm rate of GW190426\_152155 was too high to claim a confirmed detection and the uncertain nature of the secondary compact object prevents us from considering GW190814 as a NSBH merger (see Section~\ref{subsection:gw190814}). GW200105 consists of a BH with mass $m_1=\valpm{8.9}{1.2}{1.5}\,\msun{}$ and a NS of $m_2=\valpm{1.9}{0.3}{0.2}\,\msun{}$, while GW200115 hosts a BH with mass $m_1=\valpm{5.7}{1.8}{2.1}\,\msun{}$ and a NS of $m_2=\valpm{1.5}{0.7}{0.3}\,\msun{}$, \rev{using high-spin priors, i.e., $\chi_2 < 0.99$}. The effective inspiral spin parameter of GW200105 is peaked at zero ($\chi_{\rm eff}=\valpm{-0.01}{0.11}{0.15}$) with $\chi_1 = \valpm{0.08}{0.22}{0.08}$, while GW200115 has a mild preference for misaligned spins with $\chi_{\rm eff}=\valpm{-0.19}{0.23}{0.35}$ and $\chi_1 = \valpm{0.33}{0.48}{0.29}$, with $\chi_{1,z}=\valpm{-0.19}{0.24}{0.50}$.  

\citet{mandel2021_gw200115} reanalyzed the GW200115 signal using astrophysically-motivated spin priors and they obtained better constrains on the masses of the compact objects ($m_1=\valpm{7.0}{0.4}{0.4}\,\msun{}$, $m_2=\valpm{1.25}{0.09}{0.07}\,\msun{}$) and a BH spin close to zero ($\chi_{1,z}=\valpm{0.00}{0.04}{0.04}$).

The spins of the secondaries are unconstrained and no electromagnetic counterpart has been identified for both the events, in agreement with theoretical expectations for \rev{mixed} binaries with low-spinning primary BHs (e.g., \citep{fragione2021,hu2022}, but see \citep{dorazio2021}).

Taking GW200105 and GW200115 as representative of the whole NSBH population, the inferred merger rate is $\valpm{45}{75}{33}\,\gpcyr{}$ \citep{lvk_nsbh_2021}.

As possible formation channels, very similar considerations as for GW190814 apply here. Indeed, in theoretical models, GW190814 may be identified as either a NSBH or a BBH, depending on the mass threshold adopted to distinguish between BHs and NSs. The main difference with GW190814 is that the masses, mass ratios, and inferred rates of GW200105/GW200115-like events are consistent also with the isolated binary formation channel \citep{belc2002,sipior2002,belc2006,belc2016a,dominik2015,eldridge2017,kruckow2018,mapelli2018,neijssel2019,zevin2020,broekgaarden2021,santoliquido2021,giacobbo2018a,broekgaarden2021_gw200115,kinugawa2022,belc2022,zhu2021}.

\section{Summary}\label{sec:conclusions}

Merging compact-object binaries, especially BBHs, have been among the main characters of the astrophysical scene over the last ${\sim} 5$ years. This happened mainly thanks to the discovery of GWs, which gave us access to a new information medium. While the GW catalog continues its inexorable growth, our theoretical knowledge of the formation channels of merging compact objects is still very limited, and most of the parameter space which is relevant for the astrophysical interpretation of GW detections is unexplored.

In this work, we reviewed the main astrophysical processes that may drive the formation of merging compact-object binaries and the lessons learned so far through some exceptional GW events (e.g., GW190814, GW190521). We discussed the degeneracies of the astrophysical models and we showed that some of the uncertainties are large enough to prevent any conclusive statements on merging compact objects. 

The main sources of uncertainties come from: (i) our knowledge of the \textbf{evolution of massive stars} (e.g., clumpiness and porosity of stellar winds, energy transport in radiation-dominated envelopes, overshooting) -- in the next decades, the James Webb Space Telescope and the Extremely Large Telescope will provide crucial insights on the evolution of massive stars, especially those at low metallicity; (ii) the \textbf{SN explosion mechanism} (e.g., explodability dependence on progenitor's structure, fallback amount, lower mass gap, pulsational pair-instability SNe, asymmetries in the ejecta) -- improvements in self-consistent, 3D simulations will help us to shed light on the SN engine and its byproducts (e.g., compact remnants, yields); (iii) \textbf{binary evolution processes} (especially common envelope and the response of donors/accretors to mass transfer) -- comparisons with self-consistent, hydrodynamic simulations of binary stars will be crucial to calibrate the main (free) parameters in population-synthesis simulations; (iv) \textbf{direct \textit{N}-body simulations} of star clusters (e.g., they inherit the uncertainties on single and binary stellar evolution processes, and they are computationally challenging) -- new software, \textit{natively} developed for state-of-the-art computing accelerators (e.g., Graphics Processing Units), and coupled with up-to-date population-synthesis codes, will give us the opportunity to accurately study merging compact objects in very dense stellar environments, possibly up to the regime of dwarf~galaxies.

Meanwhile, detections will not stop, and a stunning collection of GW events will soon be available. The forth observing run (O4) of the LVK collaboration is planned to start later this year (December 2022) \citep{abbott2018}. Furthermore, the next years will see the birth of new, next-generation, ground-based and space-born interferometers (e.g., Einsten Telescope, Cosmic Explorer, LISA) \rev{that promise to reveal} merging compact objects throughout the entire cosmic history and across a much broader frequency range.

Thus, we are heading for exciting times in the field of GWs, with new facilities and detections that will help us to push current astrophysical models beyond the state-of-the-art. In the next decades, such a synergy will be crucial to shed light on the evolutionary histories of merging compact-object binaries across cosmic time.

\vspace{6pt}

\authorcontributions{Conceptualization: M.S. and A.A.T.; data curation: M.S., A.A.T., and M.M; writing---original draft preparation: M.S. and A.A.T.; writing---review and editing, M.S., A.A.T., and M.M.; software: M.S. and A.A.T.; supervision: M.S. All authors have read and agreed to the published version of the manuscript.}

\funding{AAT received support from JSPS KAKENHI Grant Numbers 17H06360, 19K03907 and 21K13914.}

\institutionalreview{{Not applicable.}}

\informedconsent{{Not applicable.}}

\dataavailability{{Not applicable.}} 

\acknowledgments{\rev{We are indebted to Mike Zevin for his careful reading of the review and for providing comments that helped us to improve the manuscript. We thank the anonymous referees for their careful reading of our manuscript and their insightful comments and suggestions.}}

\conflictsofinterest{The authors declare no conflict of interest.} 

\clearpage
\section*{Abbreviations}
\vspace{6pt}

\noindent
The following abbreviations are used in this manuscript:\\

\noindent 
\begin{tabular}{@{}ll}
    AGN& active galactic nuclei\\
   BBH& binary black hole\\
   BH&black hole\\
   BNS&binary neutron star\\
   CO&carbon-oxygen\\
   CE&common envelope\\
   CHE&chemical homogeneous evolution\\
   GC&globular cluster\\
   GW&gravitational wave\\
   GWTC&Gravitational Wave Transient Catalog\\
   ECSN&electron-capture supernova\\
   LIGO&Laser Interferometer Gravitational-wave Observatory\\
   LVC&LIGO-Virgo Collaboration\\
   LVK&LIGO-Virgo-KAGRA\\
   NS&neutron star\\
   NSBH&neutron star-black hole binary\\
   NSC&nuclear star cluster\\
   OC&open cluster\\
   PISN&pair-instability supernova\\
   Pop III&population III\\
   PPISN&pulsational pair-instability supernova\\
   SMBH&super-massive black hole\\
   SN&supernova\\
   TOV&Tolman–Oppenheimer–Volkoff\\
   WD&white dwarf\\
   YDSC&young dense star cluster\\
   ZAMS&zero age main sequence\\
   ZKL&von~Zeipel-Kozai-Lidov\\
\end{tabular}


\appendixtitles{no}

\begin{adjustwidth}{-\extralength}{0cm}

\printendnotes[custom] 
\reftitle{References}

\end{adjustwidth}
\end{document}